\documentclass[fleqn,usenatbib]{mnras}

\usepackage{lineno}
\usepackage{simplewick}
\usepackage{scalerel}
\usepackage[mathscr]{euscript}
\usepackage{extarrows}
\usepackage{caption}
\usepackage{multirow}

\usepackage{newtxtext,newtxmath}
\usepackage[T1]{fontenc}
\usepackage{multicol}
\usepackage{ae,aecompl}

\usepackage{graphicx}	
\usepackage{amsmath}	
\usepackage{hyperref}
\usepackage{subfigure}
\DeclareRobustCommand{\VAN}[3]{#2}
\let\VANthebibliography\thebibliography
\def\thebibliography{\DeclareRobustCommand{\VAN}[3]{##3}\VANthebibliography}

\newcommand{\ba}{\begin{aligned}}
\newcommand{\ea}{\end{aligned}}
\newcommand{\bea}{\begin{eqnarray}}
\newcommand{\eea}{\end{eqnarray}}
\newcommand{\be}{\begin{equation}}
\newcommand{\ee}{\end{equation}}
\newcommand{\ben}{\begin{enumerate}}
\newcommand{\een}{\end{enumerate}}
\newcommand{\bi}{\begin{itemize}}
\newcommand{\ei}{\end{itemize}}
\newcommand{\mr}{\mathrm}

\title{Forecasting the BAO Measurements of the CSST galaxy and AGN Spectroscopic Surveys}

\author[Miao H. et al.]{
Haitao Miao,$^{1}$
Yan Gong,$^{1,2,3}$\thanks{E-mail: gongyan@bao.ac.cn}
Xuelei Chen,$^{1,2,4,5}$
Zhiqi Huang,$^{6,7}$
Xiao-Dong Li$^{6,7}$
Hu Zhan$^{1,8}$
\\
$^{1}$National Astronomical Observatories, Chinese Academy of Sciences, Beijing 100101, People's Republic of China\\
$^{2}$University of Chinese Academy of Sciences, Beijing 100049, People's Republic of China\\
$^{3}$Science Center for China Space Station Telescope, National Astronomical Observatories, Chinese Academy of Sciences, \\20A Datun Road, Beijing 100101, China\\
$^{4}$Department of Physics, College of Sciences, Northeastern University, Shenyang 110819, China\\
$^{5}$Centre for High Energy Physics, Peking University, Beijing 100871, People's Republic of China\\
$^{6}$School of Physics and Astronomy, Sun Yat-sen University, 2 Daxue Road, Tangjia, Zhuhai, 519082, China\\
$^{7}$CSST Science Center for the Guangdong-Hongkong-Macau Greater Bay Area, SYSU, China \\
$^{8}$Kavli Institute for Astronomy and Astrophysics, Peking University, Beijing 100871, China
}

\date{Accepted XXX. Received YYY; in original form ZZZ}

\pubyear{2023}

\begin{document}

\label{firstpage}
\pagerange{\pageref{firstpage}--\pageref{lastpage}}
\maketitle

\begin{abstract}

The spectroscopic survey of the China's Space Survey Telescope (CSST) is expected to obtain a huge number of slitless spectra, including more than one hundred million galaxy spectra and millions of active galactic nuclei (AGN) spectra. By making use of these spectra, we can measure the Baryon Acoustic Oscillation (BAO) signals over large redshift ranges with excellent precisions. In this work, we predict the CSST measurements of the post-reconstruction galaxy power spectra at $0<z<1.2$ and pre-reconstruction AGN power spectra at $0<z<4$, and derive the BAO signals at different redshift bins by constraining the BAO scaling parameters using the Markov Chain Monte Carlo method. Our result shows that the CSST spectroscopic survey can provide accurate BAO measurements with precisions higher than 1\% and 3\% for the galaxy and AGN surveys, respectively. By comparing with current measurements in the same range at low redshifts, this can improve the precisions by a factor of $2\sim3$, and similar precisions can be obtained in the pessimistic case. We also investigate the constraints on the cosmological parameters using the measured BAO data by the CSST, and obtain stringent constraint results for the energy density of dark matter, Hubble constant, and equation of state of dark energy.
\end{abstract}

\begin{keywords}
cosmology: galaxy clustering - cosmology: large-scale structure of Universe - cosmology: observations - quasars: general
\end{keywords}

\section{Introduction}

Nowadays, the observations of the cosmic large-scale structure (LSS) are becoming more and more important.
Various LSS observations provide us an insight into the evolution and components of the Universe \citep{Weinberg2013}. As a main probe of the LSS observations, the Baryon Acoustic Oscillation (BAO) can be an ideal tool used to measure the geometry and expansion rate of the Universe. BAO is an imprint in the distributions of galaxies of primordial sound waves that propagate from the pre-recombination Universe \citep{Peebles1970, Sunyaev1970, Hu1996, Eisenstein1998}. The BAO feature depends on the sound scale at the radiation drag epoch, $r_{\rm drag}$, and the expansion history of the Universe. It provides a standard ruler to probe cosmic distances at different redshifts, and hence allows us to test cosmological models and make precise constraints on the cosmological parameters.

The first detections of the BAO signal were measured in the 2-degree Field Galaxy Redshift Survey \citep[2dFGRS,][]{Cole2005} and Sloan Digital Sky Survey \citep[SDSS,][]{Eisenstein2005}. Then, the BAO feature was further detected in higher precision by 6-degree Field Galaxy Survey \citep[6dFGS,][]{Jones2009}, WiggleZ Dark Energy Survey \citep[WiggleZ,][]{Parkinson2012}, the Baryon Oscillation Spectroscopic Survey \citep[BOSS,][]{Anderson2012,Anderson2014,GilMarin2016,Beutler2017,Beutler2017b}, and the extend BOSS \citep[eBOSS,][]{GilMarin2020, Bautista2021, Tamone2020,Wangy2020,Raichoor2021,deMattia2021,Neveux2020,Hou2021,Zhao2021}. Recently, the Dark Energy Spectroscopic Instrument (DESI) 
has released their data \citep{2023arXiv230606308D}, which will help us to improve the precision and extract more statistical information on the BAO. Combining with other observations, e.g. the cosmic microwave background (CMB), the BAO signal has been extensively applied to constrain the cosmological parameters \citep{Percival2007, Alam2017, Alam2021, Planck2020}.

Traditionally, the BAO signal could be found in the configuration space by the two-point correlation function with a bump, or in the Fourier space by the power spectrum with wiggles. For the clustering analysis of the power spectrum of real data, BAO information is compressed into two Alcock-Paczynski \citep[AP,][]{Alcock1979} scaling factors $\alpha_{\parallel}$ and $\alpha_{\perp}$ for the line elements along and across the line-of-sight direction, respectively. There are basically two methods for analyzing the BAO signal. The first one is based on the full shape of the power spectrum \citep{Sanchez2009,Sanchez2012}, which can make use of all the information contained in the power spectrum to perform the cosmological analysis, including BAO, AP, redshift-space distortions \citep{Ivanov2020}, and the horizon at matter-radiation equality \citep{Philcox2021b,Philcox2022b}. 
For example, the methods that use the matter-radiation equality horizon information from galaxy surveys to determine the Hubble constant \citep[e.g.][]{Philcox2021b}, and researches about full shape analysis based on the Effective Field Theory of LSS (EFTofLSS) are recently discussed \citep{dAmico2020,Philcox2022c,Zhang2022,Ivanov2022,Carrilho2023,Simon2023,Semenaite2023,zhao2023}. 

However, we should note that the BAO is also subject to some non-linear effects. Although these effects are relatively small, they can affect the accuracy of the BAO measurement, which is especially important in the next-generation Stage-IV surveys. In order to reduce the potential error introduced by the nonlinear effects and obtain a more precise measurement, the reconstruction technique was proposed \citep{Eisenstein2007,Eisenstein2007b}. This method can change the broadband shape of the power spectrum, so it is usually used for the BAO-only analysis\footnote{We note that there have been some discussions about the shortcomings of the BAO-only method \citep{Anselmi2019,ODwyer2020}. It is found that the distance uncertainties could be underestimated when the cosmological parameters are fixed to the fiducial values, and the Purely-Geometric BAO method is proposed to address this issue.} \citep{Padmanabhan2012,Anderson2012}. Although some shape information of the power spectrum is lost, the BAO-only analysis enables us to obtain more precise information about BAO. 
Besides, the joint analysis of the pre-reconstruction (i.e. full shape) and post-reconstruction was also proposed \citep{Philcox2020b}, and was further studied with different methods \citep{GilMarin2022,Chen2022b}.

As we know, galaxies are one of the most important tracers for detecting BAO signals. Given the spatial distribution of galaxies, we can derive the matter distribution of the Universe in an effective way.
In addition to galaxy, active galactic nuclei (AGN) can also be used to probe the large-scale distribution of underlying dark matter. Because of their high luminosity, AGNs can be observed at very high redshifts, and could map the matter distribution over a large redshift range given a certain number density. Owing to the enormously large and deep spatial volume that could be probed by recent surveys, AGN is becoming an increasingly important tracer in studies of the LSS, and the analysis of AGN clustering is more and more practical in recent years. In these studies, the clustering of quasars are especially explored, which are usually categorized as a subclass of the more general category of AGN. In photographic images, these objects are nearly point-like or quasi-stellar in appearance, and are classified as quasars. Since quasars are usually very luminous, their clustering can represent the clustering of AGNs, especially at high-z \citep{Ata2018,Hou2018,Smith2020,Neveux2020,Hou2021,Merz2021,Neveux2022,Chudaykin2023,Simon2022}. Especially, with the advance of Stage-IV galaxy surveys, AGN precise cosmology is coming \citep{Bargiacchi2022}.

In the next decade, ongoing or upcoming next-generation galaxy surveys such as Vera C. Rubin Observatory \citep[or LSST,][]{Ivezic19}, Nancy Grace Roman Space Telescope (RST) \citep[or WFIRST,][]{Akeson19}, Euclid \citep{Laureijs11,Amendola2018}, and China Space Station Telescope (CSST) \citep{Zhan2011SSPMA,Zhan2018,Zhan2021,Gong19,Miao2023} will perform wider and deeper observations. For instance, the CSST will cover a 17500\,$\mr{deg^2}$ survey area in ten years, and complete photometric and slitless spectroscopic surveys simultaneously with multi-band photometric imaging and slitless gratings. It has seven photometric bands ($NUV, u, g, r, i, z, y$) and three spectroscopic bands ($GU, GV, \mr{~and~} GI$), covering the wavelength range from $\sim250$-$1000$ nm. The CSST photometric bands can achieve a magnitude limit of $i\simeq26$ mag, and about 23 mag with a spectral resolution $R\simeq200$ for the three spectroscopic bands. CSST is expected to obtain more than one hundred million galaxy spectra and millions of AGN spectra, respectively. 

Besides, other surveys, such as Euclid, RST and DESI, will also perform the galaxy spectroscopic observation. The CSST will have a similar survey area as Euclid (15000 deg$^2$) and DESI (14000 deg$^2$), which is much larger than the RST (2000 deg$^2$). For the wavelength coverage, since the CSST has $NUV$ and $u$ near-ultraviolet bands, it has good complementarity with Euclid and RST that have near-infrared bands. In addition, the CSST also has more photometric bands, which is helpful for calibrating the slitless spectroscopic redshift, and it could effectively suppress systematics and obtain accurate measurement on galaxy clustering.


In this work, we explore the BAO-only method based on the post-reconstruction galaxy and pre-reconstruction AGN power spectra, and study the measurements of the scaling factors $\alpha_{\parallel}$ and $\alpha_{\perp}$, as well as the constraints on the relevant cosmological parameters by the CSST spectroscopic surveys. In Section~\ref{Sec2}, we briefly introduce the Lagrangian perturbation theory and reconstruction. The theoretical forecast of the CSST galaxy and AGN distribution is given in Section~\ref{Sec3}. We also discuss the generation of the galaxy and AGN mock data from the theoretical power spectra in this section. The Bayesian analysis of the BAO scaling parameters and cosmological parameters by those mock data are discussed in Section~\ref{Sec4}. We show our results in Section~\ref{Sec5}, and the conclusions are given in Section~\ref{Sec6}. We adopt a flat Universe with $\{\Omega_{\mathrm{c}}, \Omega_{\mathrm{b}}, h, A_{\mathrm{s}}, n_{\mathrm{s}}\} = \{0.264,\ 0.049,\ 0.673,\ 2.099\times10^{-9},\ 0.965\}$ as the fiducial cosmology \citep{Planck2020}. Note that the curvature parameter is actually important in the data analysis of BAO measurements \citep{Anselmi2023a,Anselmi2023b,Aubourg2015,Alam2021}, and here we assume it equals to zero for simplicity.

\section{Lagrangian perturbation theory and Reconstruction} \label{Sec2}

In this section, we briefly introduce the Lagrangian perturbation theory (LPT) and reconstruction method that is used to reconstruct the galaxy power spectrum for the BAO analysis.

\subsection{Background of LPT}

The LPT has been extensively applied to relevant cosmological studies \citep{Zeldovich1970,Buchert1989,Buchert1992,Hivon1995,Taylor1996,Bernardeau2002,Matsubara2008,Matsubara2008b,Matsubara2015,Carlson2013,White2014,McQuinn2016,Vlah2015b,Vlah2015,Vlah2016,Vlah2019,Tassev2014b,Tassev2014,Chen2019,Chen2019b,Chen2020,Chen2021,Schmidt2021,Chen2022,Kokron2022,DeRose2023}. In the Lagrangian scenario, the perturbation of a cosmological fluid element located at position $\mathbf{q}$ at some conformal time $t$ is described by a displacement field $\boldsymbol{\Psi}(\mathbf{q}, t)$, which maps a fluid element from initial Lagrangian coordinates $\mathbf{q}$ to Eulerian coordinates $\mathbf{x}$ by $\mathbf{x}(\mathbf{q}, t)=\mathbf{q}+\mathbf{\Psi}(\mathbf{q}, t)$. 

The dynamic of the displacement field is determined by the equation $\ddot{\boldsymbol{\Psi}}(\mathbf{q})+\mathcal{H} \dot{\boldsymbol{\Psi}}(\mathbf{q})=-\nabla_{\mathbf{x}} \Phi(\mathbf{x})$, where $\Phi(\mathbf{x})$ is the gravitational potential, $\mathcal{H}$ is the conformal Hubble parameter, and dots represent derivatives to the conformal time $t$. The gravitational potential follows the Poisson equation $\nabla^{2}\Phi(\mathbf{x},t)=\frac{3}{2}\Omega_{m}(t)\mathcal{H}^{2}(t)\delta(\mathbf{x},t)$]. The Lagrangian displacement is given by Taylor expansion $\boldsymbol{\Psi} = \sum\limits_{n=1}\boldsymbol{\Psi}^{(n)}$ in the initial overdensity $\delta_0(\mathbf{q})$, and we have the linear solution $\boldsymbol{\Psi}^{(1)} = -D(z)\nabla_{\mathbf{q}}^{-1}\delta_{0}(\mathbf{q})$, i.e. the so-called Zeldovich approximation, which only considers the linear order term of $\Psi$ but resums the effects of the displacement of all orders in a Galilean-invariant manner. For a statistically uniform initial density field, the connection of the Eulerian and Lagrangian coordinates is given by continuity relation $\rho(\mathbf{x}) d^3 \mathbf{x}= \Bar{\rho} d^3 \mathbf{q}$, where $\bar{\rho}$ represents the mean density in comoving coordinates. Based on this relation, the matter overdensity $\delta_m$ is given by 

\begin{equation}
\begin{aligned}
1+\delta_m(\mathbf{x})=\int d^3 \mathbf{q} \delta_D(\mathbf{x}-\mathbf{q}-\boldsymbol{\Psi}(\mathbf{q}))\,, \\
\quad(2 \pi)^3 \delta_D(\mathbf{k})+\tilde{\delta}_m(\mathbf{k})=\int d^3 \mathbf{q} e^{i \mathbf{k} \cdot(\mathbf{q}+\boldsymbol{\Psi})}\,.
\end{aligned}
\end{equation}

In fact, we could not directly observe the potential matter distribution, but rather the biased tracers in the non-linear density field. In the analysis of LSS, one can perturbatively expand the observed galaxy or AGN density field relying on the perturbation approach \citep{McDonald2009,Desjacques2018}. Considering biased tracers, $a$, within the Lagrangian framework, the initial overdensities are modeled as $F^{a}(\mathbf{q})=F^{a}\left[\partial^{2}\Phi(\mathbf{q}),\ldots\right]$ at Lagrangian positions $\mathbf{q}$, and the observed overdensities are given by
\begin{equation}
1+\delta^a(\mathbf{x}, t)=\int d^3 \mathbf{q} F^a(\mathbf{q}) \delta_D(\mathbf{x}-\mathbf{q}-\boldsymbol{\Psi}(\mathbf{q}, t))\,.
\end{equation}
Then the cross-power spectrum between different biased traces is given by
\be \label{eq3}
P^{ab}(k)=\int d^{3}\mathbf{q}e^{i\mathbf{k}\cdot\mathbf{q}}\langle F^{a}(\mathbf{q}_{2})F^{b}(\mathbf{q}_{1})e^{i\mathbf{k}\cdot\Delta^{ab}}\rangle_{q=|q_{2}-q_{1}|}\,,
\ee
where$\Delta^{ab}=\Psi^a(\mathbf{q}_2)-\Psi^b(\mathbf{q}_1)$, and the expectation value should only depend on $q=|q_{2}-q_{1}|$, due to the statistical isotropy.
The bias functionals, $F^{a,b}$, can be written as Taylor power in terms of bias coefficients
\be
\ba
F^a(\mathbf{q})&=1+b_1^a\delta_0(\mathbf{q})+\frac{1}{2}b_2^a\left(\delta_0(\mathbf{q})^2-\langle\delta_0^2\rangle\right) \\
&+b_s^a\left(s^2(\mathbf{q})-\langle s^2\rangle\right)+b_{\nabla^2}^a\nabla_q^2\delta_0(\mathbf{q})+\cdots\,,
\ea
\ee
where $s^{2}=s_{ij}s_{ij}$ is the square of the shear tensor. $b_{\nabla^2}^a$ is the derivative bias that corrects the bias expansion at scales close to the halo radius.

Here we consider modeling reconstruction within the Zeldovich approximation. The final expression of the cross power spectrum is given by calculating the exponential term in equation (\ref{eq3}) via the cumulant expansion and evaluating the bias expansion using functional derivatives \citep{Matsubara2008b,Carlson2013,Vlah2016b,Chen2019b}, and then we have
\bea
\begin{aligned}
P^{ab}(k)&=\int d^{3}\mathbf{q}e^{i\mathbf{k}\cdot\mathbf{q}}e^{-\frac{1}{2}k_{i}k_{j}A_{ij}^{ab}}\left[1+\alpha_{0}k^{2}+ib_{1}^{b}\mathbf{k}\cdot U^{a} \right. \\
&+ib_{1}^{a}\mathbf{k}\cdot U^{a}+b_{1}^{a}b_{1}^{b}\xi_{L}+\frac{1}{2}b_{2}^{a}b_{2}^{b}\xi_{L}^{2} \\
& -\frac{1}{2}k_{i}k_{j}(b_{2}^{b}U_{i}^{a}U_{j}^{a}+b_{2}^{a}U_{i}^{b}U_{j}^{b}+2b_{1}^{a}b_{1}^{b}U_{i}^{a}U_{j}^{b}) \\
&+ik_{i}(b_{2}^{b}b_{1}^{a}U_{i}^{a}+b_{1}^{b}b_{2}^{a}U_{i}^{b})\xi_{L} \\
&-\frac{1}{2}k_{i}k_{j}(b_{s}^{a}\Upsilon_{ij}^{b}+b_{s}^{a}\Upsilon_{ij}^{b})+ik_{i}(b_{1}^{a}b_{s}^{b}V_{i}^{ab}+b_{1}^{b}b_{s}^{a}V_{i}^{ba}) \\
&+\left.\frac{1}{2}(b_{2}^{a}b_{s}^{b}+b_{2}^{b}b_{s}^{a})\chi^{12}+b_{s}^{a}b_{s}^{b}\zeta+\cdots\right]\,,
\end{aligned}
\eea
where
\be
\ba
&A_{ij}^{ab}=\langle\Delta_i^{ab}\Delta_j^{ab}\rangle\,,U_i^b=\langle\Delta_i^{ab}\delta_0(\mathbf{q}_2)\rangle\,,\xi_L=\langle\delta_0(\mathbf{q}_2)\delta_0(\mathbf{q}_1)\rangle\,, \\
&\zeta=\langle s^2(\mathbf{q}_2)s^2(\mathbf{q}_1)\rangle\,,\Upsilon_{ij}^b=\langle\Delta_i^{ab}\Delta_j^{ab}s^2(\mathbf{q}_2)\rangle\,, \\
&V_i^{ab}=\langle\Delta_i^{ab}\delta_0(\mathbf{q}_2)s^2(\mathbf{q}_1)\rangle\,,\chi^{12}=\langle\delta_0^2(\mathbf{q}_1)s^2(\mathbf{q}_2)\rangle \,.
\ea
\ee
The two-point functions of vector and tensor defined above can be decomposed into scalar components, e.g. $A_{ij}=X(q)\delta_{ij}+Y(q)\hat{q}_{i}\hat{q}_{j}\mathrm{~and~}U_{i}=U(q)\hat{q}_{i}$, via rotational symmetry.

In redshift space, for Zeldovich approximation, the Lagrangian displacements are replaced by $\boldsymbol{\Psi}\to\boldsymbol{\Psi}+(\hat{n}\cdot\mathbf{v})\hat{n}/\mathcal{H}$, where $\hat{n}$ represents the line of sight (LOS) direction. Within the Einstein-de Sitter approximation and considering the assumption of plane parallel approximation, we could have a further simplification that
 $\boldsymbol{\Psi}_{i}\to\boldsymbol{\Psi}_{i}^{R}=R_{ij}\boldsymbol{\Psi}_{j}$, where $R_{ij}=\delta_{ij}+f\hat{n}_{i}\hat{n}_{j}$, and $f = \mathrm{d}\ln D(z)/\mathrm{d}a$ is the linear growth rate.

\subsection{Reconstruction}

Although the BAO is robust as a standard ruler to measure the expansion of the Universe, it can be affected by the non-linear structure evolution, which will degrade the BAO feature and erase the higher harmonics
in the power spectrum \citep{Meiksin1999,Springel2005,White2005,Seo2005,Jeong2006,Eisenstein2007,Crocce2006,Crocce2008,Angulo2008,Seo2008,Taruya2009,Sherwin2012,Senatore2015,Vlah2016b,Blas2016,Ding2018}. 
In order to improve the BAO measurement precision, the density field reconstruction technique \footnote{Since the reconstruction technique usually fixes the values of the linear bias and growth rate parameters, it may introduce unquantified information to the real data, and could lead to some problems in uncertainty estimation \citep{Anselmi2023b}.} was proposed \citep{Eisenstein2007b}, and then reanalyzed within the framework of the LPT \citep{Padmanabhan2009,Noh2009}.
It has been widely used for the analysis of real observational data \citep{Padmanabhan2012,Anderson2012,Anderson2014,Burden2014,Kazin2014,Ross2015,Beutler2016,GilMarin2016,Alam2017}.
There are also considerable literatures further exploring BAO reconstruction, such as a reconstruction algorithm in Eulerian frame \cite{Schmittfull2015}, the iterative methods \citep{Schmittfull2017,Zhu2017,Yu2017,Wang2017,Hada2018,Wang2020,Ota2021,Ota2023,Seo2022,ChenX2023}, the Laguerre reconstruction algorithm \citep{Nikakhtar2021}, and the optimal transport theory \citep{Hausegger2022,Nikakhtar2022,Nikakhtar2023}.

In addition, an analytical method for reconstruction built on the Zeldovich approximation was proposed \citep{White2015,Chen2019b}. Compared to other methods mentioned above, it includes a complete set of counterterms and bias terms up to quadratic order, and can derive more accurate results. Here we follow \cite{Chen2019b} and \cite{White2015}, and will only adopt the ``Rec-Sym'' method, which indicates a symmetric treatment of $\delta_\mathrm{d}$ and $\delta_\mathrm{s}$. The reconstruction is performed in the following steps \citep{Padmanabhan2009,White2015,Chen2019b}:
\begin{itemize}
    \item Smooth the density field with a kernel $\mathcal{S}(k)=\exp[-(kR_{\rm s})^{2}/2]$ to filter out small-scale modes, where $R_{\rm s}$ is a Gaussian smoothing scale and set to be $R_{\rm s} = 15$\,$h^{-1}\,{\rm Mpc}$ \citep{White2015}.
    
    \item Based on the smoothed density field, compute the shift field, $\chi$, in redshift space using the Zeldovich approximation. 
	It was calculated by dividing the smoothed galaxy density field by an Eulerian bias $b_1^{E}$ and a linear RSD factor, and then taking the inverse gradient. In Fourier space, it corresponds to 
    \begin{equation}
    \chi_{\mathbf{k}}=-\frac{i \mathbf{k}}{k^2} \mathcal{S}(k)\left(\frac{\delta_g(\mathbf{k})}{b_1^{E}+f \mu^2}\right) \approx-\mathcal{S}(k) \Psi^{(1)}(\mathbf{k})\,,
    \end{equation}
    where $\mu=\bar{n}\cdot\bar{\hat{k}}$ is the cosine of the angle of light-of-sight. 
    
    \item Shift galaxies by $\chi_{\mathrm{d}}=\mathbf{R}\chi$, where the matrix $\mathbf{R}$ means mapping the density field to redshift space and computing the displaced density field, $\delta_d$.
    
    \item The same as galaxies, shift an initially spatially uniform distribution of particles (i.e. reference field) by $\chi_{\mathrm{s}}=\mathbf{R}\chi$ to form the "shifted" density field, $\delta_{\rm s}$.

    \item The two density fields, i.e. the displaced field of galaxies and shifted field of the reference field, are derived, respectively, and the reconstructed density field is given by $\delta_{\mathrm{r}}\equiv\delta_{\rm d}-\delta_{\rm s}$ with the power spectrum $P_{\mathrm{r}}(k)\propto\langle\left|\delta_{\mathrm{r}}^{2}\right|\rangle$.
    
\end{itemize}

\subsection{Reconstructed power spectrum}\label{recon}

As shown in \cite{Chen2019b}, the reconstructed power spectrum is given by $P_{\mathrm{r}}=P^{dd}+P^{ss}-2P^{ds}$, where $P^{dd}$ and $P^{ss}$ are the auto-spectra of displaced shifted fields and $P^{ds}$ is the cross-spectra. Within the Lagrangian framework, the displaced density field is given by
\bea
\begin{aligned}
&1+\delta_{d}(\mathbf{r}) =\int d^{3}\mathbf{x}\left(1+\delta(\mathbf{x})\right)\delta_{D}\left[\mathbf{r}-\mathbf{x}-\chi_{d}(\mathbf{x})\right]  \\
&=\int d^{3}\mathbf{x}\int d^{3}\mathbf{q}F(\mathbf{q})\delta_{D}\left[\mathbf{x}-\mathbf{q}-\Psi(\mathbf{q})\right]\delta_{D}\left[\mathbf{r}-\mathbf{x}-\chi_{d}(\mathbf{x})\right] \\
&=\int d^{3}\mathbf{q}F(\mathbf{q})\delta_{D}\left[\mathbf{r}-\mathbf{q}-\Psi(\mathbf{q})-\chi_{d}(\mathbf{q}+\Psi(\mathbf{q}))\right]\,.
\end{aligned}
\eea
Here, the displacement, $\Psi$, is evaluated in the Lagrangian coordinate and the shift field is evaluated at the shifted Eulerian coordinate. When the appropriate shift field $\chi_{\mathrm{d}}$ is given, one can further generalize the above equalities in the redshift space with the map $\Psi\to\mathrm{R}\Psi$. By the Fourier transformation, it can be translated to
\bea
\begin{aligned}
&(2\pi)^{3}\delta_{D}(\mathbf{k})+\delta_{d}(\mathbf{k}) =\int d^{3}\mathbf{q}e^{-i\mathbf{k}\cdot\mathbf{q}}F(\mathbf{q})e^{-i\mathbf{k}\cdot\left[\Psi(\mathbf{q})+\chi_{d}(\mathbf{q}+\Psi(\mathbf{q}))\right]},  \\
&(2\pi)^{3}\delta_{D}(\mathbf{k})+\delta_{s}(\mathbf{k}) =\int d^{3}\mathbf{q}e^{-i\mathbf{k}\cdot\mathbf{q}}e^{-i\mathbf{k}\cdot\chi_{s}(\mathbf{q})}\,. 
\end{aligned}
\eea
When the approximation $\chi(\mathbf{q}+\boldsymbol{\Psi})\approx\chi(\mathbf{q})$ is adopted, the displaced and shifted field can be described as tracers with displacements
\begin{equation}
\qquad\qquad\qquad \boldsymbol{\Psi}^d=\boldsymbol{\Psi}+\chi_d\,, \qquad \boldsymbol{\Psi}^s=\chi_s\,.
\end{equation}

In real space, the displaced and shifted fields are moved the same smoothed negative Zeldovich displacement, i.e. $\chi_{d}=\chi_{s}=-S*\Psi$. So in Fourier space, we have
\begin{equation}
\qquad\boldsymbol{\Psi}^d(\mathbf{k})=[1-\mathcal{S}(k)] \boldsymbol{\Psi}(\mathbf{k}), \quad \boldsymbol{\Psi}^s(\mathbf{k})=-\mathcal{S}(k) \boldsymbol{\Psi}(\mathbf{k})\,.
\end{equation}
Given that the map from real space to redshift space by a matrix factor $R_{ij}$, the smoothed and displaced fields with Fourier modes in redshift space can be written as 
\be
\qquad\mathbf{\Psi}^d(\mathbf{k})=\left[1-\mathcal{S}(k)\right]\mathbf{R}\boldsymbol{\Psi}(\mathbf{k}),\quad\boldsymbol{\Psi}^s(\mathbf{k})=-\mathcal{S}(k)\mathbf{R}\boldsymbol{\Psi}(\mathbf{k}).
\ee

\begin{figure}
    \centering
    \includegraphics[width=0.5\textwidth]{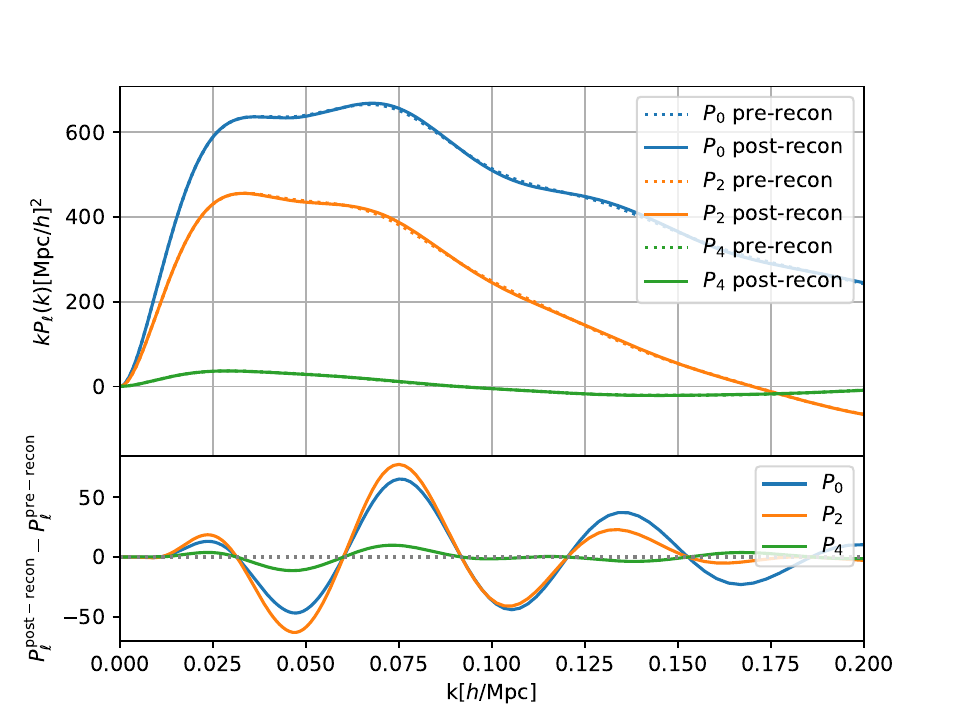}
    \caption{The multipoles of galaxy power spectra at $z=0.15$ for pre-reconstruction (dotted curves) and post-reconstruction (solid curves) cases. The differences between the two kinds of power spectra are shown in the lower panel.}
    \label{pkmodel}
\end{figure}

A complete reconstructed power spectrum calculation can be found in \cite{Chen2019b}. Here, we only adopt the form after decomposing the power spectrum into wiggle and no-wiggle parts. We use the method of splitting the power spectrum from \cite{Hinton2017}. When we get a linear power spectrum, for example, it can be split as $P_{\mathrm{L}}(k) = P_{\mathrm{nw}}(k) + P_{\mathrm{w}}(k)$. For the reconstructed power spectrum, the no-wiggle parts reproduce the broadband depending on the linear power spectrum. And we can get the wiggle parts from \cite{Chen2019b} under some approximations, that we have
\bea
\begin{aligned}
P^{dd}_{\mathrm{w}}(\mathbf{k}) \approx &
e^{-\frac{1}{2}K^{2}\Sigma_{dd}^{2}}\left[(1+f\mu^{2})(1-\mathcal{S}(k))+b_{1}\right]^{2}P_{\mathrm{w}}(k)\,, \\
P_{\mathrm{w}}^{ds}(\mathbf{k})=&-e^{-\frac{1}{2}K^{2}\Sigma_{ds}^{2}}\bigg((1+f\mu^{2})(1-\mathcal{S}(k))+b_{1}\bigg) \\
&\times(1+f\mu^{2})\mathcal{S}(k)P_{\mathrm{w}}(k)\,, \\
P_{\mathrm{w}}^{ss}(\mathbf{k})=&e^{-\frac{1}{2}K^{2}\Sigma_{ds}^{2}}(1+f\mu^{2})^{2}\mathcal{S}(k)^{2}P_{\mathrm{w}}(k)\,,
\end{aligned}
\eea
where $K^{2} = (1+f(f+2)\mu^{2})k^{2}$, the bias $b_1$ is relate to Eulerian bias by $b_1 = b_1^{E} - 1$, and $\Sigma_{dd}^{2}$, $\Sigma_{ds}^{2}$, $\Sigma_{ds}^{2}$ could also be found as \citep{Ding2018}
\begin{equation}
\begin{aligned}
\Sigma_{d d}^2(q,z)= & \frac{2}{3} \int \frac{\mathrm{d} k}{2 \pi^2}\left(1-j_0(q )\right)(1-S(k))^2 P_L(k, z)\,, \\
\Sigma_{s d}^2(q,z)= & \frac{2}{3} \int \frac{\mathrm{d} k}{2 \pi^2}\left(\frac{1}{2}\left(S(k)^2+(1-S(k))^2\right)\right. \\
& \left.+j_0(q k)(1-S(k)) S(k)\right) P_L(k, z) \,, \\
\Sigma_{s s}^2(q,z)= & \frac{2}{3} \int \frac{\mathrm{d} k}{2 \pi^2}\left(1-j_0(q k)\right) S(k)^2 P_L(k, z)\,.
\end{aligned}
\end{equation}
Then, the reconstructed power spectrum is given by 
\bea
\begin{aligned}
    P_{\mathrm{r}} = & \left[\left((1+f\mu^{2})(1-\mathcal{S}(k))+b_{1}\right)^{2}\left(P_{\mathrm{nw}}+e^{-\frac{1}{2}K^{2}\Sigma_{dd}^{2}}P_{\mathrm{w}}(k)\right) \right.\\
    & + (1+f\mu^{2})^{2}\mathcal{S}(k)^{2}\left(P_{\mathrm{nw}}+e^{-\frac{1}{2}K^{2}\Sigma_{ds}^{2}}P_{\mathrm{w}}(k)\right) \\
    & -2(1+f\mu^{2})\mathcal{S}(k)\bigg((1+f\mu^{2})(1-\mathcal{S}(k))+b_{1}\bigg) \\
    & \left. \left(P_{\mathrm{nw}}+e^{-\frac{1}{2}K^{2}\Sigma_{ds}^{2}}P_{\mathrm{w}}(k)\right)\right] \exp\left[-(k\mu\Sigma_{\mathrm{FoG}})^{2}\right].
\end{aligned}
\eea

We show the multipoles of the pre-reconstruction and post-reconstruction galaxy power spectrum at $z=0.15$ in Fig.~\ref{pkmodel}. The linear multipoles of the power spectra are calculated by CAMB \citep{Lewis2000}. Here, we also consider the Fingers-of-God (FoG) effect, and adopt a redshift-dependent value $\Sigma_{\rm FoG} = 7/(1+z)\,h^{-1}\mathrm{Mpc}$ \citep{Gong19}.

\begin{figure}
    \centering
    \includegraphics[width=0.5\textwidth]{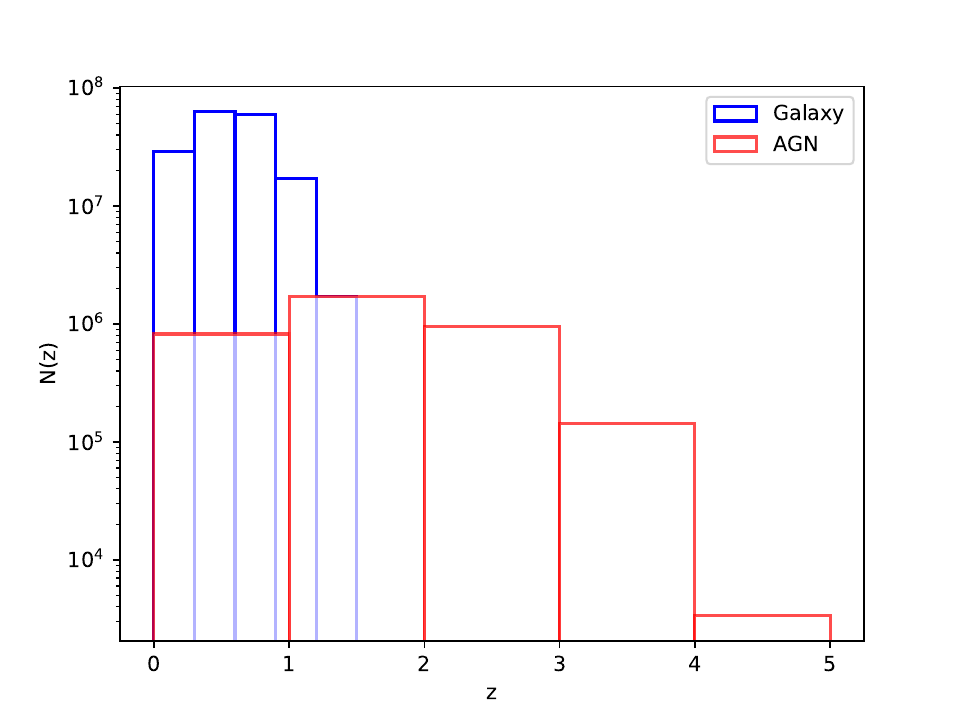}
    \centering
    \caption{The expected galaxy and AGN redshift distributions in the CSST spectroscopic survey are shown in blue and red histograms, respectively. We divide the redshift ranges into five bins for these two surveys.}
    \label{fig_distribution}
\end{figure}


\begin{table}
	\caption{The galaxy and AGN surface and volume number densities, and Eulerian biases in the five redshift bins for the CSST spectroscopic survey.}
	\begin{center}
	\scalebox{0.85}{
		\begin{tabular}{c c c c c c}
			\hline
			\hline
			$z_{\rm min}$ & $z_{\rm max}$ & $z_{\rm mean}$ & d$N$/d$\Omega{\rm[arcmin^{-2}]}$ & $\bar{n}(z)[h^3{\rm Mpc^{-3}}]$ & $b(z)$  \\
			\hline
            & & &Galaxy& & \\
            \hline
			0 & 0.3 & 0.15 & 0.46 & 2.82 $\times 10^{-2}$ & 0.91 \\
			0.3 & 0.6 & 0.45 & 1.00 & 1.17 $\times 10^{-2}$ & 1.06 \\
			0.6 & 0.9 & 0.75 & 0.95 & 5.7 $\times 10^{-3}$ & 1.23 \\
			0.9 & 1.2 & 1.05 & 0.27 & 1.17 $\times 10^{-3}$ & 1.41 \\
			1.2 & 1.5 & 1.35 & 0.03 & 9.79 $\times 10^{-5}$ & 1.60 \\
			\hline
             & & &AGN & &  \\
			\hline
			0 & 1.0 & 0.5 & 1.72 $\times 10^{-2}$ & 3.85 $\times 10^{-5}$ & 1.20 \\
			1.0 & 2.0 & 1.5 & 3.67 $\times 10^{-2}$ & 2.85 $\times 10^{-5}$ & 2.31 \\
			2.0 & 3.0 & 2.5 & 2.09 $\times 10^{-2}$ & 1.39 $\times 10^{-5}$ & 3.98 \\
			3.0 & 4.0 & 3.5 & 3.65 $\times 10^{-3}$ & 2.23 $\times 10^{-6}$ & 6.20 \\
			4.0 & 5.0 & 4.5 & 1.11 $\times 10^{-4}$ & 5.99 $\times 10^{-8}$ & 8.98 \\
			\hline
		\end{tabular}
		}
	\end{center}
	\label{number_density}
\end{table}


\section{Mock Data}\label{Sec3}

Here, we consider two tracers of the matter density field, i.e. galaxy and AGN. The CSST spectroscopic observation can measure more than one hundred million galaxy spectra as shown in previous studies \citep[e.g.][]{Gong19}, and is also expected to identify millions of AGNs covering large redshift range, based on the CSST multi-band photometric survey.

\subsection{Galaxy and AGN Mock Catalogs}

For the CSST galaxy spectroscopic survey, we take the galaxy redshift distribution given in \cite{Gong19} and \cite{Miao2023}. It is created based on the zCOSMOS catalog \citep{Lilly2007,Lilly2009} in 1.7\,$\mathrm{deg^2}$ with a magnitude limit $I_\mathrm{AB}\sim22.5$. This catalog contains more than 20,000 galaxies, and about 16,600 sources have high-quality spectroscopic redshifts (spec-$z$). The derived galaxy surface and volume number densities of the CSST spectroscopic survey are listed in Table \ref{number_density},  and the galaxy redshift distribution is plotted in Figure \ref{fig_distribution}. The galaxy distribution was given in five tomographic bins with a redshift range from 0 to 1.5. The main targets of the CSST spectroscopic survey will be the emission line galaxies (ELGs) with ${\rm H}{\alpha}$, ${\rm[OIII]}$ and [OII] emission lines.

For the AGN survey, CSST will significantly increase the number of observed AGNs by detecting and identifying them with the photometric and spectroscopic surveys. 
For calculating the expected number of AGN, we utilize the quasar luminosity function (QLF) from \cite{Palanque2016a}. The QLF is given in g-band, which is often fitted by a double power law \citep{Boyle2000,Richards2006}:
\begin{equation}
\Phi\left(M_g, z\right)=\frac{\Phi^*}{10^{0.4(\alpha+1)\left(M_g-M_g^*\right)}+10^{0.4(\beta+1)\left(M_g-M_g^*\right)}}\,,
\end{equation}
where $M_g^*$ represents a characteristic or break magnitude. The slopes $\alpha$ and $\beta$ describe the evolution of the QLF on either side of the break magnitude. The slope $\alpha$ reproduces the bright end part of the QLF, and $\beta$ is for the faint end. Here, we have chosen to convert all the AGNs and their selection functions to the absolute AB magnitude at a g-band wavelength,
\be
M_{g}\left(z\right)=m-5\log\left(\frac{d_{L}\left(z\right)}{\text{Mpc}}\right)-25-K_{m,g}\left(z\right)\,,
\ee
where 
\be
d_L(z)=(1+z)\frac{c}{H_0}\int_0^z\frac{\mathrm{d}z'}{\sqrt{\Omega_{\mathfrak{m}}(1+z')^3+\Omega_{\Lambda}}}\,,
\ee
and $K_{m,g}\left(z\right)\ $ is the $K$ correction \citep{McGreer2013,Caditz2017}. 
Considering the pure luminosity-evolution (PLE) model \citep{Croom2009}, a redshift dependence of the luminosity is introduced through an evolution in $M_g^*$ given by 
\begin{equation}
M_g^*(z)=M_g^*\left(z_{\mathrm{p}}\right)-2.5\left[k_1\left(z-z_{\mathrm{p}}\right)+k_2\left(z-z_{\mathrm{p}}\right)^2\right]\,,
\end{equation}
where $z_{\mathrm{p}} = 2.2$ is a pivot redshift. The redshift-evolution parameters ($k_1$ and $k_2$) and the slopes parameters ($\alpha$ and $\beta$) could be different on either side of the pivot redshift. The PLE model contains ten free parameters $[\Phi^{*},M_{g}^{*}(z_{\mathrm{p}}),\alpha_{l},\beta_{l},k_{1l},k_{2l},\alpha_{h},\beta_{h},k_{1h},\mathrm{~and~}k_{2h}]$,  which are fit using eBOSS data \citep{Dawson2016}. The best-fit values of these parameters are given in \cite{Palanque2016a} and corrected in \cite{Caditz2017} due to the different k-correction. We use the corrected best-fit values that are given in Table ~\ref{fits}.

\begin{table}
\caption{The values of the free parameters we adopt in the PLE models of the QLF.}
\begin{center}
\begin{tabular}{ccccc}
\hline\hline
Redshift &\multicolumn{2}{c}  { \multirow{2}{*}{Parameters} } & \\
 range\\
\hline\\[-8pt]

 &$M_g^*(z_p)$& $\log (\Phi^*)$& \\
 $0.68-5.0$ &$-26.5{\scriptstyle \pm0.04}$ & $-5.81{\scriptstyle \pm0.01}$ & & \\
 \hline
  & $\alpha$& $\beta$&$k_1$&$k_2$ \\
$0.68-2.2$ & $-3.4{\scriptstyle \pm0.19}$&$-1.53{\scriptstyle \pm0.25}$ & $-0.03{\scriptstyle \pm 0.02}$ & $-0.35{\scriptstyle \pm 0.02}$\\

 $2.2-5.0$ & $-2.62{\scriptstyle \pm 0.12}$ & $-1.48{\scriptstyle \pm 0.05}$ & $-0.36{\scriptstyle \pm 0.04}$ &$0.01{\scriptstyle \pm 0.03}$ \\
 
\hline
  \end{tabular}
  \end{center}
\label{fits}
\end{table}

Based on the above QLF, one can assess the number of observed AGNs by
\be
N_{\rm AGN} = \int {\rm d}M_g \int{\rm d}z \,\Phi\,\frac{{\rm d}V}{{\rm d}z}\,,
\ee
where $M_g$ is the magnitude in the g-band. 
The comoving volume element $\frac{{\rm d}V}{{\rm d}z}$ is given by 
\begin{equation}
\frac{\mathrm{d} V}{\mathrm{~d} z}=\frac{\mathrm{d} V}{\mathrm{~d} z \mathrm{~d} \Omega} \times A \times \frac{4 \pi}{41253}\,,
\end{equation}
where A is the survey area in $\mathrm{deg}^2$, and 
\be
\frac{\mathrm{d}V}{\mathrm{d}z\mathrm{d}\Omega}=\frac{c}{H_0}\frac{d_L^2\left(z\right)}{\left(1+z\right)^2\left[\Omega_\mathrm{m}\left(1+z\right)^3+\Omega_\Lambda\right]^{1/2}}\,,
\ee
represents the comoving volume element per unit solid angle. We find that more than four million AGNs can be identified by the CSST, and the AGN redshift distribution is plotted in Figure \ref{fig_distribution}. The AGN surface and volume number densities at different redshift bins from $z=0$ to 4 are given in Table \ref{number_density}.

\begin{figure*}

	\centering
	\subfigure{
		\centering
		\includegraphics[width=0.49\textwidth]{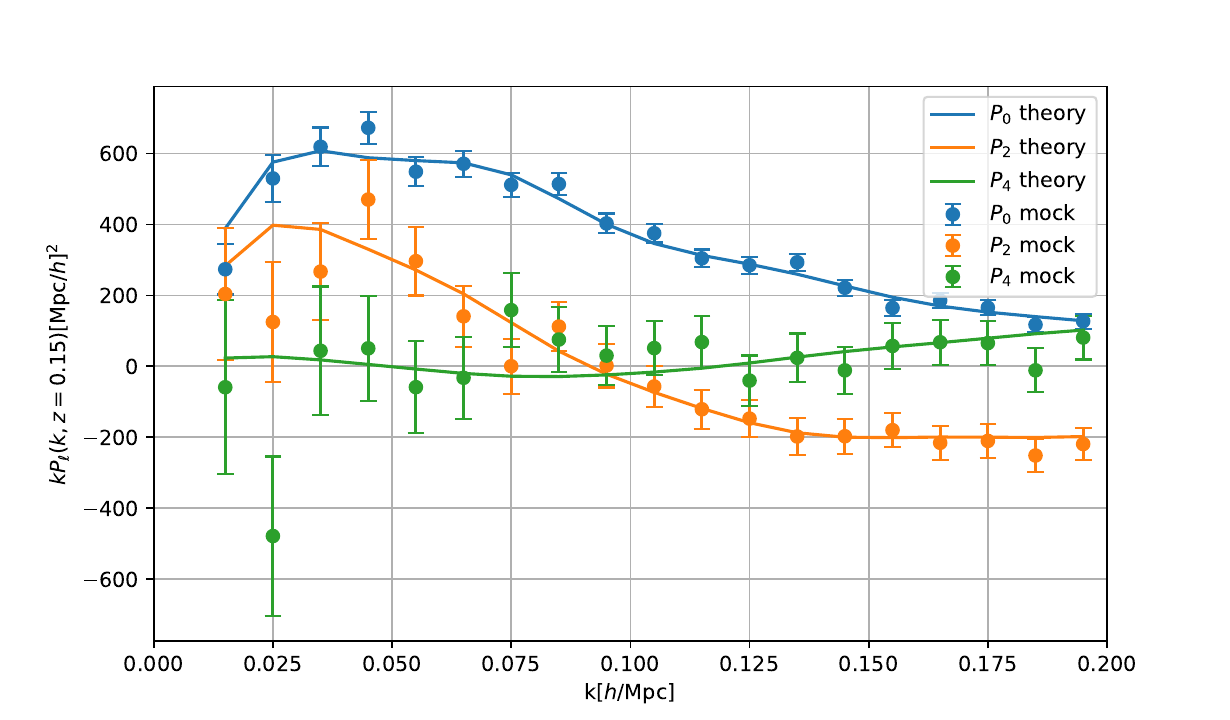}
	}
	\centering
	\subfigure{
		\centering
		\includegraphics[width=0.49\textwidth]{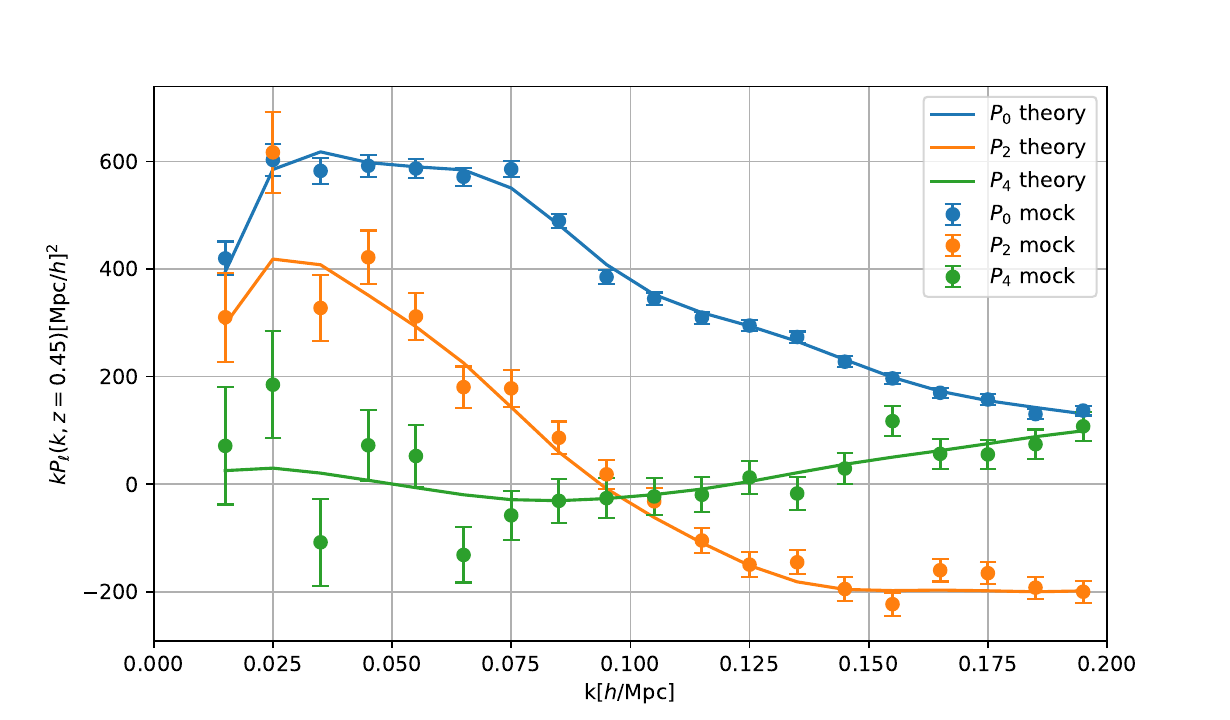}
        }
    \centering
	\subfigure{
		\centering
		\includegraphics[width=0.49\textwidth]{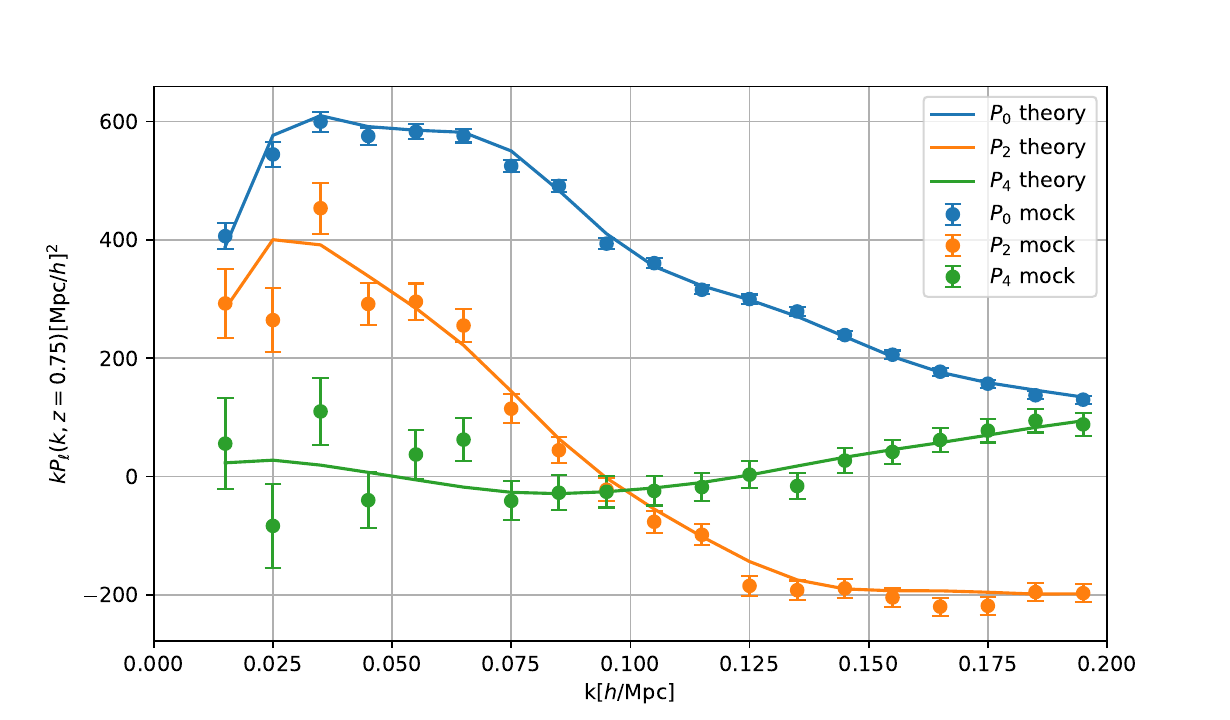}
        }
    \centering
	\subfigure{
		\centering
		\includegraphics[width=0.49\textwidth]{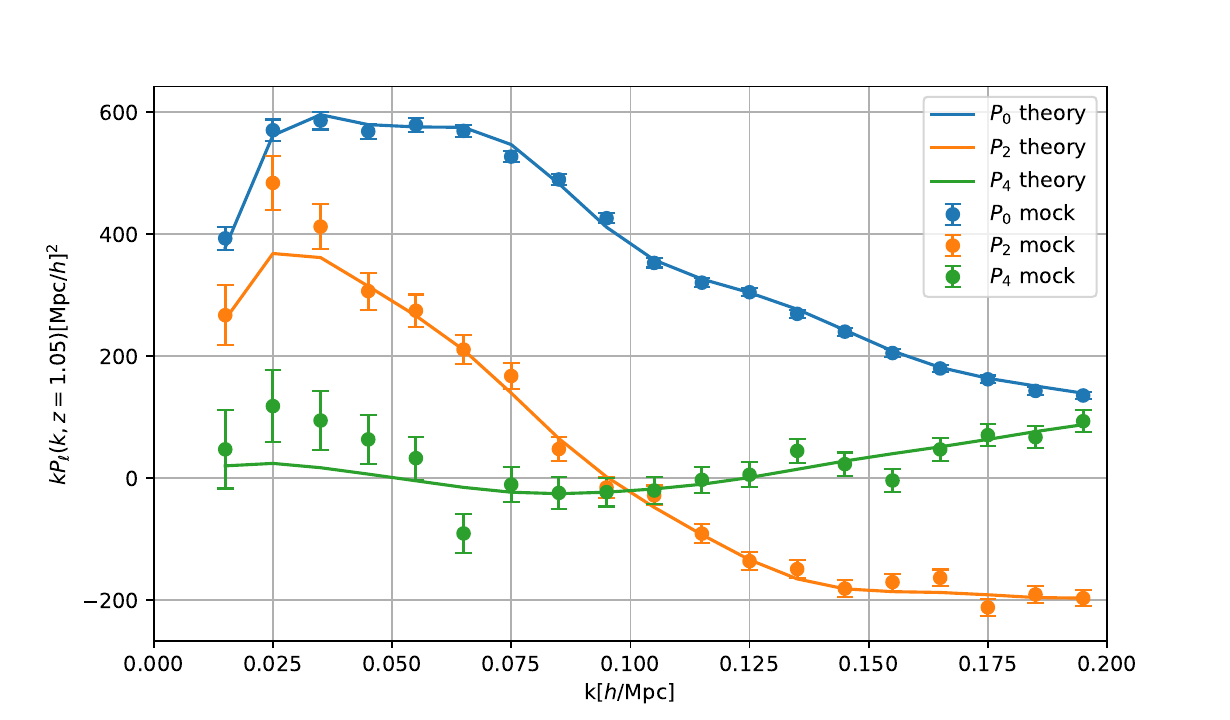}
        }
	\centering
	\caption{The mock data of multipoles of the galaxy post-reconstruction power spectra in four redshift bins from $z=0$ to 1.2 for the CSST spectroscopic survey. The corresponding theoretical power spectra are also shown as solid curves for comparison. }
 \label{datagalaxy}
\end{figure*}

\begin{figure*}
	\centering
	\subfigure{
		\centering
		\includegraphics[width=0.49\textwidth]{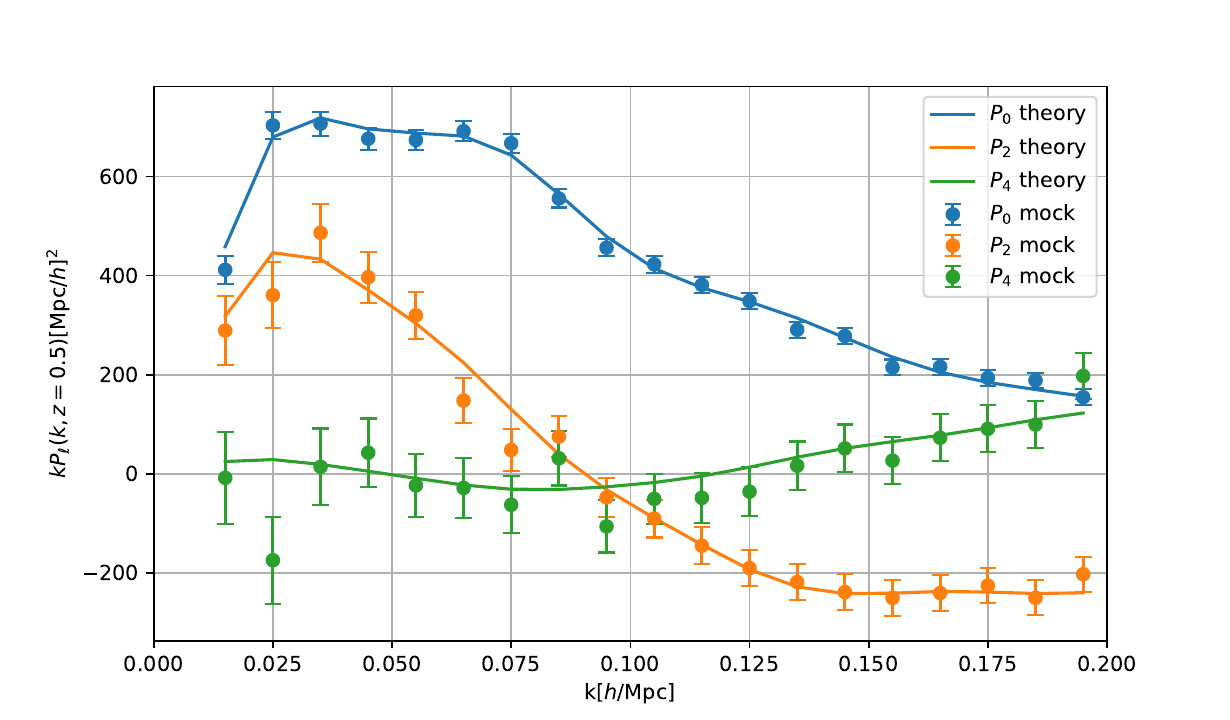}
	}
	\centering
	\subfigure{
		\centering
		\includegraphics[width=0.49\textwidth]{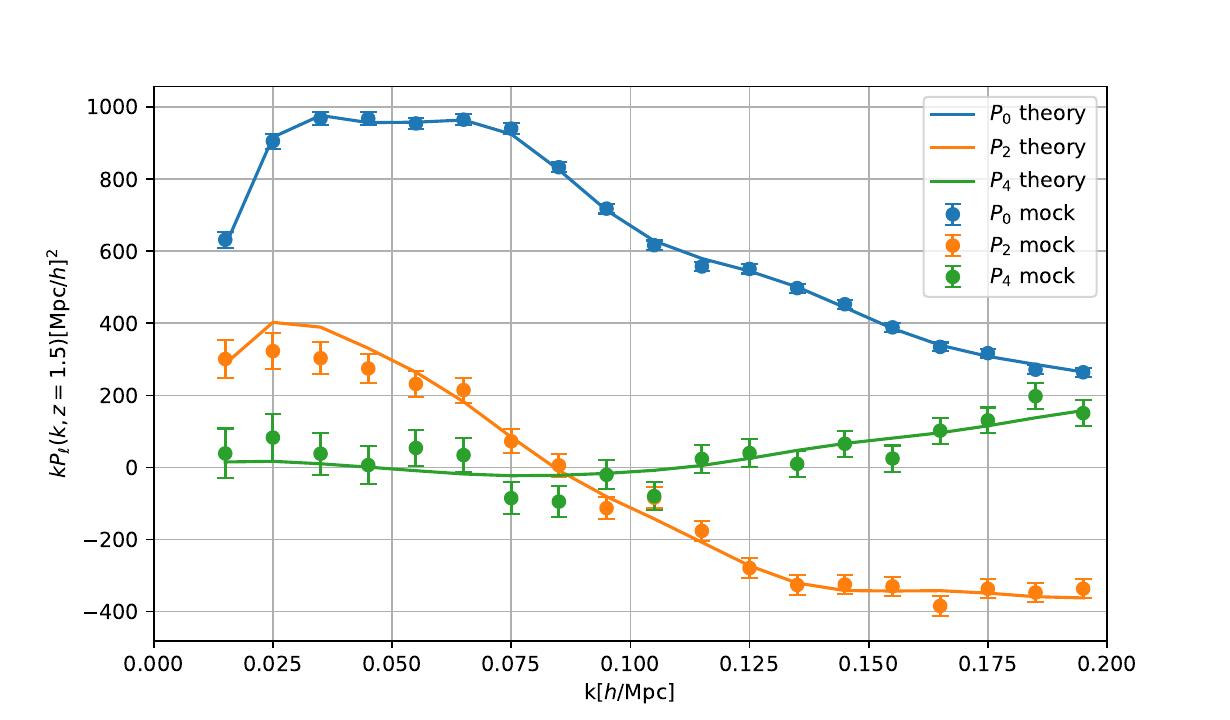}
        }
	\centering
	\subfigure{
		\centering
		\includegraphics[width=0.49\textwidth]{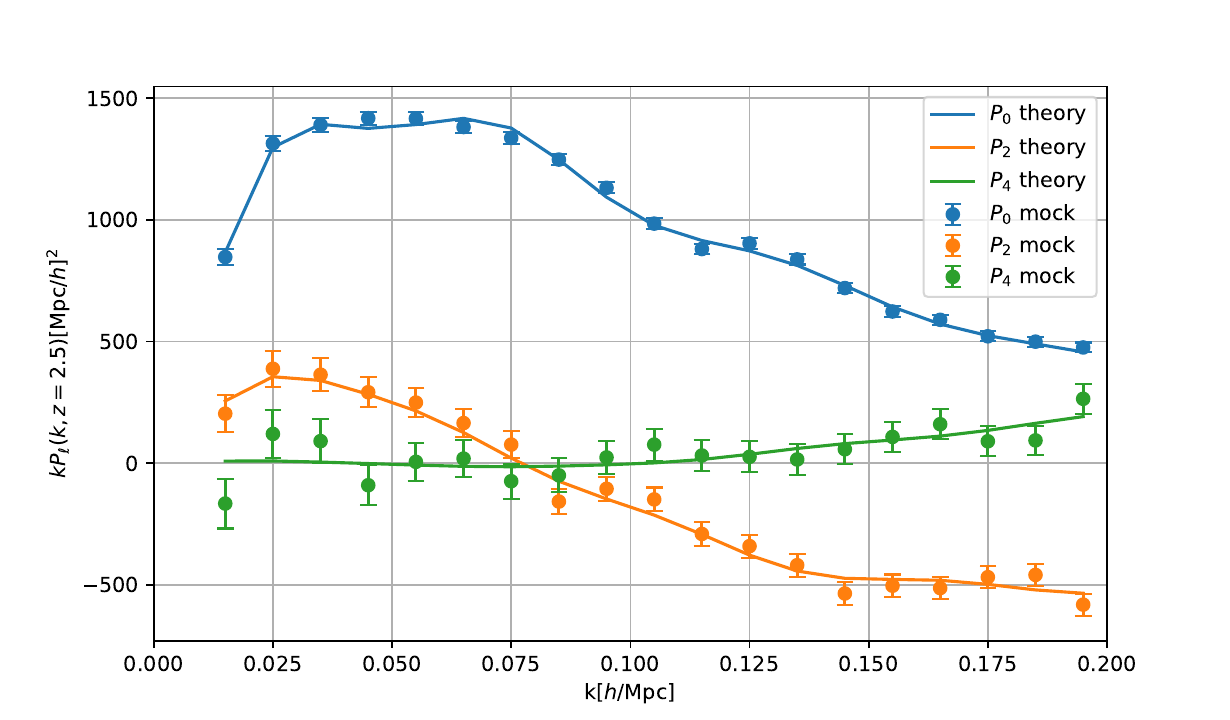}
	}
    \centering
	\subfigure{
		\centering
		\includegraphics[width=0.49\textwidth]{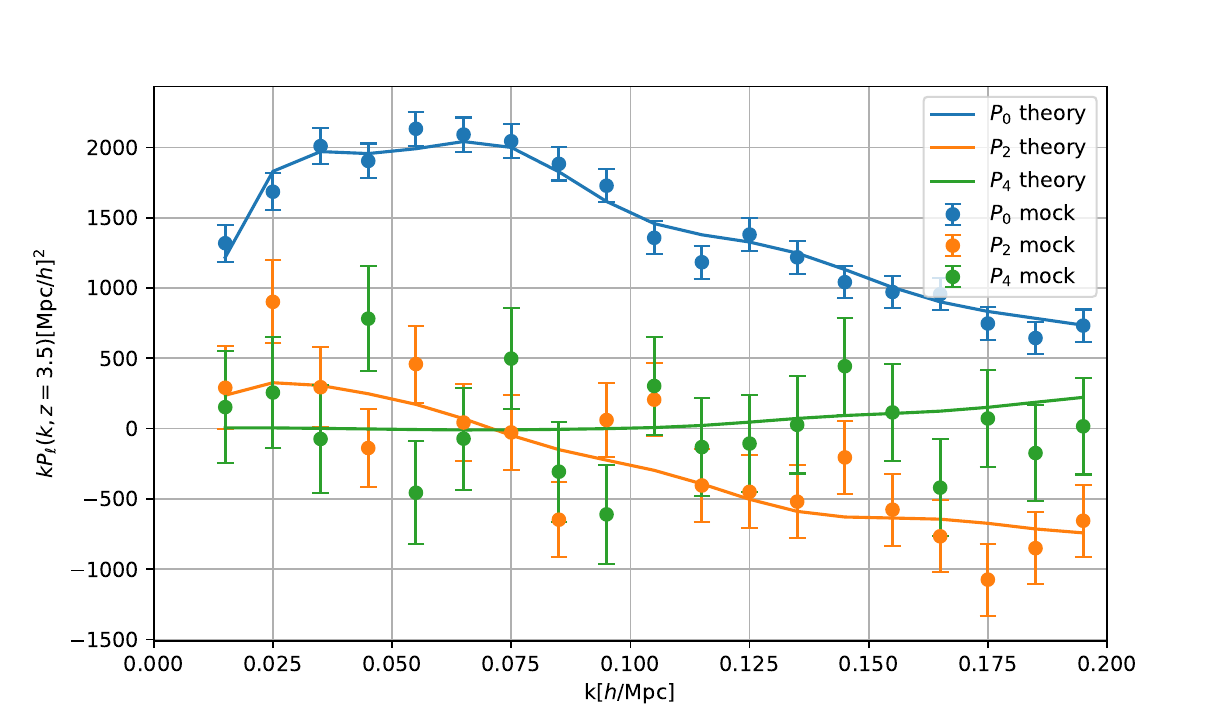}
	}
    \centering
	\centering
	\caption{The mock data of the multipoles of the AGN power spectra in four redshift bins from $z=0$ to 4 for the CSST spectroscopic survey. The corresponding theoretical power spectra are also shown in solid curves for comparison.}
    \label{dataagn}
\end{figure*}

\subsection{BAO template}

The BAO information can be derived by fitting the mock data of the galaxy and AGN power spectra in redshift space.
For the CSST galaxy survey, we adopt the reconstructed power spectrum discussed in Sec.~\ref{recon}, and the smearing factor caused by the low spectral resolution of the slitless spectroscopic survey is also considered. The final galaxy power spectrum is then given by
\be
P(k,\mu) = \exp\left[-(k\mu\sigma_{\mathrm{zerr}})^{2}\right]P_{\mathrm{r}}(k,\mu)\,,
\ee
where $\sigma_{\mathrm{zerr}} = \frac{c}{H(z)} \left(1+z\right) \sigma_{0,z}$, and $\sigma_{0,z} = 0.002$ denotes the redshift accuracy of spectral calibration \citep{Gong19}{\footnote{We also test the result with $\sigma_{0,z} = 0.005$, that is assuming a large redshift error caused by the spectral calibration in slitless spectroscopic survey. We find that the measurement of the scaling factors, especially $\alpha_{\parallel}$ in the radial direction, can be significantly affected in this case, which can result in deviations of the best-fits of some cosmological parameters from their fiducial values more than 1$\sigma$ confidence level. This implies that to obtain a reliable result, it needs to suppress the redshift error of the spectral calibration less than 0.005 for the BAO analysis in the CSST slitless spectroscopic surveys.}}. Since the CSST mainly targets the ELGs, we make use of the galaxy bias given by \cite{DESI2016}, and we have
\be
b_{\mathrm{g}}(z) = \frac{0.84}{D(z)}\,,
\ee
where $D(z)$ is the growth factor. Note that $b_{\mathrm{g}}$ is the Eulerian bias, i.e. $b_{\mathrm{g}}=b_1^{E}$. We have listed the values of the galaxy biases in the five redshift bins in Table~\ref{number_density}, and set them as free parameters in the fitting process.

For the AGN observation, as shown in Table~\ref{number_density}, we can find that the volume number density is always $< 10^{-4}\,h^3{\rm Mpc^{-3}}$ in the CSST spectroscopic survey. Given such low AGN number density, the reconstruction method probably cannot be used to restore the BAO feature as the case in the CSST galaxy survey \citep{Neveux2020}. So we would not adopt the reconstruction method in the AGN analysis, and then the AGN power spectrum can be modeled by 
\be
\begin{aligned} 
P(k, \mu)=& (b_A+f\mu^2)^2\exp\left[-(k\mu\sigma_{\mathrm{zerr}})^{2}\right]\exp\left[-(k\mu\Sigma_{\mathrm{FoG}})^{2}\right] \\
& \left[P_{\mathrm{nw}}(k, \mu)+P_{\mathrm{w}}(k, \mu) e^{-\Sigma_{\mathrm{nl}} k^2}\right]\,,
\end{aligned}
\ee
where $\Sigma_{\mathrm{nl}}=\left(1-\mu^2\right) \Sigma_{\perp}^2 / 2+\mu^2 \Sigma_{\parallel}^2 / 2$ denotes the anisotropic non-linear damping effect of the BAO, and we fix $\Sigma_{\perp} = 8\,h^{-1}\mathrm{Mpc}$ and $ \Sigma_{\parallel} = 3\,h^{-1}\mathrm{Mpc}$ \citep{Neveux2020}. $b_A$ is the bias of AGN,  and we take the form given in \cite{Laurent2017},
\bea
b_{\mathrm{A}}(z) = \alpha[(1+z)^2-6.565]+\beta \,,
\eea
with $\alpha = 0.278$ and $\beta = 2.3993$. The values of $b_{\mathrm{A}}$ in the five redshift bins can be found in Table~\ref{number_density}, and we set them as free parameters when extracting the BAO signal.

To perform the measurements of the BAO scaling parameters, in general, a fiducial cosmology is adopted to measure the distances in radial and transverse directions. Then we can introduce the two scaling parameters in the two directions as
\bea
\ba\label{eq:alpha}
\alpha_{\parallel}(z)=\frac{D_{\mathrm{H}}(z)r_{\mathrm{drag}}^{\mathrm{ref}}}{D_{\mathrm{H}}^{\mathrm{ref}}(z)r_{\mathrm{drag}}}\,,   \\
\alpha_{\perp}(z)=\frac{D_{\mathrm{M}}(z)r_{\mathrm{drag}}^{\mathrm{ref}}}{D_{\mathrm{M}}^{\mathrm{ref}}(z)r_{\mathrm{drag}}}\,.
\ea
\eea
Here $D_{\mathrm{H}}\equiv c/H(z)$, $D_{\mathrm{M}}(z)$ is the comoving angular diameter distance, and `ref' superscript represents the reference cosmology. The sound horizon, $r_\mathrm{drag}$, is determined by early-time physics and given by \citep{Brieden2023}.
\begin{equation}
\begin{aligned}\label{eq:r_drag}
r_{\mathrm{drag}}&=\int_{\infty}^{z_\mathrm{drag}} \frac{c_s(z)}{H(z)} d z \\
&\simeq \frac{147.05}{\mathrm{Mpc}}\left(\frac{\Omega_{\rm m} h^2}{0.1432}\right)^{-0.23}\left(\frac{N_{\mathrm{eff}}}{3.04}\right)^{-0.1}\left(\frac{\Omega_{\mathrm{b}} h^2}{0.02235}\right)^{-0.13}\,,
\end{aligned}
\end{equation}
where $N_{\mathrm{eff}}$ is the effective number of neutrino species.

When we assume a reference cosmology that is different from the true cosmology, it will produce additional anisotropies, which is known as the AP effect \citep{Alcock1979}. It can be parametrized as 
\begin{equation}
F_{\mathrm{AP}}(z)=F_\epsilon^{-1}(z) D_{\mathrm{M}}(z)^{\mathrm{ref}} / D_{\mathrm{H}}(z)^{\mathrm{ref}}\,,
\end{equation}
where $F_{\epsilon}=\alpha_{\parallel}/\alpha_{\perp}$. Although the sound horizon $r_{\mathrm{drag}}$ is used as a reference scale, $F_{\mathrm{AP}}(z)$ is not dependent on it. Since the AP effect can distort the true wavenumbers of the power spectrum, the true wavenumbers $k_{\parallel}^{\prime}$ and $k_{\perp}^{\prime}$ are then related to the observed wavenumbers $k$ by $k_{\parallel}^{\prime}=k_\parallel/\alpha_\parallel$ and $k_{\perp}^{\prime}=k_{\perp}/\alpha_{\perp}$. Given the total wavenumber $k^{\prime} = \sqrt{k^{\prime}{}_{\parallel}^{2}+{k^{\prime}}_{\perp}^{2}}$ and the cosine of the angle to the line-of-sight $\mu$, we can write the relations between the true ($k^{\prime}, \mu^{\prime}$) and observed values ($k, \mu$), that we have \citep{Ballinger1996}
\bea
\ba
k^{\prime}=\frac{k}{\alpha_{\perp}}\left[1+\mu^{2}\left(\frac{1}{F_{\epsilon}^{2}}-1\right)\right]^{1/2}\,, \\
\mu^{\prime}=\frac{\mu}{F_{\epsilon}}\left[1+\mu^{2}\left(\frac{1}{F_{\epsilon}^{2}}-1\right)\right]^{-1/2}\,. 
\ea
\eea
Finally, the multipoles of the power spectrum are given by \citep{GilMarin2020}
\be \label{eq:Pl}
P_{\ell}(k)=\frac{2\ell+1}{2}\int_{-1}^{1}\mathrm{d}L_{\ell}(\mu)P[k'(k,\mu),\mu'(\mu)]+\sum_{i=1}^{n}A_{\ell,i}k^{i}\, .
\ee
where $L_{\ell}(\mu)$ is the Legendre polynomial of order $\ell$, the last term denotes a polynomial added to fit the broadband power spectrum, and we find 5th-order is good enough for fitting. Since we only focus on the BAO signal, we set the linear bias and the linear growth rate as free parameters, and the extra normalization factor  $\frac{1}{\alpha_{\parallel}\alpha_{\perp}^2}$ is absorbed into the amplitude of the broadband power spectrum. Then we can generate the mock data of the power spectrum for monopole $P_0$, quadrupole $P_2$, and hexadecapole $P_4$. We create our mock data based on the Gaussian distribution. The mean value of the  Gaussian distribution function is given by the theoretical value of the multipole of the power spectrum derived from Eq.~(\ref{eq:Pl}), and the sigma is obtained by the square root of the diagonal elements of the covariance from Sec.~\ref{covariance}.

\subsection{Covariance Matrix}\label{covariance}

\begin{figure}
    \centering
    \includegraphics[width=0.5\textwidth]{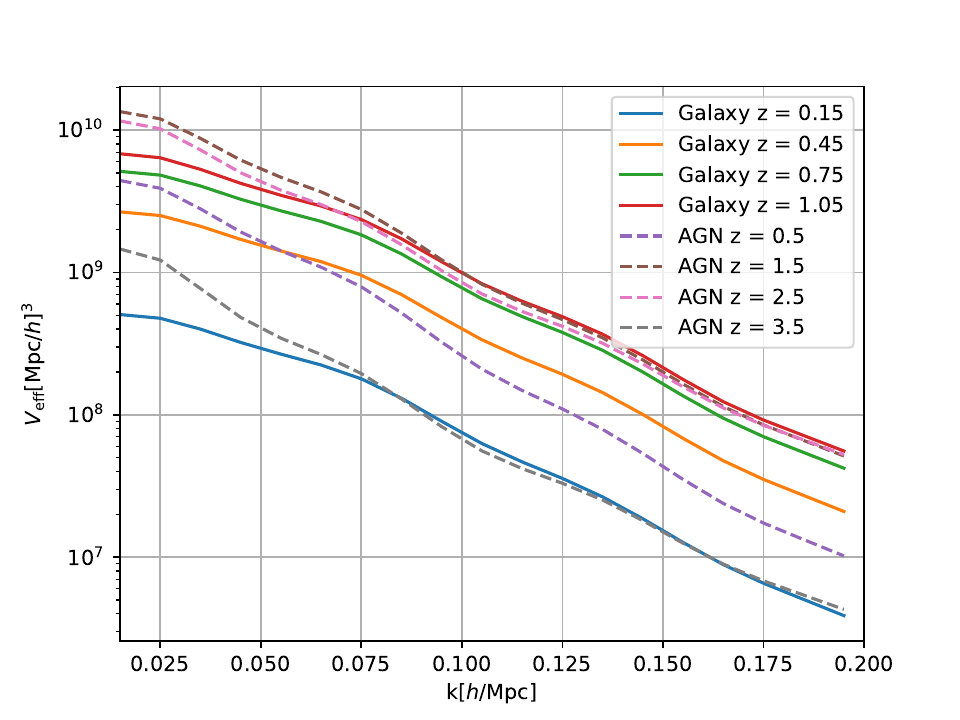}
    \caption{The effective volumes of the $P_0$ for the galaxy (solid curves) and AGN  (dashed curves) in different redshift bins in the CSST spectroscopic surveys. }
    \label{veff}
\end{figure}

Generally, the covariance matrix can be estimated using analytical computation, simulation or observational data \citep{Hamilton2006,Takahashi2009,Mohammed2014,Mohammed2017,OConnell2019,Philcox2020,Chudaykin2019,Wadekar2020,Wadekar2020b,Hikage2020,Taruya2021,Mohammad2022,Philcox2022,Hou2022,DingJ2023}.
Here we adopt the analytical computation method and estimate the covariance matrix by \citep{Wadekar2020,Chudaykin2019},
\begin{equation}\label{cov}
\begin{aligned}
C_{k_{i}k_{j}}^{(\ell_{1}\ell_{2})}&=\delta_{ij}\frac{2}{N_{k}}\frac{(2\ell_{1}+1)}{2}\left(2\ell_{2}+1\right) \\ 
& \times \int_{-1}^{1}d\mu L_{\ell_{1}}(\mu)L_{\ell_{2}}(\mu)\left[P(k_{i},\mu)+\frac{1}{\bar{n}} + N_{\mathrm{sys}}\right]^{2}\,,
\end{aligned}
\end{equation}
where $N_{k}=V_{\mathrm{survey}}k^{2}dk/(4\pi^{2})$ is the number of modes, and $\bar{n}$ is the average source volume density. We calculate our covariance matrix for both pre-reconstruction and post-reconstruction power spectrum by this formula. The covariance matrices are dominated by the Gaussian part in the linear and quasi-linear regions. For the pre-reconstruction power spectrum, since it is nearly Gaussian in these regions, we could ignore the mode coupling caused by non-linearity. It has been tested in simulation using linear and non-linear input power spectra \citep{Grieb2016}. We use Eq.~(\ref{cov}) also for estimating the  covariance matrix of the post-reconstruction power spectrum, even if a careful assessment of its validity still needs to be done. A potential systematical noise term $N_{\mathrm{sys}}$ is also considered \citep{Gong19}, which can include the instrumental effects of the CSST slitless gratings, e.g. the success rate of achieving the required spec-$z$ accuracy. We will explore the results assuming $N_{\mathrm{sys}} = 0$ and $10^4\,h^{-3}\,\mathrm{Mpc^3}$ as the optimistic and pessimistic cases, respectively.

In Figure~\ref{datagalaxy} and Figure~\ref{dataagn}, we show the mock data of the multipoles of the galaxy post-reconstruction and AGN power spectra in different redshift bins for the CSST spectroscopic survey. Note that we do not use the mock data in the last redshift bins of the CSST galaxy and AGN surveys in the fitting process, since the number densities of galaxy and AGN are quite low in the two bins, as shown in Table~\ref{number_density}. For the galaxy survey, the reconstruction cannot be performed in the redshift bin of $z$=1.2-1.5 with $\bar{n}_{\rm g}<10^{-4}\ h^3{\rm Mpc}^{-3}$. For the AGN survey, the density is less than $10^{-7}\ h^3{\rm Mpc}^{-3}$, and there is no effective measurement on the BAO signal in the redshift bin of $z$=3-4.
For each of the other four redshift bins, we generate 19 data points from $k = 0.015\,h/\mathrm{Mpc}$ to $0.195\,h/\mathrm{Mpc}$ with $\mathrm{d}k = 0.01\,h/\mathrm{Mpc}$, and a random shift are added to each data point which is generated from a Gaussian distribution based on the covariance matrix.

We also calculate the effective volumes for the galaxy and AGN surveys in different redshift bins. The corresponding effective volume $V_{\mathrm{eff}} \equiv V_{\mathrm{survey}}\frac{P_0^2}{C^{(00)}}$ for the galaxies and AGNs, where $C^{(00)}$ is the diagonal element of the covariance matrix which contains the cosmic variance, shot noise and systematical noise. In Figure~\ref{veff}, we plot the corresponding effective volume for both the CSST galaxy and AGN spectroscopic surveys. We can see that our result is consistent with \cite{Gong19} for the galaxy survey, and it is basically a factor of 3 larger compared to eBOSS surveys \citep{Foroozan2021}. On the other hand, the effective volume of AGN is comparable to the CSST galaxy survey, and even higher at large scales with $k<0.05\ h {\rm Mpc}^{-1}$. Therefore, we also expect to obtain precise BAO measurements in the CSST AGN spectroscopic survey.


\begin{figure*}

	\centering
	\subfigure{
		\centering
		\includegraphics[height=3in,width=3in]{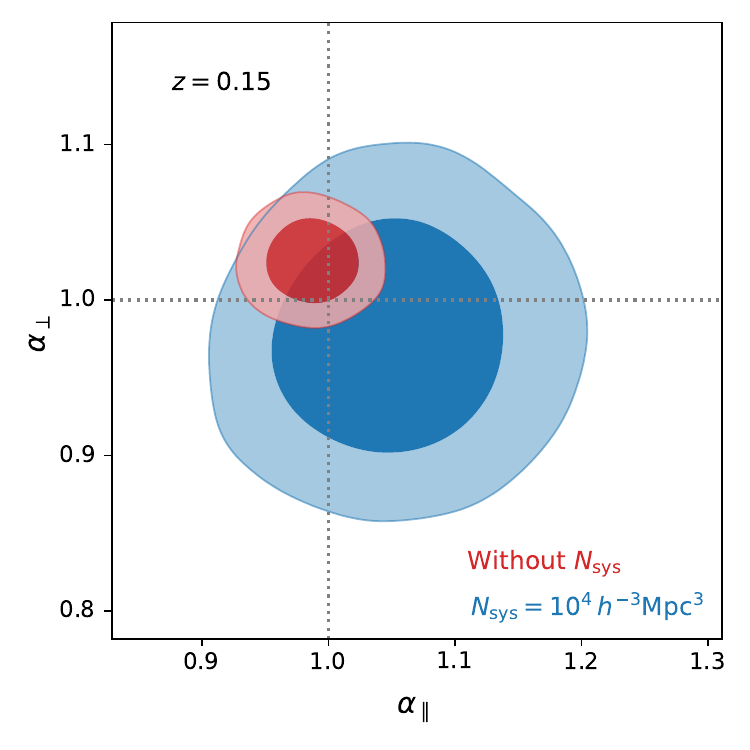}
	}
	\centering
	\subfigure{
		\centering
        \includegraphics[height=3in,width=3in]{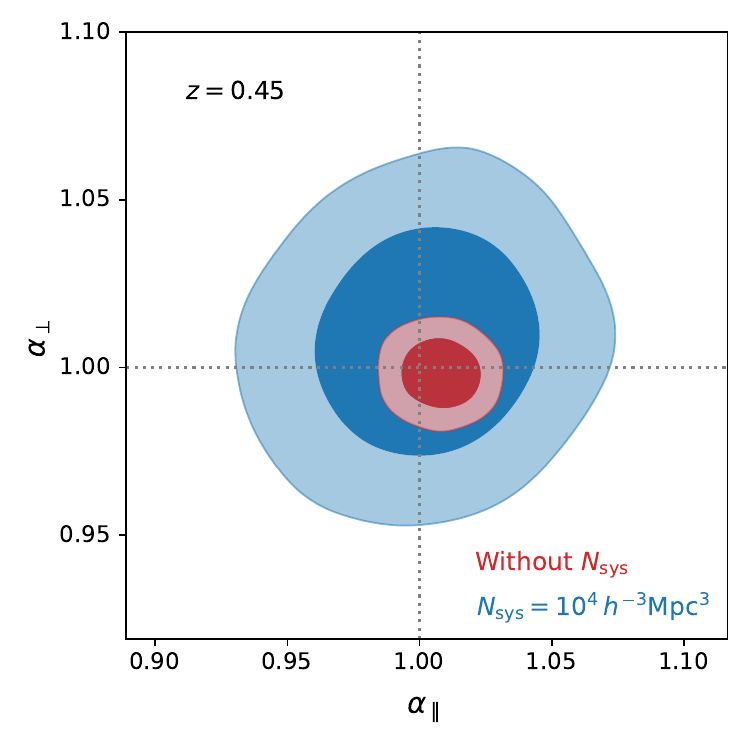}
        }
	\centering
	\subfigure{
		\centering
        \includegraphics[height=3in,width=3in]{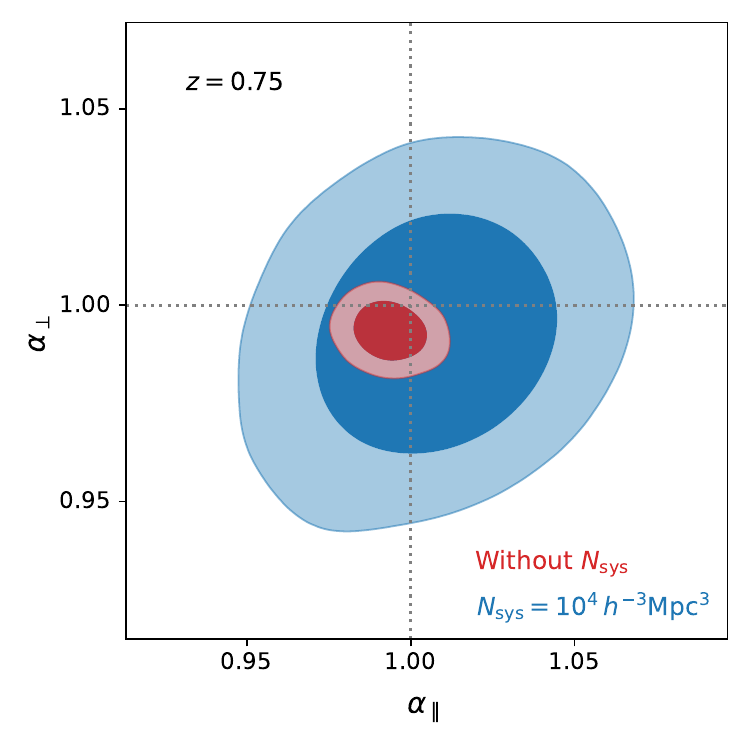}
	}
    \centering
	\subfigure{
		\centering
        \includegraphics[height=3in,width=3in]{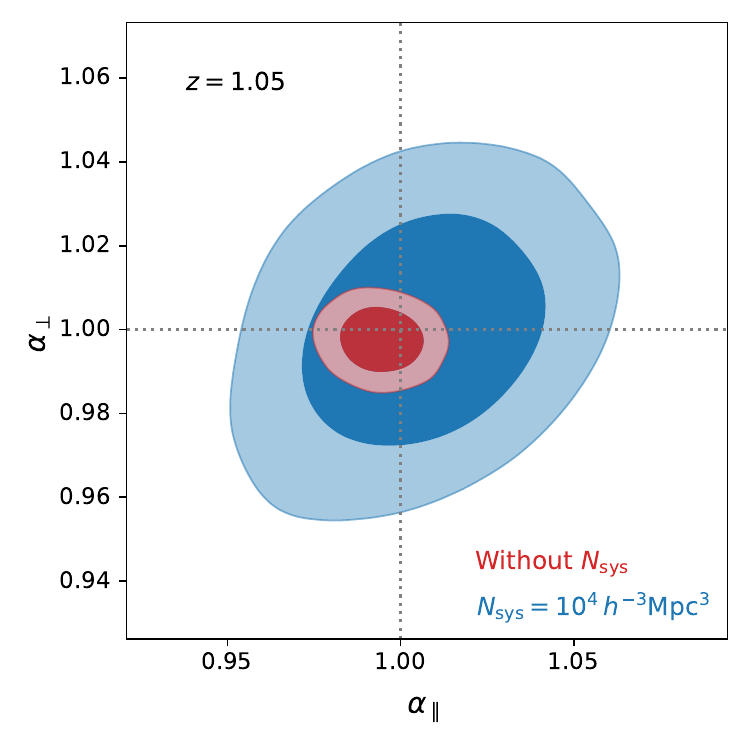}
	}
	\centering
	\caption{The predicted constraint results on the BAO scaling parameters in the CSST galaxy spectroscopic survey in the four redshift bins. The red and blue contours show the results assuming $N_{\rm sys}=0$ and $10^4\,h^{-3}\,\mathrm{Mpc^3}$ as the optimistic and pessimistic cases, respectively. The gray dotted lines denote the fiducial values.}
 \label{galaxyresults}
\end{figure*}

\section{Bayesian analysis}\label{Sec4}

After obtaining the mock data of the multipole power spectra for the galaxy and AGN in the CSST spectroscopic surveys, we use the Markov Chain Monte Carlo (MCMC) method to fit the mock data and extract the BAO information.
The Gaussian likelihood function can be written as
\be
\mathcal{L}_{G}(\theta)\propto\mathrm{e}^{-\chi^{2}(\theta)/2}\,,
\ee
where $\theta$ denotes the model parameters, and $\chi^{2}(\theta)$ is given by 
\be
\chi^{2}(\theta)\equiv (P_{\ell}^{\mathrm{data}} - P_{\ell}^{\mathrm{model}})^\mathrm{T}C_{\rm P}^{-1}(P_{\ell}^{\mathrm{data}} - P_{\ell}^{\mathrm{model}})\,.
\ee
Here $C_{\rm P}$ is the covariance matrix of the mock data, which is given by Eq.~(\ref{cov}), and the model and mock data vectors are composed of $\{P_{0}\,, P_{2}\,, P_{4}\}$, and the parameter vector $\theta$ stands for the 14 parameters in each redshift bin. Here we consider the two physical parameters, $\{\alpha_{\parallel}, \alpha_{\perp}\}$, and 12 nuisance parameters,$\{b_{\rm g}\ {\rm or}\ b_{\rm A}, f, A_{i}^{0}, A_{i}^{2}\}$, where $f$ is the growth rate and $i = 1,...5$ denotes the order of the polynomial. Note that we only consider the polynomial term in Eq.~(\ref{eq:Pl}) for $P_0$ and $P_2$ here, and ignore it for $P_4$ since it is relatively small compared to the monopole and quadrupole power spectra.


Then, based on the extracted BAO information, i.e. $\alpha_{\parallel}$ and $\alpha_{\perp}$ derived from the above MCMC results, we also perform the Bayesian analysis of the cosmological parameters for exploring the constraint power. Here we investigate the constraints on the cosmological parameters for the $\Lambda$CDM and $w$CDM models. The $\chi^2$ is given by 
\begin{equation}
    \chi^{2}(\theta)\equiv \Sigma_{i}(D_{\mathrm{data}} - D_{\mathrm{model}})^\mathrm{T}C_{\rm D}^{-1}(D_{\mathrm{data}} - D_{\mathrm{model}})\,,
\end{equation}
where $\Sigma_{i}$ represents the sum of all redshift bins, and the parameter vector $\theta$ stands for the cosmological parameters i.e. $\theta=\{h, \Omega_\mathrm{m}, w, \Omega_{\mathrm{b}}h^2\}$. For the sound horizon given by Eq.~(\ref{eq:r_drag}), we set  $N_{\mathrm{eff}} = 3.04$ and adopt the prior information of baryon density from Big Bang nucleosynthesis (BBN), i.e. $\Omega_{\mathrm{b}}h^2 = 0.02235\pm0.00037$ in our fitting process \citep{Schoneberg2019,Schoneberg2022}. The data vector $D_{\mathrm{data}} = \{\alpha_{\parallel}, \alpha_{\perp}\}$, and the $D_{\mathrm{model}}$ can be estimated by Eq.~(\ref{eq:alpha}). $C_{\rm D}$ is the covariance matrix for $\alpha_{\parallel}$ and $\alpha_{\perp}$ which can be derived from the MCMC chains. We use {\tt Cobaya} \citep{Torrado2021} to complete the Bayesian inference, and set $R-1 = 0.005$ for the stopping criterion when generating chains. The first 30 percent of the chain points are removed in our analysis, and the rest chain points are used to generate the probability distribution of the parameters.

\section{Results}\label{Sec5}

\begin{figure*}

	\centering
	\subfigure{
		\centering
		\includegraphics[height=3in,width=3.in]{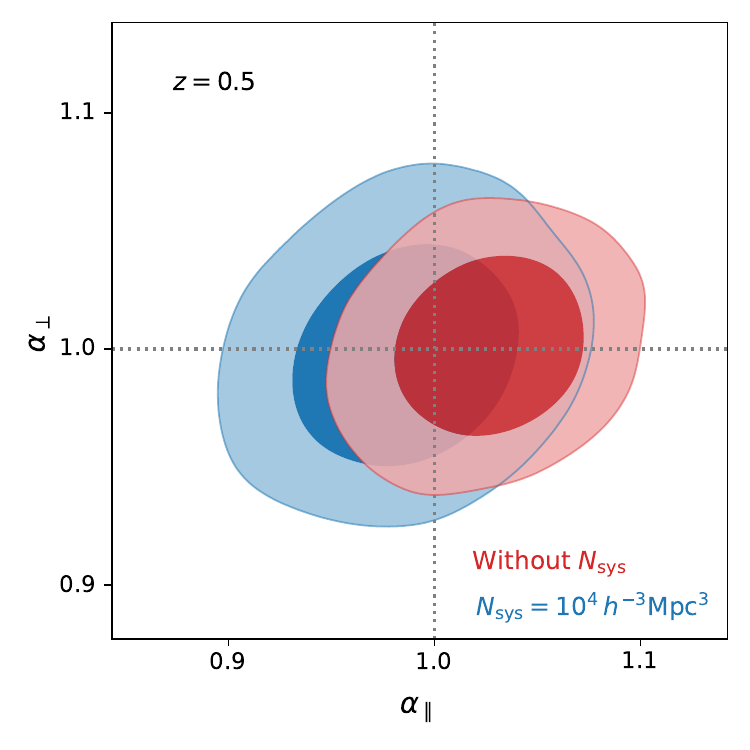}
	}
	\centering
	\subfigure{
		\centering
        \includegraphics[height=3in,width=3.in]{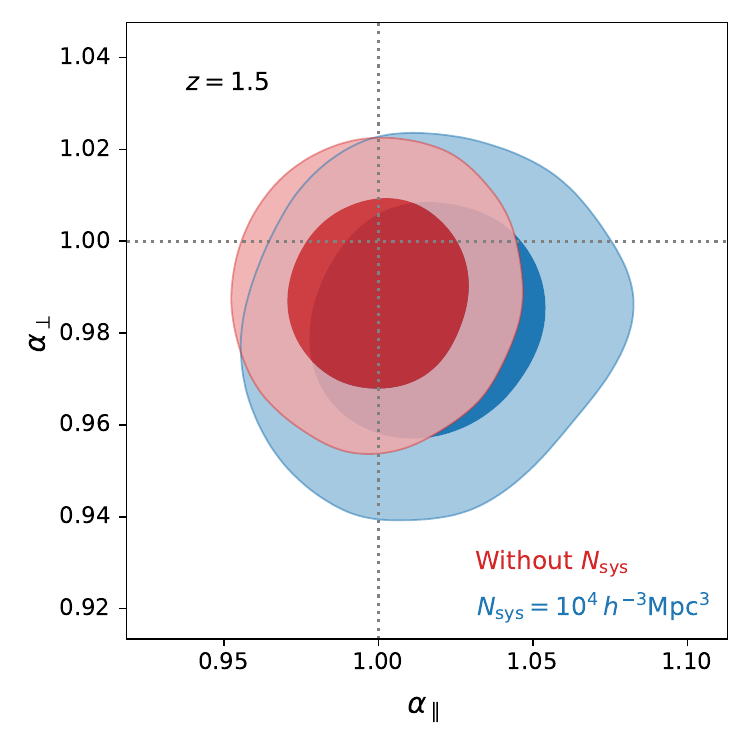}
        }
	\centering
	\subfigure{
		\centering
        \includegraphics[height=3in,width=3.in]{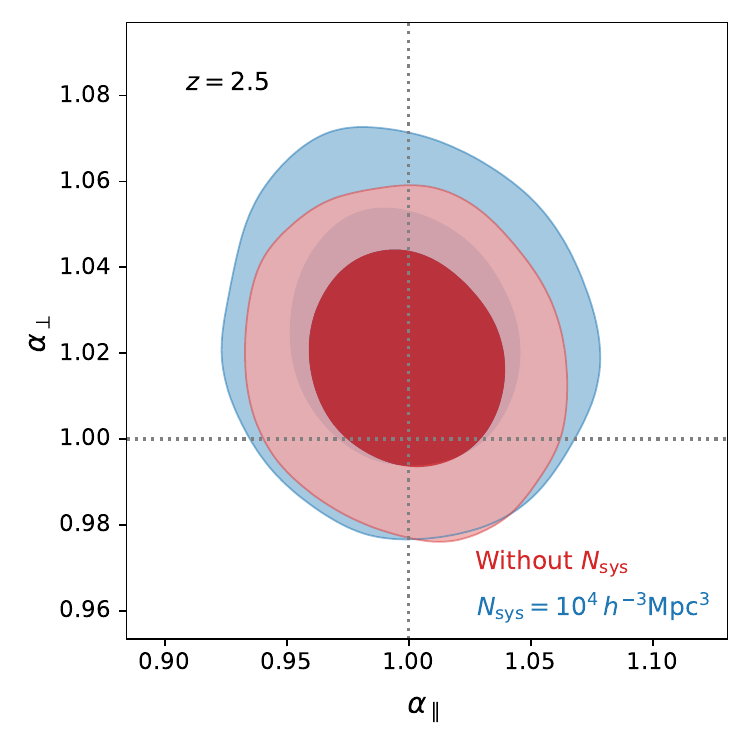}
	}
    \centering
	\subfigure{
		\centering
        \includegraphics[height=3in,width=3.in]{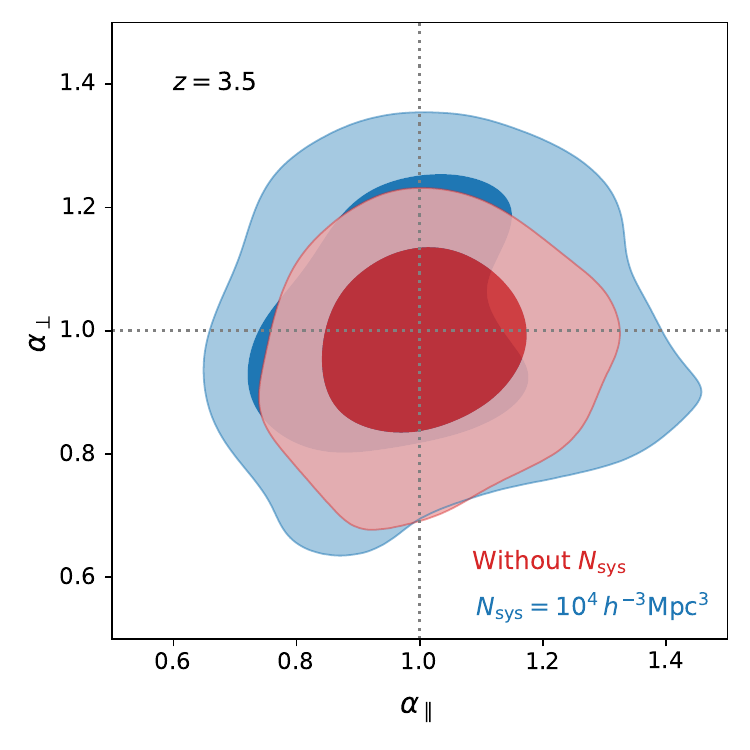}
	}
	\centering
	\caption{The predicted constraint results on the BAO scaling parameters in the CSST AGN spectroscopic survey in the four redshift bins. The red and blue contours show the results assuming $N_{\rm sys}=0$ and $10^4\,h^{-3}\,\mathrm{Mpc^3}$ as the optimistic and pessimistic cases, respectively. The gray dotted lines denote the fiducial values.}
     \label{agnresults}
\end{figure*}

\begin{table*}
	\caption{The predicted best-fit values and 1$\sigma$ errors of $\alpha_{\parallel}$ and $\alpha_{\perp}$ from the CSST galaxy and AGN spectroscopic surveys. We have considered both the optimistic and pessimistic cases with $N_{\rm sys}=0$ and $10^4\,h^{-3}\,\mathrm{Mpc^3}$, respectively. The derived $D_\mathrm{H}/r_\mathrm{drag}$,  $D_\mathrm{M}/r_\mathrm{drag}$, and reduce $\chi^2$ (57 data points and 14 parameters in each redshift bin) are also listed.}
	\begin{center}
	\scalebox{1.}{
		\begin{tabular}{c c c c c c c c c c c}
			\hline
			\hline
			& $z_{\rm min}$ & $z_{\rm max}$ & $z_{\rm mean}$ &  $\alpha_{\parallel}$ (precision) & $\alpha_{\perp}$ (precision) & $D_\mathrm{H}/r_\mathrm{drag}$ & $D_\mathrm{M}/r_\mathrm{drag}$ & reduced $\chi^2$ \\
			\hline
             Galaxy \\
            \multirow{4}*{$N_{\mathrm{sys}} = 0$} & 0 & 0.3 & 0.15 & $0.987\pm0.024$ (2.4\%) & $1.025\pm0.018$ (1.8\%) & $27.69\pm0.67$ & $4.48\pm0.079$ & 1.28 \\
            & 0.3 & 0.6 & 0.45 & $1.0079\pm0.0098$ (0.97\%) & $0.9983\pm0.0069$ (0.69\%) & $23.79\pm0.23$ & $12.10\pm0.084$ & 1.17 \\
			& 0.6 & 0.9 & 0.75 & $0.9936\pm0.0074$ (0.74\%) & $0.9934\pm0.0050$ (0.5\%) & $19.53\pm0.15$ & $18.47\pm0.093$ & 1.51 \\
			& 0.9 & 1.2 & 1.05 & $0.9944\pm0.0080$ (0.8\%) & $0.9976\pm0.0051$ (0.51\%) & $16.34\pm0.13$ & $23.93\pm0.122$ & 1.44 \\
			\hline
			\multirow{4}*{$N_{\mathrm{sys}} = 10^4$} & 0 & 0.3 & 0.15 & $1.050\pm0.061$ (5.8\%) & $0.976\pm0.050$ (5.1\%) & $29.45\pm1.71$ & $4.27\pm0.22$ & 1.36 \\
            & 0.3 & 0.6 & 0.45 & $1.003\pm0.028$ (2.8\%) & $1.008\pm0.022$ (2.2\%) & $23.68\pm0.66$ & $12.21\pm0.27$ & 1.20 \\
		     & 0.6 & 0.9 & 0.75 & $1.007\pm0.025$ (2.5\%) & $0.993\pm0.020$ (2.0\%) & $19.80\pm0.49$ & $18.50\pm0.37$ & 1.21 \\
			& 0.9 & 1.2 & 1.05 & $1.006\pm0.023$ (2.3\%) & $0.9996\pm0.018$ (1.8\%) & $16.53\pm0.38$ & $23.98\pm0.43$ & 1.43 \\
                \hline
                \hline
            AGN \\
            \multirow{4}*{$N_{\mathrm{sys}} = 0$} & 0 & 1.0 & 0.5 & $1.026\pm0.031$ (3.0\%) & $1.001\pm0.025$ (2.5\%) & $23.50\pm0.71$ & $13.29\pm0.33$ & 1.04  \\
			& 1.0 & 2.0 & 1.5 & $0.9999\pm0.019$ (1.9\%) & $0.989\pm0.014$ (1.4\%) & $12.79\pm0.24$ & $30.18\pm0.43$ & 1.43  \\
			& 2.0 & 3.0 & 2.5 & $0.999\pm0.027$ (2.7\%) & $1.018\pm0.017$ (1.7\%) & $8.03\pm0.22$ & $41.38\pm0.69$ & 1.10 \\
			& 3.0 & 4.0 & 3.5 & $1.007^{+0.097}_{-0.12}$ (11.4\%) & $0.973^{+0.11}_{-0.089}$ (11.2\%) & $5.63^{+0.54}_{-0.67}$ & $46.07^{+5.21}_{-4.21}$ & 1.12 \\
                \hline
			\multirow{4}*{$N_{\mathrm{sys}} = 10^4$} & 0 & 1.0 & 0.5 & $0.986\pm0.036$ (3.7\%) & $0.998^{+0.029}_{-0.032}$ (3.1\%) & $22.58\pm0.83$ & $13.26\pm0.39$ & 1.15  \\
			& 1.0 & 2.0 & 1.5 & $1.016\pm0.025$ (2.5\%) & $0.982\pm0.017$ (1.7\%) & $13.0\pm0.32$ & $29.96\pm0.52$ & 1.48 \\
			& 2.0 & 3.0 & 2.5 & $0.998\pm0.031$ (3.1\%) & $1.024\pm0.02$ (2\%) & $8.03\pm0.25$ & $41.62\pm0.81$ & 1.26 \\
			& 3.0 & 4.0 & 3.5 & $0.97^{+0.11}_{-0.19}$ (16.0\%) & $0.99^{+0.12}_{-0.16}$ (14.7\%) & $5.42^{+0.61}_{-1.06}$ & $46.87^{+5.68}_{-7.57}$ & 1.31 \\
			\hline
                \hline
		\end{tabular}
		}
	\end{center}
	\label{gabao}
\end{table*}

\begin{figure*}

	\centering
	\subfigure{
		\centering
           \includegraphics[height=3.4in,width=3.4in]{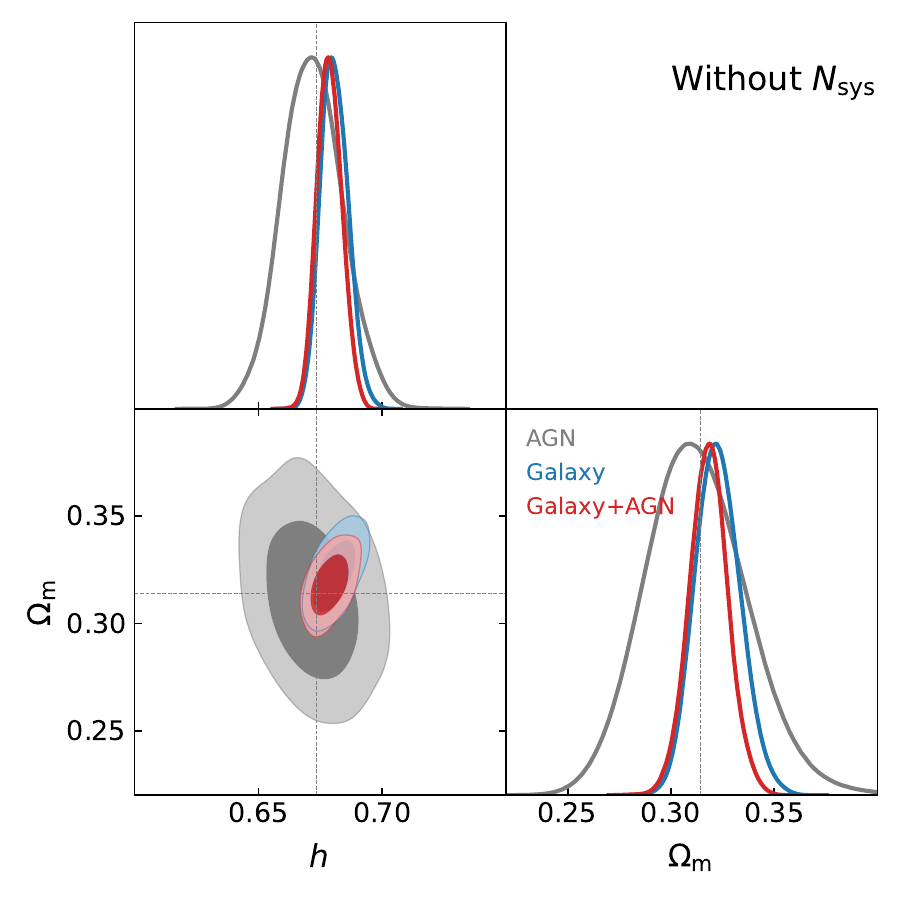}
	}
     \centering
	\subfigure{
		\centering
           \includegraphics[height=3.4in,width=3.4in]{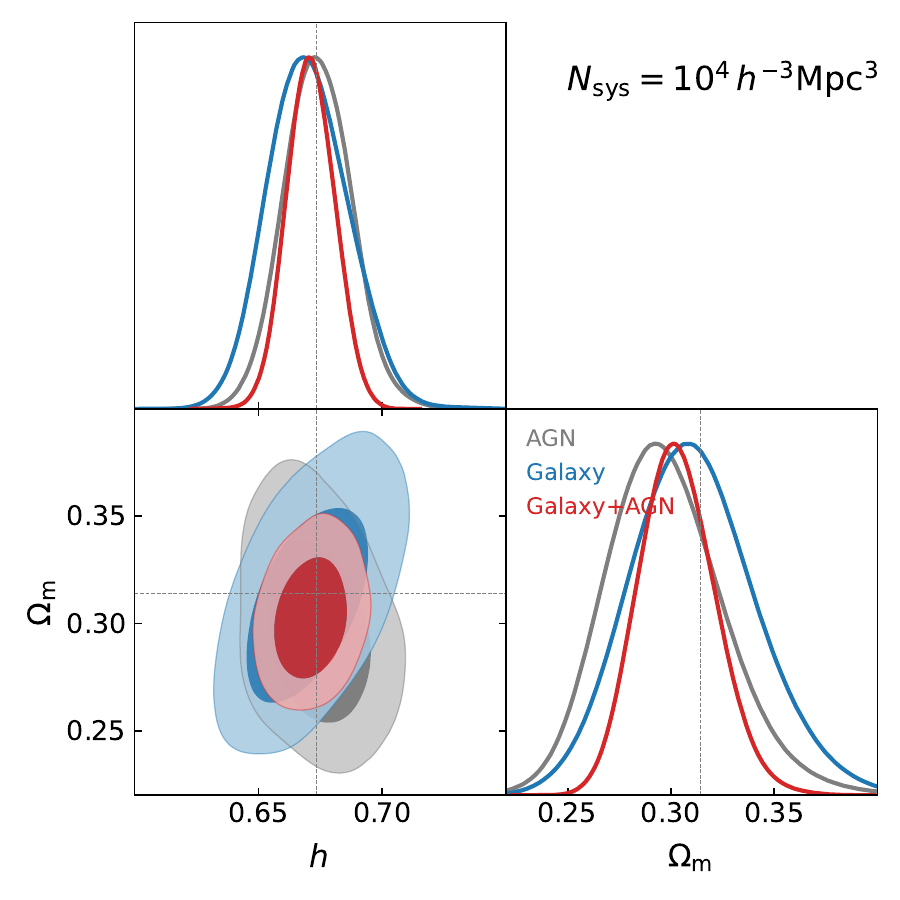}
	}
	\centering
	\caption{The predicted contour maps and 1D PDFs of $\Omega_{\rm m}$ and $h$ in the $\Lambda$CDM model for the CSST galaxy and AGN spectroscopic surveys. The constraint results by assuming $N_{\rm sys}=0$ and $10^4\,h^{-3}\,\mathrm{Mpc^3}$ are shown in the left and right panels, respectively. The gray, blue, and red contours and PDFs denote the results from the CSST galaxy, AGN, and joint surveys, respectively.}
 \label{lcdmresults}
\end{figure*}

\begin{figure*}

     \centering
	\subfigure{
		\centering
        \includegraphics[width=0.49\textwidth]{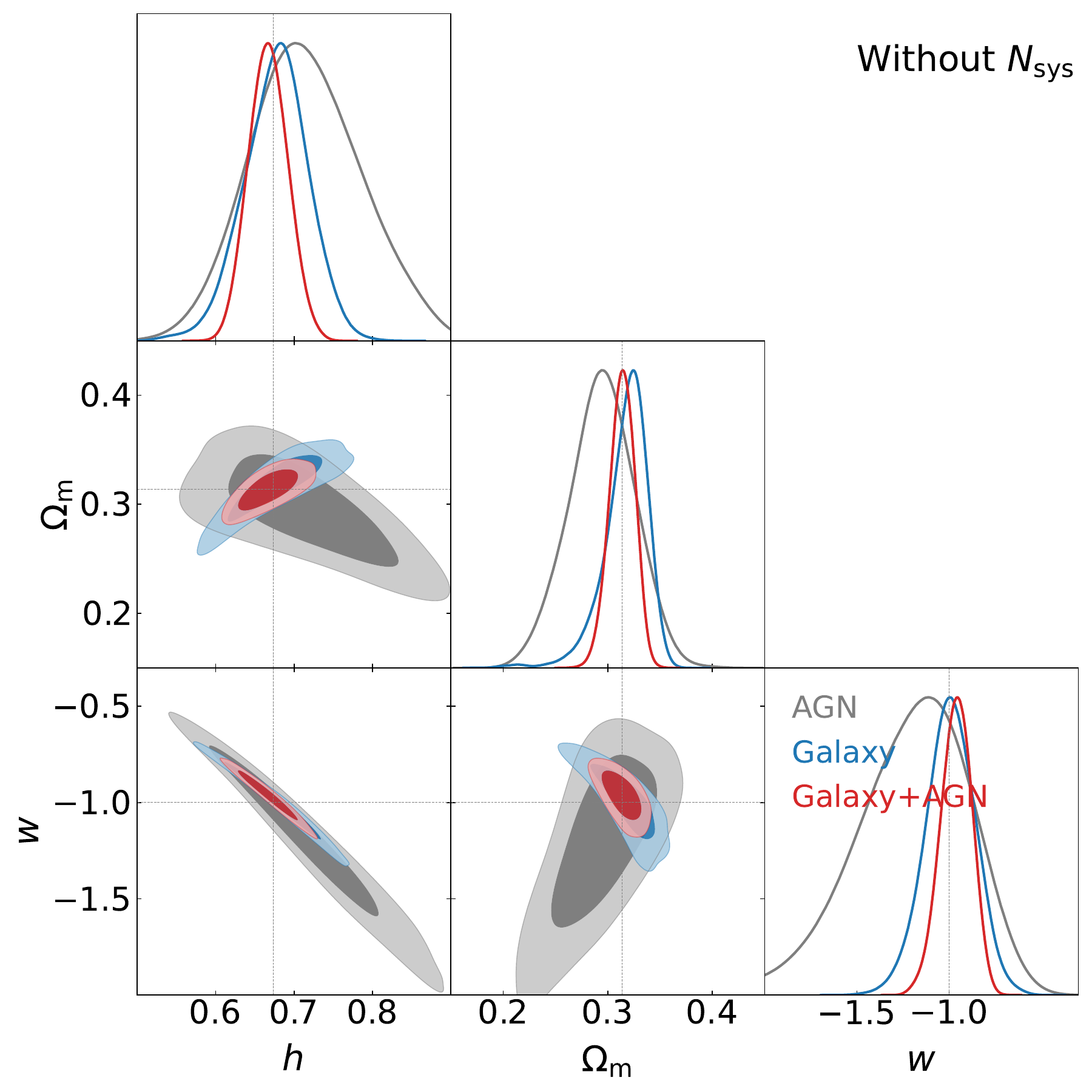}
        }
	\centering
	\subfigure{
		\centering
        \includegraphics[width=0.49\textwidth]{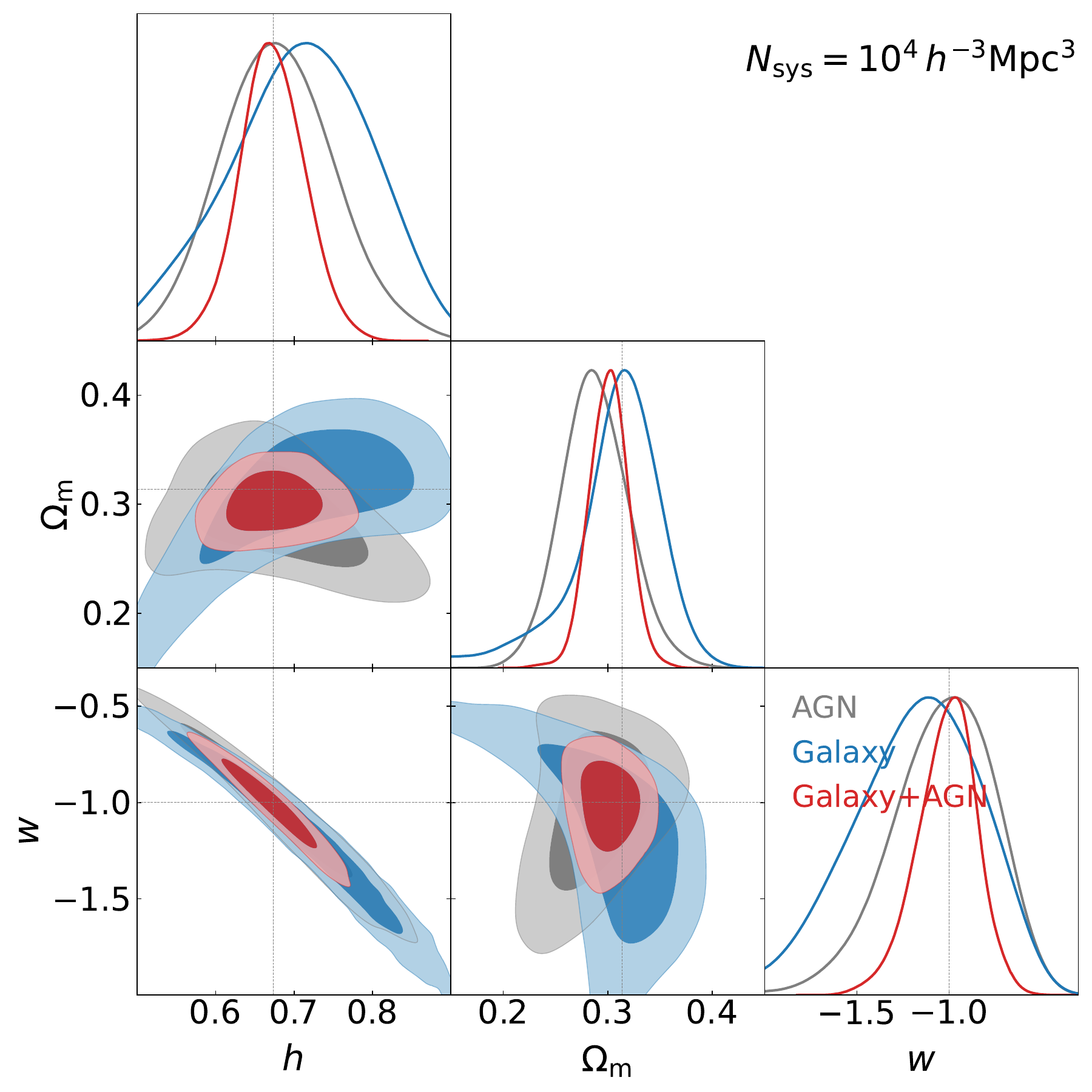}
        }
	\centering
	\caption{The predicted contour maps and 1D PDFs of $\Omega_{\rm m}$, $h$ and $w$ in the $w$CDM model for the CSST galaxy and AGN spectroscopic surveys. The constraint results by assuming $N_{\rm sys}=0$ and $10^4\,h^{-3}\,\mathrm{Mpc^3}$ are shown in the left and right panels, respectively. The gray, blue, and red contours and PDFs denote the results from the CSST galaxy, AGN, and joint surveys, respectively.}
 \label{wcdmresults}
\end{figure*}

\begin{table*}
	\caption{The best-fit values and 1$\sigma$ errors of the relevant cosmological parameters in the $\Lambda$CDM and $w$CDM models derived from the mock BAO data ($\alpha_{\parallel}$ and $\alpha_{\perp}$) for the CSST galaxy and AGN spectroscopic surveys. The constraint results assuming $N_{\rm sys}=0$ and $10^4\,h^{-3}\,\mathrm{Mpc^3}$ are listed.}
	\begin{center}
	\scalebox{1.2}{
		\begin{tabular}{c c c c c c c}
			\hline
			\hline
			& & $\Omega_{\mathrm{m}}$ (precision) & $H_0$ (precision) & w (precision) & \\
			\hline
            $\Lambda$CDM \\

			\multirow{3}*{$N_{\mathrm{sys}} = 0$} & Galaxy BAO + BBN & $0.323^{+0.010}_{-0.012}$ (3.4\%) & $68.07^{+0.51}_{-0.59}$ (0.81\%) & - \\
            & AGN BAO + BBN & $0.312^{+0.023}_{-0.026}$ (7.9\%) & $67.2^{+1.1}_{-1.3}$ (1.9\%) & - \\
			& All BAO + BBN & $0.3182\pm0.0094$ (3.0\%) & $67.88^{+0.47}_{-0.53}$ (0.75\%) & - \\
			\hline
			\multirow{3}*{$N_{\mathrm{sys}} = 10^4$} & Galaxy BAO + BBN & $0.311^{+0.028}_{-0.032}$ (9.7\%)& $67.0^{+1.5}_{-1.7}$ (2.4\%) & - \\
            & AGN BAO + BBN & $0.294^{+0.025}_{-0.030}$ (9.5\%) & $67.4\pm1.4$ (2.1\%) & - \\
			& All BAO + BBN & $0.302^{+0.016}_{-0.019}$ (5.9\%) & $67.0\pm0.99$ (1.5\%) & - \\
                \hline
                \hline 
            $w$CDM \\
			\multirow{3}*{$N_{\mathrm{sys}} = 0$} & Galaxy BAO + BBN & $0.317^{+0.024}_{-0.013}$ (6.8\%)& $67.7^{+4.3}_{-3.8}$ (6\%) & $-0.99\pm0.14$ (14\%) \\
            & AGN BAO + BBN & $0.295\pm0.032$ (10.8\%) & $71.5^{+6.6}_{-7.6}$ (9.8\%) & $-1.19^{+0.34}_{-0.26}$ (24.4\%)  \\
			& All BAO + BBN & $0.314^{+0.013}_{-0.011}$ (3.9\%) & $66.7\pm2.5$ (3.7\%) & $-0.963^{+0.091}_{-0.079}$ (8.7\%) \\
                \hline
                \multirow{3}*{$N_{\mathrm{sys}} = 10^4$} & Galaxy BAO + BBN & $0.306^{+0.050}_{-0.026}$ (16.3\%) & $70.6^{+10}_{-8.2}$ (13.2\%) & $-1.16^{+0.38}_{-0.31}$ (28.6\%) \\
			& AGN BAO + BBN & $0.286\pm{0.032}$ (11.5\%) & $67.9^{+6.4}_{-7.6}$ (10.7\%) & $-1.03^{+0.31}_{-0.22}$ (26.7\%) \\
			& All BAO + BBN & $0.299\pm0.019$ (6.5\%) & $66.7\pm4.1$ (6.2\%) & $-0.995^{+0.17}_{-0.13}$ (16\%) \\
			\hline
            \hline
		\end{tabular}
		}
	\end{center}
	\label{cosmological parameters}
\end{table*}

We show the constraint results of $\alpha_{\parallel}$ and $\alpha_{\perp}$ for the CSST galaxy and AGN spectroscopic surveys in Figure~\ref{galaxyresults} and \ref{agnresults}, respectively. The best-fit values and 1$\sigma$ error of  $\alpha_{\parallel}$, $\alpha_{\perp}$, and derived $D_\mathrm{H}/r_\mathrm{drag}$ and $D_\mathrm{M}/r_\mathrm{drag}$ in each redshift bin are also listed in Table~\ref{gabao}. We investigate the constraint power with $N_{\rm sys}=0$ (red contours) and $10^4\,h^{-3}\,\mathrm{Mpc^3}$ (blue contours) as the optimistic and pessimistic cases, respectively. 

We find that the constraint precision of the BAO scaling parameters can be higher than 1\% and 3\% at $0.3<z<1.2$ for the optimistic (without $N_{\rm sys}$) and pessimistic ($N_{\rm sys}=10^4\,h^{-3}\,\mathrm{Mpc^3}$) cases in the CSST galaxy spectroscopic survey. In $z<0.3$, the precision becomes lower by a factor of $\sim2$ because of the small effective volume as shown in Figure~\ref{veff}. The constraint on $\alpha_{\perp}$ is basically better than $\alpha_{\parallel}$, due to the relatively low precision of spectroscopic redshift measured by the CSST slitless gratings. Our result is also consistent with that given by \cite{Ding2023}, where they predict the measurement precisions of $H(z)r_\mathrm{drag}$ and $D_\mathrm{M}/r_\mathrm{drag}$ in the CSST spectroscopic and photometric surveys. Compared to the current measurements, e.g. eBOSS \citep{GilMarin2020}, our result indicates that the CSST galaxy spectroscopic survey can effectively probe the BAO signal at higher redshifts up to $z\simeq1.2$. Besides, it could improve the precision of the BAO measurement by a factor of $\sim$3 at least in the optimistic case, and can achieve similar precision in the pessimistic case. 

We notice that, as shown in Figure~\ref{galaxyresults}, there is about 1$\sigma$ deviation from the fiducial values for the best-fits of the BAO scaling parameters in some redshift bins in the optimistic case (red contours). This is due to the Gaussian random shifts we add to the mock data (as shown in Figure~\ref{datagalaxy}) and degeneracies with the nuisance parameters, such as $b_{\rm g}$ and $A_{i}$. In the following discussions, we can see that this deviation is not statistically large enough to affect the constraints on the cosmological parameters.

In the CSST AGN spectroscopic survey,  the precisions of the BAO measurements can be higher than 3\% and 4\% at $0<z<3$ for $N_{\rm sys}=0$ and $10^4\,h^{-3}\,\mathrm{Mpc^3}$, respectively. We can see that the effect of including $N_{\rm sys}$ is not as large as that in the CSST galaxy survey. This is because that the AGN power spectrum is higher than the galaxy power spectrum due to the larger AGN bias as shown in Figure~\ref{dataagn}, and it cannot significantly affect the AGN power spectrum by adding a $N_{\rm sys}$ around $10^4\,h^{-3}\,\mathrm{Mpc^3}$. At $3<z<4$, the constraint precision becomes much worse by a factor of $\sim4$, which is due to a much lower AGN number density and effective volume as shown in Table~\ref{number_density} and Figure~\ref{veff}. When comparing our result with the current eBOSS measurements \citep{Neveux2020}, we can find that the CSST AGN spectroscopic survey can reach much higher redshift up to $z\sim4$, and could improve the constraints on both $\alpha_{\parallel}$ and $\alpha_{\perp}$ by a factor of $\sim2$ at least.

By making use of the mean values and covariance matrices of $\alpha_{\parallel}$ and $\alpha_{\perp}$ derived from the MCMC chains as the mock BAO data, we also explore the constraint power on the cosmological parameters in the $\Lambda$CDM and $w$CDM models. In Figure~\ref{lcdmresults} and Figure~\ref{wcdmresults}, we show the contour maps and 1D probability distribution functions (PDFs) of the relevant cosmological parameters of the two models for $N_{\rm sys}=0$ and $10^4\,h^{-3}\,\mathrm{Mpc^3}$ cases in the CSST galaxy and AGN spectroscopic surveys. 

For the $\Lambda$CDM model without considering $N_{\rm sys}$, we find that the constraint precisions of $\Omega_{\mathrm{m}}$ and $H_0$ can reach 3.4\% and 0.81\% in the CSST galaxy survey, respectively, and they are about 8\% and 2\% for the AGN survey. When combining both the galaxy and AGN data, we can further improve the constraint precisions to be 3\% and 0.75\%. After including $N_{\rm sys}=10^4\,h^{-3}\,\mathrm{Mpc^3}$, the parameter constraints become worse by a factor of $\sim$3 for the galaxy survey, $\sim$20\% worse for the AGN survey, and a factor of $\sim$2 for the joint survey.  

For the $w$CDM model, the constraint power becomes weaker for the cosmological parameters, especially for $H_0$, since an additional parameter $w$ is included. We find that, without $N_{\rm sys}$, the constraint precisions of $\Omega_{\mathrm{m}}$ and $H_0$ are 7\% and 6\% for the galaxy survey, 11\% and 10\% for the AGN survey, and 4\% and 4\% for the joint survey, respectively. When considering $N_{\rm sys}10^4\,h^{-3}\,\mathrm{Mpc^3}$, the precisions are worse by a factor of $\sim$2 for the galaxy survey, similar for the AGN survey, and 50\% worse for the joint survey. The constraint precisions of $w$ are 14\%, 24\% and 9\% for the galaxy, AGN and joint surveys, respectively, in the optimistic case without $N_{\rm sys}$, and 29\%, 27\% and 16\% when considering $N_{\rm sys}=10^4\,h^{-3}\,\mathrm{Mpc^3}$ as the pessimistic case.

We also notice that the degeneracy directions of $\Omega_{\mathrm{m}}$ vs. $H_0$ are different for the CSST galaxy and AGN surveys in both Figure~\ref{lcdmresults} and Figure~\ref{wcdmresults}. This is because that these two surveys mainly explore different redshift ranges, that $0<z\lesssim1.2$ for the galaxy survey and $0<z\lesssim4$ for the AGN survey, and the expansion of the Universe at $z\lesssim1$ and $z>1$ are different, which are basically dominated by dark energy and dark matter, respectively. This can affect the degeneracy of $\Omega_{\mathrm{m}}$ vs. $H_0$, which is also indicated by previous studies \citep[e.g.][]{Alam2021}. This indicates that the joint analysis of galaxies and AGNs can somehow break the degeneracy between parameters $w$ and $\Omega_{\mathrm{m}}$, and effectively improve the constraints. 

\section{Conclusions}\label{Sec6}

In this work, we have studied the constraints on the BAO scaling parameters $\alpha_{\parallel}$ and $\alpha_{\perp}$ from the CSST galaxy and AGN spectroscopic surveys. We first forecast and generate the observed galaxy and AGN mock data of the multipole power spectra. For the galaxy survey, we perform reconstruction to reduce the non-linear effect on the BAO signal at $0<z<1.2$. For the AGN survey, we adopt the pre-reconstruction power spectra at $0<z<4$, since the AGN number density is too low to perform the reconstruction. We equally divided the redshift range into four bins for both of the CSST galaxy and AGN surveys. We find that more than one hundred million galaxies and four million AGNs will be observed by the CSST spectroscopic surveys.

Then we applied those mock power spectrum data to derive the BAO signal based on the BAO-only analysis by constraining the BAO scaling parameters, i.e. $\alpha_{\parallel}$ and $\alpha_{\perp}$. The MCMC technique is adopted to derive the constraint results. We explore the constraint power in the optimistic and pessimistic cases by excluding and including a systematical error $N_{\rm sys}$, which can account for the instrumental effects including the success rate of spec-$z$ accuracy. We find that the constraint precisions of $\alpha_{\parallel}$ and $\alpha_{\perp}$ can be higher than 1\% and 3\% for the galaxy and AGN surveys, respectively, in the optimistic case, which can improve the precision by a factor of 2-3 compared to the current measurements in the same redshift range. On the other hand, the CSST can provide similar constraint power as the current result in the pessimistic case, but can cover higher redshifts.

We also investigate the constraints on the cosmological parameters in the $\Lambda$CDM and $w$CDM models using the derived BAO data of $\alpha_{\parallel}$ and $\alpha_{\perp}$. We find that, for the $\Lambda$CDM model, the CSST joint galaxy and AGN spectroscopic surveys can provide stringent constraints on $\Omega_{\mathrm{m}}$ and $H_0$ with precisions $\sim$3\% and 1\%, respectively, in the optimistic case, and $\sim$6\% and 1.5\% in the pessimistic case. For the $w$CDM model including the dark energy equation of state $w$, the constraint precision becomes lower, especially for $H_0$ which is worse by a factor of $\sim$4, and it is 9\% and 16\% for $w$ in the optimistic and pessimistic cases, respectively. 
In addition, we only consider the case of CSST BAO measurements assisted by a prior of $\Omega_{\rm b}h^2$ from the BBN measurements. However, the constraint precision can be further improved by incorporating additional measurements, such as those from the CMB and the other CSST cosmological probes. This would enable unprecedented precise probing of the LSS as well as the properties of dark matter and dark energy.


\section*{Acknowledgements}

We thank Gongbo Zhao and Yuting Wang for helpful discussions. HM and YG acknowledge the support from National Key R\&D Program of China grant Nos. 2022YFF0503404, 2020SKA0110402, and the CAS Project for Young Scientists in Basic Research (No. YSBR-092). XC acknowledges the support of the National Natural
Science Foundation of China through Grant Nos. 11473044 and
11973047, and the Chinese Academy of Science grants ZDKYYQ20200008, QYZDJ-SSW-SLH017, XDB 23040100, and XDA15020200. ZH acknowledges the support of the National Key R\&D Program of China
(Grant No. 2020YFC2201600), National SKA Program of China No.
2020SKA0110402, National Natural Science Foundation of China
(NSFC) under Grant No. 12073088, and Guangdong Major Project
of Basic and Applied Basic Research (Grant No. 2019B030302001).
XL acknowledges support from an NSFC grant (No. 11803094) and
the Science and Technology Program of Guangzhou, China (No.
202002030360). This work is also supported by science research
grants from the China Manned Space Project with Grant Nos. CMS-CSST-2021-B01 and CMS-CSST-2021-A01. 
The authors also acknowledge the Beijing Super Cloud
Center (BSCC) for providing HPC resources that have contributed to
the research results reported within this work.

\section*{Data Availability}

The data that support the findings of this study are available from the
corresponding author upon reasonable request.



\bibliographystyle{mnras}
\bibliography{bao} 

\begin{thebibliography}{}
\makeatletter
\relax
\def\mn@urlcharsother{\let\do\@makeother \do\$\do\&\do\#\do\^\do\_\do\%\do\~}
\def\mn@doi{\begingroup\mn@urlcharsother \@ifnextchar [ {\mn@doi@}
  {\mn@doi@[]}}
\def\mn@doi@[#1]#2{\def\@tempa{#1}\ifx\@tempa\@empty \href
  {http://dx.doi.org/#2} {doi:#2}\else \href {http://dx.doi.org/#2} {#1}\fi
  \endgroup}
\def\mn@eprint#1#2{\mn@eprint@#1:#2::\@nil}
\def\mn@eprint@arXiv#1{\href {http://arxiv.org/abs/#1} {{\tt arXiv:#1}}}
\def\mn@eprint@dblp#1{\href {http://dblp.uni-trier.de/rec/bibtex/#1.xml}
  {dblp:#1}}
\def\mn@eprint@#1:#2:#3:#4\@nil{\def\@tempa {#1}\def\@tempb {#2}\def\@tempc
  {#3}\ifx \@tempc \@empty \let \@tempc \@tempb \let \@tempb \@tempa \fi \ifx
  \@tempb \@empty \def\@tempb {arXiv}\fi \@ifundefined
  {mn@eprint@\@tempb}{\@tempb:\@tempc}{\expandafter \expandafter \csname
  mn@eprint@\@tempb\endcsname \expandafter{\@tempc}}}

\bibitem[\protect\citeauthoryear{{Akeson} et~al.,}{{Akeson}
  et~al.}{2019}]{Akeson19}
{Akeson} R.,  et~al., 2019, arXiv e-prints, \href
  {https://ui.adsabs.harvard.edu/abs/2019arXiv190205569A} {p. arXiv:1902.05569}

\bibitem[\protect\citeauthoryear{{Alam} et~al.,}{{Alam}
  et~al.}{2017}]{Alam2017}
{Alam} S.,  et~al., 2017, \mn@doi [\mnras] {10.1093/mnras/stx721}, \href
  {https://ui.adsabs.harvard.edu/abs/2017MNRAS.470.2617A} {470, 2617}

\bibitem[\protect\citeauthoryear{{Alam} et~al.,}{{Alam}
  et~al.}{2021}]{Alam2021}
{Alam} S.,  et~al., 2021, \mn@doi [\prd] {10.1103/PhysRevD.103.083533}, \href
  {https://ui.adsabs.harvard.edu/abs/2021PhRvD.103h3533A} {103, 083533}

\bibitem[\protect\citeauthoryear{{Alcock} \& {Paczynski}}{{Alcock} \&
  {Paczynski}}{1979}]{Alcock1979}
{Alcock} C.,  {Paczynski} B.,  1979, \mn@doi [\nat] {10.1038/281358a0}, \href
  {https://ui.adsabs.harvard.edu/abs/1979Natur.281..358A} {281, 358}

\bibitem[\protect\citeauthoryear{{Amendola} et~al.,}{{Amendola}
  et~al.}{2018}]{Amendola2018}
{Amendola} L.,  et~al., 2018, \mn@doi [Living Reviews in Relativity]
  {10.1007/s41114-017-0010-3}, \href
  {https://ui.adsabs.harvard.edu/abs/2018LRR....21....2A} {21, 2}

\bibitem[\protect\citeauthoryear{{Anderson} et~al.,}{{Anderson}
  et~al.}{2012}]{Anderson2012}
{Anderson} L.,  et~al., 2012, \mn@doi [\mnras]
  {10.1111/j.1365-2966.2012.22066.x}, \href
  {https://ui.adsabs.harvard.edu/abs/2012MNRAS.427.3435A} {427, 3435}

\bibitem[\protect\citeauthoryear{{Anderson} et~al.,}{{Anderson}
  et~al.}{2014}]{Anderson2014}
{Anderson} L.,  et~al., 2014, \mn@doi [\mnras] {10.1093/mnras/stu523}, \href
  {https://ui.adsabs.harvard.edu/abs/2014MNRAS.441...24A} {441, 24}

\bibitem[\protect\citeauthoryear{{Angulo}, {Baugh}, {Frenk}  \&
  {Lacey}}{{Angulo} et~al.}{2008}]{Angulo2008}
{Angulo} R.~E.,  {Baugh} C.~M.,  {Frenk} C.~S.,   {Lacey} C.~G.,  2008, \mn@doi
  [\mnras] {10.1111/j.1365-2966.2007.12587.x}, \href
  {https://ui.adsabs.harvard.edu/abs/2008MNRAS.383..755A} {383, 755}

\bibitem[\protect\citeauthoryear{{Anselmi}, {Corasaniti}, {Sanchez},
  {Starkman}, {Sheth}  \& {Zehavi}}{{Anselmi} et~al.}{2019}]{Anselmi2019}
{Anselmi} S.,  {Corasaniti} P.-S.,  {Sanchez} A.~G.,  {Starkman} G.~D.,
  {Sheth} R.~K.,   {Zehavi} I.,  2019, \mn@doi [\prd]
  {10.1103/PhysRevD.99.123515}, \href
  {https://ui.adsabs.harvard.edu/abs/2019PhRvD..99l3515A} {99, 123515}

\bibitem[\protect\citeauthoryear{{Anselmi}, {Starkman}  \& {Renzi}}{{Anselmi}
  et~al.}{2023a}]{Anselmi2023b}
{Anselmi} S.,  {Starkman} G.~D.,   {Renzi} A.,  2023a, \mn@doi [\prd]
  {10.1103/PhysRevD.107.123506}, \href
  {https://ui.adsabs.harvard.edu/abs/2022arXiv220509098A} {107, 123506}

\bibitem[\protect\citeauthoryear{{Anselmi}, {Carney}, {Giblin}, {Kumar},
  {Mertens}, {O'Dwyer}, {Starkman}  \& {Tian}}{{Anselmi}
  et~al.}{2023b}]{Anselmi2023a}
{Anselmi} S.,  {Carney} M.~F.,  {Giblin} J.~T.,  {Kumar} S.,  {Mertens} J.~B.,
  {O'Dwyer} M.,  {Starkman} G.~D.,   {Tian} C.,  2023b, \mn@doi [\jcap]
  {10.1088/1475-7516/2023/02/049}, \href
  {https://ui.adsabs.harvard.edu/abs/2023JCAP...02..049A} {2023, 049}

\bibitem[\protect\citeauthoryear{{Ata} et~al.,}{{Ata} et~al.}{2018}]{Ata2018}
{Ata} M.,  et~al., 2018, \mn@doi [\mnras] {10.1093/mnras/stx2630}, \href
  {https://ui.adsabs.harvard.edu/abs/2018MNRAS.473.4773A} {473, 4773}

\bibitem[\protect\citeauthoryear{{Aubourg} et~al.,}{{Aubourg}
  et~al.}{2015}]{Aubourg2015}
{Aubourg} {\'E}.,  et~al., 2015, \mn@doi [\prd] {10.1103/PhysRevD.92.123516},
  \href {https://ui.adsabs.harvard.edu/abs/2015PhRvD..92l3516A} {92, 123516}

\bibitem[\protect\citeauthoryear{{Ballinger}, {Peacock}  \&
  {Heavens}}{{Ballinger} et~al.}{1996}]{Ballinger1996}
{Ballinger} W.~E.,  {Peacock} J.~A.,   {Heavens} A.~F.,  1996, \mn@doi [\mnras]
  {10.1093/mnras/282.3.877}, \href
  {https://ui.adsabs.harvard.edu/abs/1996MNRAS.282..877B} {282, 877}

\bibitem[\protect\citeauthoryear{{Bargiacchi}, {Benetti}, {Capozziello},
  {Lusso}, {Risaliti}  \& {Signorini}}{{Bargiacchi}
  et~al.}{2022}]{Bargiacchi2022}
{Bargiacchi} G.,  {Benetti} M.,  {Capozziello} S.,  {Lusso} E.,  {Risaliti} G.,
    {Signorini} M.,  2022, \mn@doi [\mnras] {10.1093/mnras/stac1941}, \href
  {https://ui.adsabs.harvard.edu/abs/2022MNRAS.515.1795B} {515, 1795}

\bibitem[\protect\citeauthoryear{{Bautista} et~al.,}{{Bautista}
  et~al.}{2021}]{Bautista2021}
{Bautista} J.~E.,  et~al., 2021, \mn@doi [\mnras] {10.1093/mnras/staa2800},
  \href {https://ui.adsabs.harvard.edu/abs/2021MNRAS.500..736B} {500, 736}

\bibitem[\protect\citeauthoryear{{Bernardeau}, {Colombi}, {Gazta{\~n}aga}  \&
  {Scoccimarro}}{{Bernardeau} et~al.}{2002}]{Bernardeau2002}
{Bernardeau} F.,  {Colombi} S.,  {Gazta{\~n}aga} E.,   {Scoccimarro} R.,  2002,
  \mn@doi [\physrep] {10.1016/S0370-1573(02)00135-7}, \href
  {https://ui.adsabs.harvard.edu/abs/2002PhR...367....1B} {367, 1}

\bibitem[\protect\citeauthoryear{{Beutler}, {Blake}, {Koda}, {Mar{\'\i}n},
  {Seo}, {Cuesta}  \& {Schneider}}{{Beutler} et~al.}{2016}]{Beutler2016}
{Beutler} F.,  {Blake} C.,  {Koda} J.,  {Mar{\'\i}n} F.~A.,  {Seo} H.-J.,
  {Cuesta} A.~J.,   {Schneider} D.~P.,  2016, \mn@doi [\mnras]
  {10.1093/mnras/stv1943}, \href
  {https://ui.adsabs.harvard.edu/abs/2016MNRAS.455.3230B} {455, 3230}

\bibitem[\protect\citeauthoryear{{Beutler} et~al.,}{{Beutler}
  et~al.}{2017a}]{Beutler2017}
{Beutler} F.,  et~al., 2017a, \mn@doi [\mnras] {10.1093/mnras/stw2373}, \href
  {https://ui.adsabs.harvard.edu/abs/2017MNRAS.464.3409B} {464, 3409}

\bibitem[\protect\citeauthoryear{{Beutler} et~al.,}{{Beutler}
  et~al.}{2017b}]{Beutler2017b}
{Beutler} F.,  et~al., 2017b, \mn@doi [\mnras] {10.1093/mnras/stw3298}, \href
  {https://ui.adsabs.harvard.edu/abs/2017MNRAS.466.2242B} {466, 2242}

\bibitem[\protect\citeauthoryear{{Blas}, {Garny}, {Ivanov}  \&
  {Sibiryakov}}{{Blas} et~al.}{2016}]{Blas2016}
{Blas} D.,  {Garny} M.,  {Ivanov} M.~M.,   {Sibiryakov} S.,  2016, \mn@doi
  [\jcap] {10.1088/1475-7516/2016/07/028}, \href
  {https://ui.adsabs.harvard.edu/abs/2016JCAP...07..028B} {2016, 028}

\bibitem[\protect\citeauthoryear{{Boyle}, {Shanks}, {Croom}, {Smith}, {Miller},
  {Loaring}  \& {Heymans}}{{Boyle} et~al.}{2000}]{Boyle2000}
{Boyle} B.~J.,  {Shanks} T.,  {Croom} S.~M.,  {Smith} R.~J.,  {Miller} L.,
  {Loaring} N.,   {Heymans} C.,  2000, \mn@doi [\mnras]
  {10.1046/j.1365-8711.2000.03730.x}, \href
  {https://ui.adsabs.harvard.edu/abs/2000MNRAS.317.1014B} {317, 1014}

\bibitem[\protect\citeauthoryear{{Brieden}, {Gil-Mar{\'\i}n}  \&
  {Verde}}{{Brieden} et~al.}{2023}]{Brieden2023}
{Brieden} S.,  {Gil-Mar{\'\i}n} H.,   {Verde} L.,  2023, \mn@doi [\jcap]
  {10.1088/1475-7516/2023/04/023}, \href
  {https://ui.adsabs.harvard.edu/abs/2023JCAP...04..023B} {2023, 023}

\bibitem[\protect\citeauthoryear{{Buchert}}{{Buchert}}{1989}]{Buchert1989}
{Buchert} T.,  1989, \aap, \href
  {https://ui.adsabs.harvard.edu/abs/1989A&A...223....9B} {223, 9}

\bibitem[\protect\citeauthoryear{{Buchert}}{{Buchert}}{1992}]{Buchert1992}
{Buchert} T.,  1992, \mn@doi [\mnras] {10.1093/mnras/254.4.729}, \href
  {https://ui.adsabs.harvard.edu/abs/1992MNRAS.254..729B} {254, 729}

\bibitem[\protect\citeauthoryear{{Burden}, {Percival}, {Manera}, {Cuesta},
  {Vargas Magana}  \& {Ho}}{{Burden} et~al.}{2014}]{Burden2014}
{Burden} A.,  {Percival} W.~J.,  {Manera} M.,  {Cuesta} A.~J.,  {Vargas Magana}
  M.,   {Ho} S.,  2014, \mn@doi [\mnras] {10.1093/mnras/stu1965}, \href
  {https://ui.adsabs.harvard.edu/abs/2014MNRAS.445.3152B} {445, 3152}

\bibitem[\protect\citeauthoryear{{Caditz}}{{Caditz}}{2017}]{Caditz2017}
{Caditz} D.~M.,  2017, \mn@doi [\aap] {10.1051/0004-6361/201731850}, \href
  {https://ui.adsabs.harvard.edu/abs/2017A&A...608A..64C} {608, A64}

\bibitem[\protect\citeauthoryear{{Carlson}, {Reid}  \& {White}}{{Carlson}
  et~al.}{2013}]{Carlson2013}
{Carlson} J.,  {Reid} B.,   {White} M.,  2013, \mn@doi [\mnras]
  {10.1093/mnras/sts457}, \href
  {https://ui.adsabs.harvard.edu/abs/2013MNRAS.429.1674C} {429, 1674}

\bibitem[\protect\citeauthoryear{{Carrilho}, {Moretti}  \&
  {Pourtsidou}}{{Carrilho} et~al.}{2023}]{Carrilho2023}
{Carrilho} P.,  {Moretti} C.,   {Pourtsidou} A.,  2023, \mn@doi [\jcap]
  {10.1088/1475-7516/2023/01/028}, \href
  {https://ui.adsabs.harvard.edu/abs/2023JCAP...01..028C} {2023, 028}

\bibitem[\protect\citeauthoryear{{Chen} \& {Padmanabhan}}{{Chen} \&
  {Padmanabhan}}{2023}]{ChenX2023}
{Chen} X.,  {Padmanabhan} N.,  2023, \mn@doi [arXiv e-prints]
  {10.48550/arXiv.2311.09531}, \href
  {https://ui.adsabs.harvard.edu/abs/2023arXiv231109531C} {p. arXiv:2311.09531}

\bibitem[\protect\citeauthoryear{{Chen}, {Castorina}  \& {White}}{{Chen}
  et~al.}{2019a}]{Chen2019}
{Chen} S.-F.,  {Castorina} E.,   {White} M.,  2019a, \mn@doi [\jcap]
  {10.1088/1475-7516/2019/06/006}, \href
  {https://ui.adsabs.harvard.edu/abs/2019JCAP...06..006C} {2019, 006}

\bibitem[\protect\citeauthoryear{{Chen}, {Vlah}  \& {White}}{{Chen}
  et~al.}{2019b}]{Chen2019b}
{Chen} S.-F.,  {Vlah} Z.,   {White} M.,  2019b, \mn@doi [\jcap]
  {10.1088/1475-7516/2019/09/017}, \href
  {https://ui.adsabs.harvard.edu/abs/2019JCAP...09..017C} {2019, 017}

\bibitem[\protect\citeauthoryear{{Chen}, {Vlah}  \& {White}}{{Chen}
  et~al.}{2020}]{Chen2020}
{Chen} S.-F.,  {Vlah} Z.,   {White} M.,  2020, \mn@doi [\jcap]
  {10.1088/1475-7516/2020/07/062}, \href
  {https://ui.adsabs.harvard.edu/abs/2020JCAP...07..062C} {2020, 062}

\bibitem[\protect\citeauthoryear{{Chen}, {Vlah}, {Castorina}  \&
  {White}}{{Chen} et~al.}{2021}]{Chen2021}
{Chen} S.-F.,  {Vlah} Z.,  {Castorina} E.,   {White} M.,  2021, \mn@doi [\jcap]
  {10.1088/1475-7516/2021/03/100}, \href
  {https://ui.adsabs.harvard.edu/abs/2021JCAP...03..100C} {2021, 100}

\bibitem[\protect\citeauthoryear{{Chen}, {Vlah}  \& {White}}{{Chen}
  et~al.}{2022a}]{Chen2022b}
{Chen} S.-F.,  {Vlah} Z.,   {White} M.,  2022a, \mn@doi [\jcap]
  {10.1088/1475-7516/2022/02/008}, \href
  {https://ui.adsabs.harvard.edu/abs/2022JCAP...02..008C} {2022, 008}

\bibitem[\protect\citeauthoryear{{Chen}, {White}, {DeRose}  \& {Kokron}}{{Chen}
  et~al.}{2022b}]{Chen2022}
{Chen} S.-F.,  {White} M.,  {DeRose} J.,   {Kokron} N.,  2022b, \mn@doi [\jcap]
  {10.1088/1475-7516/2022/07/041}, \href
  {https://ui.adsabs.harvard.edu/abs/2022JCAP...07..041C} {2022, 041}

\bibitem[\protect\citeauthoryear{{Chudaykin} \& {Ivanov}}{{Chudaykin} \&
  {Ivanov}}{2019}]{Chudaykin2019}
{Chudaykin} A.,  {Ivanov} M.~M.,  2019, \mn@doi [\jcap]
  {10.1088/1475-7516/2019/11/034}, \href
  {https://ui.adsabs.harvard.edu/abs/2019JCAP...11..034C} {2019, 034}

\bibitem[\protect\citeauthoryear{{Chudaykin} \& {Ivanov}}{{Chudaykin} \&
  {Ivanov}}{2023}]{Chudaykin2023}
{Chudaykin} A.,  {Ivanov} M.~M.,  2023, \mn@doi [\prd]
  {10.1103/PhysRevD.107.043518}, \href
  {https://ui.adsabs.harvard.edu/abs/2023PhRvD.107d3518C} {107, 043518}

\bibitem[\protect\citeauthoryear{{Cole} et~al.,}{{Cole}
  et~al.}{2005}]{Cole2005}
{Cole} S.,  et~al., 2005, \mn@doi [\mnras] {10.1111/j.1365-2966.2005.09318.x},
  \href {https://ui.adsabs.harvard.edu/abs/2005MNRAS.362..505C} {362, 505}

\bibitem[\protect\citeauthoryear{{Crocce} \& {Scoccimarro}}{{Crocce} \&
  {Scoccimarro}}{2006}]{Crocce2006}
{Crocce} M.,  {Scoccimarro} R.,  2006, \mn@doi [\prd]
  {10.1103/PhysRevD.73.063520}, \href
  {https://ui.adsabs.harvard.edu/abs/2006PhRvD..73f3520C} {73, 063520}

\bibitem[\protect\citeauthoryear{{Crocce} \& {Scoccimarro}}{{Crocce} \&
  {Scoccimarro}}{2008}]{Crocce2008}
{Crocce} M.,  {Scoccimarro} R.,  2008, \mn@doi [\prd]
  {10.1103/PhysRevD.77.023533}, \href
  {https://ui.adsabs.harvard.edu/abs/2008PhRvD..77b3533C} {77, 023533}

\bibitem[\protect\citeauthoryear{{Croom} et~al.,}{{Croom}
  et~al.}{2009}]{Croom2009}
{Croom} S.~M.,  et~al., 2009, \mn@doi [\mnras]
  {10.1111/j.1365-2966.2008.14052.x}, \href
  {https://ui.adsabs.harvard.edu/abs/2009MNRAS.392...19C} {392, 19}

\bibitem[\protect\citeauthoryear{{DESI Collaboration} et~al.,}{{DESI
  Collaboration} et~al.}{2016}]{DESI2016}
{DESI Collaboration} et~al., 2016, \mn@doi [arXiv e-prints]
  {10.48550/arXiv.1611.00036}, \href
  {https://ui.adsabs.harvard.edu/abs/2016arXiv161100036D} {p. arXiv:1611.00036}

\bibitem[\protect\citeauthoryear{{DESI Collaboration} et~al.,}{{DESI
  Collaboration} et~al.}{2023}]{2023arXiv230606308D}
{DESI Collaboration} et~al., 2023, \mn@doi [arXiv e-prints]
  {10.48550/arXiv.2306.06308}, \href
  {https://ui.adsabs.harvard.edu/abs/2023arXiv230606308D} {p. arXiv:2306.06308}

\bibitem[\protect\citeauthoryear{{Dawson} et~al.,}{{Dawson}
  et~al.}{2016}]{Dawson2016}
{Dawson} K.~S.,  et~al., 2016, \mn@doi [\aj] {10.3847/0004-6256/151/2/44},
  \href {https://ui.adsabs.harvard.edu/abs/2016AJ....151...44D} {151, 44}

\bibitem[\protect\citeauthoryear{{DeRose}, {Chen}, {Kokron}  \&
  {White}}{{DeRose} et~al.}{2023}]{DeRose2023}
{DeRose} J.,  {Chen} S.-F.,  {Kokron} N.,   {White} M.,  2023, \mn@doi [\jcap]
  {10.1088/1475-7516/2023/02/008}, \href
  {https://ui.adsabs.harvard.edu/abs/2023JCAP...02..008D} {2023, 008}

\bibitem[\protect\citeauthoryear{{Desjacques}, {Jeong}  \&
  {Schmidt}}{{Desjacques} et~al.}{2018}]{Desjacques2018}
{Desjacques} V.,  {Jeong} D.,   {Schmidt} F.,  2018, \mn@doi [\physrep]
  {10.1016/j.physrep.2017.12.002}, \href
  {https://ui.adsabs.harvard.edu/abs/2018PhR...733....1D} {733, 1}

\bibitem[\protect\citeauthoryear{{Ding}, {Seo}, {Vlah}, {Feng}, {Schmittfull}
  \& {Beutler}}{{Ding} et~al.}{2018}]{Ding2018}
{Ding} Z.,  {Seo} H.-J.,  {Vlah} Z.,  {Feng} Y.,  {Schmittfull} M.,   {Beutler}
  F.,  2018, \mn@doi [\mnras] {10.1093/mnras/sty1413}, \href
  {https://ui.adsabs.harvard.edu/abs/2018MNRAS.479.1021D} {479, 1021}

\bibitem[\protect\citeauthoryear{{Ding}, {Li}, {Zheng}, {Luo}, {Zhang}  \&
  {Li}}{{Ding} et~al.}{2024a}]{DingJ2023}
{Ding} J.,  {Li} S.,  {Zheng} Y.,  {Luo} X.,  {Zhang} L.,   {Li} X.-D.,  2024a,
  \mn@doi [\apjs] {10.3847/1538-4365/ad0c5b}, \href
  {https://ui.adsabs.harvard.edu/abs/2024ApJS..270...25D} {270, 25}

\bibitem[\protect\citeauthoryear{{Ding}, {Yu}  \& {Zhang}}{{Ding}
  et~al.}{2024b}]{Ding2023}
{Ding} Z.,  {Yu} Y.,   {Zhang} P.,  2024b, \mn@doi [\mnras]
  {10.1093/mnras/stad3379}, \href
  {https://ui.adsabs.harvard.edu/abs/2024MNRAS.527.3728D} {527, 3728}

\bibitem[\protect\citeauthoryear{{Eisenstein} \& {Hu}}{{Eisenstein} \&
  {Hu}}{1998}]{Eisenstein1998}
{Eisenstein} D.~J.,  {Hu} W.,  1998, \mn@doi [\apj] {10.1086/305424}, \href
  {https://ui.adsabs.harvard.edu/abs/1998ApJ...496..605E} {496, 605}

\bibitem[\protect\citeauthoryear{{Eisenstein} et~al.,}{{Eisenstein}
  et~al.}{2005}]{Eisenstein2005}
{Eisenstein} D.~J.,  et~al., 2005, \mn@doi [\apj] {10.1086/466512}, \href
  {https://ui.adsabs.harvard.edu/abs/2005ApJ...633..560E} {633, 560}

\bibitem[\protect\citeauthoryear{{Eisenstein}, {Seo}  \& {White}}{{Eisenstein}
  et~al.}{2007a}]{Eisenstein2007}
{Eisenstein} D.~J.,  {Seo} H.-J.,   {White} M.,  2007a, \mn@doi [\apj]
  {10.1086/518755}, \href
  {https://ui.adsabs.harvard.edu/abs/2007ApJ...664..660E} {664, 660}

\bibitem[\protect\citeauthoryear{{Eisenstein}, {Seo}, {Sirko}  \&
  {Spergel}}{{Eisenstein} et~al.}{2007b}]{Eisenstein2007b}
{Eisenstein} D.~J.,  {Seo} H.-J.,  {Sirko} E.,   {Spergel} D.~N.,  2007b,
  \mn@doi [\apj] {10.1086/518712}, \href
  {https://ui.adsabs.harvard.edu/abs/2007ApJ...664..675E} {664, 675}

\bibitem[\protect\citeauthoryear{{Foroozan}, {Krolewski}  \&
  {Percival}}{{Foroozan} et~al.}{2021}]{Foroozan2021}
{Foroozan} S.,  {Krolewski} A.,   {Percival} W.~J.,  2021, \mn@doi [\jcap]
  {10.1088/1475-7516/2021/10/044}, \href
  {https://ui.adsabs.harvard.edu/abs/2021JCAP...10..044F} {2021, 044}

\bibitem[\protect\citeauthoryear{{Gil-Mar{\'\i}n}}{{Gil-Mar{\'\i}n}}{2022}]{GilMarin2022}
{Gil-Mar{\'\i}n} H.,  2022, \mn@doi [\jcap] {10.1088/1475-7516/2022/05/040},
  \href {https://ui.adsabs.harvard.edu/abs/2022JCAP...05..040G} {2022, 040}

\bibitem[\protect\citeauthoryear{{Gil-Mar{\'\i}n} et~al.,}{{Gil-Mar{\'\i}n}
  et~al.}{2016}]{GilMarin2016}
{Gil-Mar{\'\i}n} H.,  et~al., 2016, \mn@doi [\mnras] {10.1093/mnras/stw1264},
  \href {https://ui.adsabs.harvard.edu/abs/2016MNRAS.460.4210G} {460, 4210}

\bibitem[\protect\citeauthoryear{{Gil-Mar{\'\i}n} et~al.,}{{Gil-Mar{\'\i}n}
  et~al.}{2020}]{GilMarin2020}
{Gil-Mar{\'\i}n} H.,  et~al., 2020, \mn@doi [\mnras] {10.1093/mnras/staa2455},
  \href {https://ui.adsabs.harvard.edu/abs/2020MNRAS.498.2492G} {498, 2492}

\bibitem[\protect\citeauthoryear{{Gong} et~al.,}{{Gong} et~al.}{2019}]{Gong19}
{Gong} Y.,  et~al., 2019, \mn@doi [\apj] {10.3847/1538-4357/ab391e}, \href
  {https://ui.adsabs.harvard.edu/abs/2019ApJ...883..203G} {883, 203}

\bibitem[\protect\citeauthoryear{{Grieb}, {S{\'a}nchez}, {Salazar-Albornoz}  \&
  {Dalla Vecchia}}{{Grieb} et~al.}{2016}]{Grieb2016}
{Grieb} J.~N.,  {S{\'a}nchez} A.~G.,  {Salazar-Albornoz} S.,   {Dalla Vecchia}
  C.,  2016, \mn@doi [\mnras] {10.1093/mnras/stw065}, \href
  {https://ui.adsabs.harvard.edu/abs/2016MNRAS.457.1577G} {457, 1577}

\bibitem[\protect\citeauthoryear{{Hada} \& {Eisenstein}}{{Hada} \&
  {Eisenstein}}{2018}]{Hada2018}
{Hada} R.,  {Eisenstein} D.~J.,  2018, \mn@doi [\mnras]
  {10.1093/mnras/sty1203}, \href
  {https://ui.adsabs.harvard.edu/abs/2018MNRAS.478.1866H} {478, 1866}

\bibitem[\protect\citeauthoryear{{Hamilton}, {Rimes}  \&
  {Scoccimarro}}{{Hamilton} et~al.}{2006}]{Hamilton2006}
{Hamilton} A. J.~S.,  {Rimes} C.~D.,   {Scoccimarro} R.,  2006, \mn@doi
  [\mnras] {10.1111/j.1365-2966.2006.10709.x}, \href
  {https://ui.adsabs.harvard.edu/abs/2006MNRAS.371.1188H} {371, 1188}

\bibitem[\protect\citeauthoryear{{Hikage}, {Takahashi}  \& {Koyama}}{{Hikage}
  et~al.}{2020}]{Hikage2020}
{Hikage} C.,  {Takahashi} R.,   {Koyama} K.,  2020, \mn@doi [\prd]
  {10.1103/PhysRevD.102.083514}, \href
  {https://ui.adsabs.harvard.edu/abs/2020PhRvD.102h3514H} {102, 083514}

\bibitem[\protect\citeauthoryear{{Hinton} et~al.,}{{Hinton}
  et~al.}{2017}]{Hinton2017}
{Hinton} S.~R.,  et~al., 2017, \mn@doi [\mnras] {10.1093/mnras/stw2725}, \href
  {https://ui.adsabs.harvard.edu/abs/2017MNRAS.464.4807H} {464, 4807}

\bibitem[\protect\citeauthoryear{{Hivon}, {Bouchet}, {Colombi}  \&
  {Juszkiewicz}}{{Hivon} et~al.}{1995}]{Hivon1995}
{Hivon} E.,  {Bouchet} F.~R.,  {Colombi} S.,   {Juszkiewicz} R.,  1995, \mn@doi
  [\aap] {10.48550/arXiv.astro-ph/9407049}, \href
  {https://ui.adsabs.harvard.edu/abs/1995A&A...298..643H} {298, 643}

\bibitem[\protect\citeauthoryear{{Hou} et~al.,}{{Hou} et~al.}{2018}]{Hou2018}
{Hou} J.,  et~al., 2018, \mn@doi [\mnras] {10.1093/mnras/sty1984}, \href
  {https://ui.adsabs.harvard.edu/abs/2018MNRAS.480.2521H} {480, 2521}

\bibitem[\protect\citeauthoryear{{Hou} et~al.,}{{Hou} et~al.}{2021}]{Hou2021}
{Hou} J.,  et~al., 2021, \mn@doi [\mnras] {10.1093/mnras/staa3234}, \href
  {https://ui.adsabs.harvard.edu/abs/2021MNRAS.500.1201H} {500, 1201}

\bibitem[\protect\citeauthoryear{{Hou}, {Cahn}, {Philcox}  \& {Slepian}}{{Hou}
  et~al.}{2022}]{Hou2022}
{Hou} J.,  {Cahn} R.~N.,  {Philcox} O. H.~E.,   {Slepian} Z.,  2022, \mn@doi
  [\prd] {10.1103/PhysRevD.106.043515}, \href
  {https://ui.adsabs.harvard.edu/abs/2022PhRvD.106d3515H} {106, 043515}

\bibitem[\protect\citeauthoryear{{Hu} \& {Sugiyama}}{{Hu} \&
  {Sugiyama}}{1996}]{Hu1996}
{Hu} W.,  {Sugiyama} N.,  1996, \mn@doi [\apj] {10.1086/177989}, \href
  {https://ui.adsabs.harvard.edu/abs/1996ApJ...471..542H} {471, 542}

\bibitem[\protect\citeauthoryear{{Ivanov}}{{Ivanov}}{2022}]{Ivanov2022}
{Ivanov} M.~M.,  2022, \mn@doi [arXiv e-prints] {10.48550/arXiv.2212.08488},
  \href {https://ui.adsabs.harvard.edu/abs/2022arXiv221208488I} {p.
  arXiv:2212.08488}

\bibitem[\protect\citeauthoryear{{Ivanov}, {Simonovi{\'c}}  \&
  {Zaldarriaga}}{{Ivanov} et~al.}{2020}]{Ivanov2020}
{Ivanov} M.~M.,  {Simonovi{\'c}} M.,   {Zaldarriaga} M.,  2020, \mn@doi [\jcap]
  {10.1088/1475-7516/2020/05/042}, \href
  {https://ui.adsabs.harvard.edu/abs/2020JCAP...05..042I} {2020, 042}

\bibitem[\protect\citeauthoryear{{Ivezi{\'c}} et~al.,}{{Ivezi{\'c}}
  et~al.}{2019}]{Ivezic19}
{Ivezi{\'c}} {\v{Z}}.,  et~al., 2019, \mn@doi [\apj]
  {10.3847/1538-4357/ab042c}, \href
  {https://ui.adsabs.harvard.edu/abs/2019ApJ...873..111I} {873, 111}

\bibitem[\protect\citeauthoryear{{Jeong} \& {Komatsu}}{{Jeong} \&
  {Komatsu}}{2006}]{Jeong2006}
{Jeong} D.,  {Komatsu} E.,  2006, \mn@doi [\apj] {10.1086/507781}, \href
  {https://ui.adsabs.harvard.edu/abs/2006ApJ...651..619J} {651, 619}

\bibitem[\protect\citeauthoryear{{Jones} et~al.,}{{Jones}
  et~al.}{2009}]{Jones2009}
{Jones} D.~H.,  et~al., 2009, \mn@doi [\mnras]
  {10.1111/j.1365-2966.2009.15338.x}, \href
  {https://ui.adsabs.harvard.edu/abs/2009MNRAS.399..683J} {399, 683}

\bibitem[\protect\citeauthoryear{{Kazin} et~al.,}{{Kazin}
  et~al.}{2014}]{Kazin2014}
{Kazin} E.~A.,  et~al., 2014, \mn@doi [\mnras] {10.1093/mnras/stu778}, \href
  {https://ui.adsabs.harvard.edu/abs/2014MNRAS.441.3524K} {441, 3524}

\bibitem[\protect\citeauthoryear{{Kokron}, {Chen}, {White}, {DeRose}  \&
  {Maus}}{{Kokron} et~al.}{2022}]{Kokron2022}
{Kokron} N.,  {Chen} S.-F.,  {White} M.,  {DeRose} J.,   {Maus} M.,  2022,
  \mn@doi [\jcap] {10.1088/1475-7516/2022/09/059}, \href
  {https://ui.adsabs.harvard.edu/abs/2022JCAP...09..059K} {2022, 059}

\bibitem[\protect\citeauthoryear{{Laureijs} et~al.,}{{Laureijs}
  et~al.}{2011}]{Laureijs11}
{Laureijs} R.,  et~al., 2011, arXiv e-prints, \href
  {https://ui.adsabs.harvard.edu/abs/2011arXiv1110.3193L} {p. arXiv:1110.3193}

\bibitem[\protect\citeauthoryear{{Laurent} et~al.,}{{Laurent}
  et~al.}{2017}]{Laurent2017}
{Laurent} P.,  et~al., 2017, \mn@doi [\jcap] {10.1088/1475-7516/2017/07/017},
  \href {https://ui.adsabs.harvard.edu/abs/2017JCAP...07..017L} {2017, 017}

\bibitem[\protect\citeauthoryear{{Lewis}, {Challinor}  \& {Lasenby}}{{Lewis}
  et~al.}{2000}]{Lewis2000}
{Lewis} A.,  {Challinor} A.,   {Lasenby} A.,  2000, \mn@doi [\apj]
  {10.1086/309179}, \href
  {https://ui.adsabs.harvard.edu/abs/2000ApJ...538..473L} {538, 473}

\bibitem[\protect\citeauthoryear{{Lilly} et~al.,}{{Lilly}
  et~al.}{2007}]{Lilly2007}
{Lilly} S.~J.,  et~al., 2007, \mn@doi [\apjs] {10.1086/516589}, \href
  {https://ui.adsabs.harvard.edu/abs/2007ApJS..172...70L} {172, 70}

\bibitem[\protect\citeauthoryear{{Lilly} et~al.,}{{Lilly}
  et~al.}{2009}]{Lilly2009}
{Lilly} S.~J.,  et~al., 2009, \mn@doi [\apjs] {10.1088/0067-0049/184/2/218},
  \href {https://ui.adsabs.harvard.edu/abs/2009ApJS..184..218L} {184, 218}

\bibitem[\protect\citeauthoryear{{Matsubara}}{{Matsubara}}{2008a}]{Matsubara2008}
{Matsubara} T.,  2008a, \mn@doi [\prd] {10.1103/PhysRevD.77.063530}, \href
  {https://ui.adsabs.harvard.edu/abs/2008PhRvD..77f3530M} {77, 063530}

\bibitem[\protect\citeauthoryear{{Matsubara}}{{Matsubara}}{2008b}]{Matsubara2008b}
{Matsubara} T.,  2008b, \mn@doi [\prd] {10.1103/PhysRevD.78.083519}, \href
  {https://ui.adsabs.harvard.edu/abs/2008PhRvD..78h3519M} {78, 083519}

\bibitem[\protect\citeauthoryear{{Matsubara}}{{Matsubara}}{2015}]{Matsubara2015}
{Matsubara} T.,  2015, \mn@doi [\prd] {10.1103/PhysRevD.92.023534}, \href
  {https://ui.adsabs.harvard.edu/abs/2015PhRvD..92b3534M} {92, 023534}

\bibitem[\protect\citeauthoryear{{McDonald} \& {Roy}}{{McDonald} \&
  {Roy}}{2009}]{McDonald2009}
{McDonald} P.,  {Roy} A.,  2009, \mn@doi [\jcap]
  {10.1088/1475-7516/2009/08/020}, \href
  {https://ui.adsabs.harvard.edu/abs/2009JCAP...08..020M} {2009, 020}

\bibitem[\protect\citeauthoryear{{McGreer} et~al.,}{{McGreer}
  et~al.}{2013}]{McGreer2013}
{McGreer} I.~D.,  et~al., 2013, \mn@doi [\apj] {10.1088/0004-637X/768/2/105},
  \href {https://ui.adsabs.harvard.edu/abs/2013ApJ...768..105M} {768, 105}

\bibitem[\protect\citeauthoryear{{McQuinn} \& {White}}{{McQuinn} \&
  {White}}{2016}]{McQuinn2016}
{McQuinn} M.,  {White} M.,  2016, \mn@doi [\jcap]
  {10.1088/1475-7516/2016/01/043}, \href
  {https://ui.adsabs.harvard.edu/abs/2016JCAP...01..043M} {2016, 043}

\bibitem[\protect\citeauthoryear{{Meiksin}, {White}  \& {Peacock}}{{Meiksin}
  et~al.}{1999}]{Meiksin1999}
{Meiksin} A.,  {White} M.,   {Peacock} J.~A.,  1999, \mn@doi [\mnras]
  {10.1046/j.1365-8711.1999.02369.x}, \href
  {https://ui.adsabs.harvard.edu/abs/1999MNRAS.304..851M} {304, 851}

\bibitem[\protect\citeauthoryear{{Merz} et~al.,}{{Merz}
  et~al.}{2021}]{Merz2021}
{Merz} G.,  et~al., 2021, \mn@doi [\mnras] {10.1093/mnras/stab1887}, \href
  {https://ui.adsabs.harvard.edu/abs/2021MNRAS.506.2503M} {506, 2503}

\bibitem[\protect\citeauthoryear{{Miao}, {Gong}, {Chen}, {Huang}, {Li}  \&
  {Zhan}}{{Miao} et~al.}{2023}]{Miao2023}
{Miao} H.,  {Gong} Y.,  {Chen} X.,  {Huang} Z.,  {Li} X.-D.,   {Zhan} H.,
  2023, \mn@doi [\mnras] {10.1093/mnras/stac3583}, \href
  {https://ui.adsabs.harvard.edu/abs/2023MNRAS.519.1132M} {519, 1132}

\bibitem[\protect\citeauthoryear{{Mohammad} \& {Percival}}{{Mohammad} \&
  {Percival}}{2022}]{Mohammad2022}
{Mohammad} F.~G.,  {Percival} W.~J.,  2022, \mn@doi [\mnras]
  {10.1093/mnras/stac1458}, \href
  {https://ui.adsabs.harvard.edu/abs/2022MNRAS.514.1289M} {514, 1289}

\bibitem[\protect\citeauthoryear{{Mohammed} \& {Seljak}}{{Mohammed} \&
  {Seljak}}{2014}]{Mohammed2014}
{Mohammed} I.,  {Seljak} U.,  2014, \mn@doi [\mnras] {10.1093/mnras/stu1972},
  \href {https://ui.adsabs.harvard.edu/abs/2014MNRAS.445.3382M} {445, 3382}

\bibitem[\protect\citeauthoryear{{Mohammed}, {Seljak}  \& {Vlah}}{{Mohammed}
  et~al.}{2017}]{Mohammed2017}
{Mohammed} I.,  {Seljak} U.,   {Vlah} Z.,  2017, \mn@doi [\mnras]
  {10.1093/mnras/stw3196}, \href
  {https://ui.adsabs.harvard.edu/abs/2017MNRAS.466..780M} {466, 780}

\bibitem[\protect\citeauthoryear{{Neveux} et~al.,}{{Neveux}
  et~al.}{2020}]{Neveux2020}
{Neveux} R.,  et~al., 2020, \mn@doi [\mnras] {10.1093/mnras/staa2780}, \href
  {https://ui.adsabs.harvard.edu/abs/2020MNRAS.499..210N} {499, 210}

\bibitem[\protect\citeauthoryear{{Neveux} et~al.,}{{Neveux}
  et~al.}{2022}]{Neveux2022}
{Neveux} R.,  et~al., 2022, \mn@doi [\mnras] {10.1093/mnras/stac2114}, \href
  {https://ui.adsabs.harvard.edu/abs/2022MNRAS.516.1910N} {516, 1910}

\bibitem[\protect\citeauthoryear{{Nikakhtar}, {Sheth}  \& {Zehavi}}{{Nikakhtar}
  et~al.}{2021}]{Nikakhtar2021}
{Nikakhtar} F.,  {Sheth} R.~K.,   {Zehavi} I.,  2021, \mn@doi [\prd]
  {10.1103/PhysRevD.104.043530}, \href
  {https://ui.adsabs.harvard.edu/abs/2021PhRvD.104d3530N} {104, 043530}

\bibitem[\protect\citeauthoryear{{Nikakhtar}, {Sheth}, {L{\'e}vy}  \&
  {Mohayaee}}{{Nikakhtar} et~al.}{2022}]{Nikakhtar2022}
{Nikakhtar} F.,  {Sheth} R.~K.,  {L{\'e}vy} B.,   {Mohayaee} R.,  2022, \mn@doi
  [\prl] {10.1103/PhysRevLett.129.251101}, \href
  {https://ui.adsabs.harvard.edu/abs/2022PhRvL.129y1101N} {129, 251101}

\bibitem[\protect\citeauthoryear{{Nikakhtar}, {Padmanabhan}, {L{\'e}vy},
  {Sheth}  \& {Mohayaee}}{{Nikakhtar} et~al.}{2023}]{Nikakhtar2023}
{Nikakhtar} F.,  {Padmanabhan} N.,  {L{\'e}vy} B.,  {Sheth} R.~K.,   {Mohayaee}
  R.,  2023, \mn@doi [\prd] {10.1103/PhysRevD.108.083534}, \href
  {https://ui.adsabs.harvard.edu/abs/2023PhRvD.108h3534N} {108, 083534}

\bibitem[\protect\citeauthoryear{{Noh}, {White}  \& {Padmanabhan}}{{Noh}
  et~al.}{2009}]{Noh2009}
{Noh} Y.,  {White} M.,   {Padmanabhan} N.,  2009, \mn@doi [\prd]
  {10.1103/PhysRevD.80.123501}, \href
  {https://ui.adsabs.harvard.edu/abs/2009PhRvD..80l3501N} {80, 123501}

\bibitem[\protect\citeauthoryear{{O'Connell} \& {Eisenstein}}{{O'Connell} \&
  {Eisenstein}}{2019}]{OConnell2019}
{O'Connell} R.,  {Eisenstein} D.~J.,  2019, \mn@doi [\mnras]
  {10.1093/mnras/stz1359}, \href
  {https://ui.adsabs.harvard.edu/abs/2019MNRAS.487.2701O} {487, 2701}

\bibitem[\protect\citeauthoryear{{O'Dwyer}, {Anselmi}, {Starkman},
  {Corasaniti}, {Sheth}  \& {Zehavi}}{{O'Dwyer} et~al.}{2020}]{ODwyer2020}
{O'Dwyer} M.,  {Anselmi} S.,  {Starkman} G.~D.,  {Corasaniti} P.-S.,  {Sheth}
  R.~K.,   {Zehavi} I.,  2020, \mn@doi [\prd] {10.1103/PhysRevD.101.083517},
  \href {https://ui.adsabs.harvard.edu/abs/2020PhRvD.101h3517O} {101, 083517}

\bibitem[\protect\citeauthoryear{{Ota}, {Seo}, {Saito}  \& {Beutler}}{{Ota}
  et~al.}{2021}]{Ota2021}
{Ota} A.,  {Seo} H.-J.,  {Saito} S.,   {Beutler} F.,  2021, \mn@doi [\prd]
  {10.1103/PhysRevD.104.123508}, \href
  {https://ui.adsabs.harvard.edu/abs/2021PhRvD.104l3508O} {104, 123508}

\bibitem[\protect\citeauthoryear{{Ota}, {Seo}, {Saito}  \& {Beutler}}{{Ota}
  et~al.}{2023}]{Ota2023}
{Ota} A.,  {Seo} H.-J.,  {Saito} S.,   {Beutler} F.,  2023, \mn@doi [\prd]
  {10.1103/PhysRevD.107.123523}, \href
  {https://ui.adsabs.harvard.edu/abs/2023PhRvD.107l3523O} {107, 123523}

\bibitem[\protect\citeauthoryear{{Padmanabhan}, {White}  \&
  {Cohn}}{{Padmanabhan} et~al.}{2009}]{Padmanabhan2009}
{Padmanabhan} N.,  {White} M.,   {Cohn} J.~D.,  2009, \mn@doi [\prd]
  {10.1103/PhysRevD.79.063523}, \href
  {https://ui.adsabs.harvard.edu/abs/2009PhRvD..79f3523P} {79, 063523}

\bibitem[\protect\citeauthoryear{{Padmanabhan}, {Xu}, {Eisenstein}, {Scalzo},
  {Cuesta}, {Mehta}  \& {Kazin}}{{Padmanabhan} et~al.}{2012}]{Padmanabhan2012}
{Padmanabhan} N.,  {Xu} X.,  {Eisenstein} D.~J.,  {Scalzo} R.,  {Cuesta} A.~J.,
   {Mehta} K.~T.,   {Kazin} E.,  2012, \mn@doi [\mnras]
  {10.1111/j.1365-2966.2012.21888.x}, \href
  {https://ui.adsabs.harvard.edu/abs/2012MNRAS.427.2132P} {427, 2132}

\bibitem[\protect\citeauthoryear{{Palanque-Delabrouille}
  et~al.,}{{Palanque-Delabrouille} et~al.}{2016}]{Palanque2016a}
{Palanque-Delabrouille} N.,  et~al., 2016, \mn@doi [\aap]
  {10.1051/0004-6361/201527392}, \href
  {https://ui.adsabs.harvard.edu/abs/2016A&A...587A..41P} {587, A41}

\bibitem[\protect\citeauthoryear{{Parkinson} et~al.,}{{Parkinson}
  et~al.}{2012}]{Parkinson2012}
{Parkinson} D.,  et~al., 2012, \mn@doi [\prd] {10.1103/PhysRevD.86.103518},
  \href {https://ui.adsabs.harvard.edu/abs/2012PhRvD..86j3518P} {86, 103518}

\bibitem[\protect\citeauthoryear{{Peebles} \& {Yu}}{{Peebles} \&
  {Yu}}{1970}]{Peebles1970}
{Peebles} P.~J.~E.,  {Yu} J.~T.,  1970, \mn@doi [\apj] {10.1086/150713}, \href
  {https://ui.adsabs.harvard.edu/abs/1970ApJ...162..815P} {162, 815}

\bibitem[\protect\citeauthoryear{{Percival}, {Cole}, {Eisenstein}, {Nichol},
  {Peacock}, {Pope}  \& {Szalay}}{{Percival} et~al.}{2007}]{Percival2007}
{Percival} W.~J.,  {Cole} S.,  {Eisenstein} D.~J.,  {Nichol} R.~C.,  {Peacock}
  J.~A.,  {Pope} A.~C.,   {Szalay} A.~S.,  2007, \mn@doi [\mnras]
  {10.1111/j.1365-2966.2007.12268.x}, \href
  {https://ui.adsabs.harvard.edu/abs/2007MNRAS.381.1053P} {381, 1053}

\bibitem[\protect\citeauthoryear{{Philcox} \& {Ivanov}}{{Philcox} \&
  {Ivanov}}{2022}]{Philcox2022c}
{Philcox} O. H.~E.,  {Ivanov} M.~M.,  2022, \mn@doi [\prd]
  {10.1103/PhysRevD.105.043517}, \href
  {https://ui.adsabs.harvard.edu/abs/2022PhRvD.105d3517P} {105, 043517}

\bibitem[\protect\citeauthoryear{{Philcox} \& {Slepian}}{{Philcox} \&
  {Slepian}}{2022}]{Philcox2022}
{Philcox} O. H.~E.,  {Slepian} Z.,  2022, \mn@doi [Proceedings of the National
  Academy of Science] {10.1073/pnas.2111366119}, \href
  {https://ui.adsabs.harvard.edu/abs/2022PNAS..11911366P} {119, e2111366119}

\bibitem[\protect\citeauthoryear{{Philcox}, {Eisenstein}, {O'Connell}  \&
  {Wiegand}}{{Philcox} et~al.}{2020a}]{Philcox2020}
{Philcox} O. H.~E.,  {Eisenstein} D.~J.,  {O'Connell} R.,   {Wiegand} A.,
  2020a, \mn@doi [\mnras] {10.1093/mnras/stz3218}, \href
  {https://ui.adsabs.harvard.edu/abs/2020MNRAS.491.3290P} {491, 3290}

\bibitem[\protect\citeauthoryear{{Philcox}, {Ivanov}, {Simonovi{\'c}}  \&
  {Zaldarriaga}}{{Philcox} et~al.}{2020b}]{Philcox2020b}
{Philcox} O. H.~E.,  {Ivanov} M.~M.,  {Simonovi{\'c}} M.,   {Zaldarriaga} M.,
  2020b, \mn@doi [\jcap] {10.1088/1475-7516/2020/05/032}, \href
  {https://ui.adsabs.harvard.edu/abs/2020JCAP...05..032P} {2020, 032}

\bibitem[\protect\citeauthoryear{{Philcox}, {Sherwin}, {Farren}  \&
  {Baxter}}{{Philcox} et~al.}{2021}]{Philcox2021b}
{Philcox} O. H.~E.,  {Sherwin} B.~D.,  {Farren} G.~S.,   {Baxter} E.~J.,  2021,
  \mn@doi [\prd] {10.1103/PhysRevD.103.023538}, \href
  {https://ui.adsabs.harvard.edu/abs/2021PhRvD.103b3538P} {103, 023538}

\bibitem[\protect\citeauthoryear{{Philcox}, {Farren}, {Sherwin}, {Baxter}  \&
  {Brout}}{{Philcox} et~al.}{2022}]{Philcox2022b}
{Philcox} O. H.~E.,  {Farren} G.~S.,  {Sherwin} B.~D.,  {Baxter} E.~J.,
  {Brout} D.~J.,  2022, \mn@doi [\prd] {10.1103/PhysRevD.106.063530}, \href
  {https://ui.adsabs.harvard.edu/abs/2022PhRvD.106f3530P} {106, 063530}

\bibitem[\protect\citeauthoryear{{Planck Collaboration} et~al.,}{{Planck
  Collaboration} et~al.}{2020}]{Planck2020}
{Planck Collaboration} et~al., 2020, \mn@doi [\aap]
  {10.1051/0004-6361/201833910}, \href
  {https://ui.adsabs.harvard.edu/abs/2020A&A...641A...6P} {641, A6}

\bibitem[\protect\citeauthoryear{{Raichoor} et~al.,}{{Raichoor}
  et~al.}{2021}]{Raichoor2021}
{Raichoor} A.,  et~al., 2021, \mn@doi [\mnras] {10.1093/mnras/staa3336}, \href
  {https://ui.adsabs.harvard.edu/abs/2021MNRAS.500.3254R} {500, 3254}

\bibitem[\protect\citeauthoryear{{Richards} et~al.,}{{Richards}
  et~al.}{2006}]{Richards2006}
{Richards} G.~T.,  et~al., 2006, \mn@doi [\aj] {10.1086/503559}, \href
  {https://ui.adsabs.harvard.edu/abs/2006AJ....131.2766R} {131, 2766}

\bibitem[\protect\citeauthoryear{{Ross}, {Samushia}, {Howlett}, {Percival},
  {Burden}  \& {Manera}}{{Ross} et~al.}{2015}]{Ross2015}
{Ross} A.~J.,  {Samushia} L.,  {Howlett} C.,  {Percival} W.~J.,  {Burden} A.,
  {Manera} M.,  2015, \mn@doi [\mnras] {10.1093/mnras/stv154}, \href
  {https://ui.adsabs.harvard.edu/abs/2015MNRAS.449..835R} {449, 835}

\bibitem[\protect\citeauthoryear{{S{\'a}nchez}, {Crocce}, {Cabr{\'e}}, {Baugh}
  \& {Gazta{\~n}aga}}{{S{\'a}nchez} et~al.}{2009}]{Sanchez2009}
{S{\'a}nchez} A.~G.,  {Crocce} M.,  {Cabr{\'e}} A.,  {Baugh} C.~M.,
  {Gazta{\~n}aga} E.,  2009, \mn@doi [\mnras]
  {10.1111/j.1365-2966.2009.15572.x}, \href
  {https://ui.adsabs.harvard.edu/abs/2009MNRAS.400.1643S} {400, 1643}

\bibitem[\protect\citeauthoryear{{S{\'a}nchez} et~al.,}{{S{\'a}nchez}
  et~al.}{2012}]{Sanchez2012}
{S{\'a}nchez} A.~G.,  et~al., 2012, \mn@doi [\mnras]
  {10.1111/j.1365-2966.2012.21502.x}, \href
  {https://ui.adsabs.harvard.edu/abs/2012MNRAS.425..415S} {425, 415}

\bibitem[\protect\citeauthoryear{{Schmidt}}{{Schmidt}}{2021}]{Schmidt2021}
{Schmidt} F.,  2021, \mn@doi [\jcap] {10.1088/1475-7516/2021/04/033}, \href
  {https://ui.adsabs.harvard.edu/abs/2021JCAP...04..033S} {2021, 033}

\bibitem[\protect\citeauthoryear{{Schmittfull}, {Feng}, {Beutler}, {Sherwin}
  \& {Chu}}{{Schmittfull} et~al.}{2015}]{Schmittfull2015}
{Schmittfull} M.,  {Feng} Y.,  {Beutler} F.,  {Sherwin} B.,   {Chu} M.~Y.,
  2015, \mn@doi [\prd] {10.1103/PhysRevD.92.123522}, \href
  {https://ui.adsabs.harvard.edu/abs/2015PhRvD..92l3522S} {92, 123522}

\bibitem[\protect\citeauthoryear{{Schmittfull}, {Baldauf}  \&
  {Zaldarriaga}}{{Schmittfull} et~al.}{2017}]{Schmittfull2017}
{Schmittfull} M.,  {Baldauf} T.,   {Zaldarriaga} M.,  2017, \mn@doi [\prd]
  {10.1103/PhysRevD.96.023505}, \href
  {https://ui.adsabs.harvard.edu/abs/2017PhRvD..96b3505S} {96, 023505}

\bibitem[\protect\citeauthoryear{{Sch{\"o}neberg}, {Lesgourgues}  \&
  {Hooper}}{{Sch{\"o}neberg} et~al.}{2019}]{Schoneberg2019}
{Sch{\"o}neberg} N.,  {Lesgourgues} J.,   {Hooper} D.~C.,  2019, \mn@doi
  [\jcap] {10.1088/1475-7516/2019/10/029}, \href
  {https://ui.adsabs.harvard.edu/abs/2019JCAP...10..029S} {2019, 029}

\bibitem[\protect\citeauthoryear{{Sch{\"o}neberg}, {Verde}, {Gil-Mar{\'\i}n}
  \& {Brieden}}{{Sch{\"o}neberg} et~al.}{2022}]{Schoneberg2022}
{Sch{\"o}neberg} N.,  {Verde} L.,  {Gil-Mar{\'\i}n} H.,   {Brieden} S.,  2022,
  \mn@doi [\jcap] {10.1088/1475-7516/2022/11/039}, \href
  {https://ui.adsabs.harvard.edu/abs/2022JCAP...11..039S} {2022, 039}

\bibitem[\protect\citeauthoryear{{Semenaite} et~al.,}{{Semenaite}
  et~al.}{2023}]{Semenaite2023}
{Semenaite} A.,  et~al., 2023, \mn@doi [\mnras] {10.1093/mnras/stad849}, \href
  {https://ui.adsabs.harvard.edu/abs/2023MNRAS.521.5013S} {521, 5013}

\bibitem[\protect\citeauthoryear{{Senatore} \& {Zaldarriaga}}{{Senatore} \&
  {Zaldarriaga}}{2015}]{Senatore2015}
{Senatore} L.,  {Zaldarriaga} M.,  2015, \mn@doi [\jcap]
  {10.1088/1475-7516/2015/02/013}, \href
  {https://ui.adsabs.harvard.edu/abs/2015JCAP...02..013S} {2015, 013}

\bibitem[\protect\citeauthoryear{{Seo} \& {Eisenstein}}{{Seo} \&
  {Eisenstein}}{2005}]{Seo2005}
{Seo} H.-J.,  {Eisenstein} D.~J.,  2005, \mn@doi [\apj] {10.1086/491599}, \href
  {https://ui.adsabs.harvard.edu/abs/2005ApJ...633..575S} {633, 575}

\bibitem[\protect\citeauthoryear{{Seo}, {Siegel}, {Eisenstein}  \&
  {White}}{{Seo} et~al.}{2008}]{Seo2008}
{Seo} H.-J.,  {Siegel} E.~R.,  {Eisenstein} D.~J.,   {White} M.,  2008, \mn@doi
  [\apj] {10.1086/589921}, \href
  {https://ui.adsabs.harvard.edu/abs/2008ApJ...686...13S} {686, 13}

\bibitem[\protect\citeauthoryear{{Seo}, {Ota}, {Schmittfull}, {Saito}  \&
  {Beutler}}{{Seo} et~al.}{2022}]{Seo2022}
{Seo} H.-J.,  {Ota} A.,  {Schmittfull} M.,  {Saito} S.,   {Beutler} F.,  2022,
  \mn@doi [\mnras] {10.1093/mnras/stac082}, \href
  {https://ui.adsabs.harvard.edu/abs/2022MNRAS.511.1557S} {511, 1557}

\bibitem[\protect\citeauthoryear{{Sherwin} \& {Zaldarriaga}}{{Sherwin} \&
  {Zaldarriaga}}{2012}]{Sherwin2012}
{Sherwin} B.~D.,  {Zaldarriaga} M.,  2012, \mn@doi [\prd]
  {10.1103/PhysRevD.85.103523}, \href
  {https://ui.adsabs.harvard.edu/abs/2012PhRvD..85j3523S} {85, 103523}

\bibitem[\protect\citeauthoryear{{Simon}, {Zhang}, {Poulin}  \&
  {Smith}}{{Simon} et~al.}{2023a}]{Simon2023}
{Simon} T.,  {Zhang} P.,  {Poulin} V.,   {Smith} T.~L.,  2023a, \mn@doi [\prd]
  {10.1103/PhysRevD.107.123530}, \href
  {https://ui.adsabs.harvard.edu/abs/2023PhRvD.107l3530S} {107, 123530}

\bibitem[\protect\citeauthoryear{{Simon}, {Zhang}  \& {Poulin}}{{Simon}
  et~al.}{2023b}]{Simon2022}
{Simon} T.,  {Zhang} P.,   {Poulin} V.,  2023b, \mn@doi [\jcap]
  {10.1088/1475-7516/2023/07/041}, \href
  {https://ui.adsabs.harvard.edu/abs/2023JCAP...07..041S} {2023, 041}

\bibitem[\protect\citeauthoryear{{Smith} et~al.,}{{Smith}
  et~al.}{2020}]{Smith2020}
{Smith} A.,  et~al., 2020, \mn@doi [\mnras] {10.1093/mnras/staa2825}, \href
  {https://ui.adsabs.harvard.edu/abs/2020MNRAS.499..269S} {499, 269}

\bibitem[\protect\citeauthoryear{{Springel} et~al.,}{{Springel}
  et~al.}{2005}]{Springel2005}
{Springel} V.,  et~al., 2005, \mn@doi [\nat] {10.1038/nature03597}, \href
  {https://ui.adsabs.harvard.edu/abs/2005Natur.435..629S} {435, 629}

\bibitem[\protect\citeauthoryear{{Sunyaev} \& {Zeldovich}}{{Sunyaev} \&
  {Zeldovich}}{1970}]{Sunyaev1970}
{Sunyaev} R.~A.,  {Zeldovich} Y.~B.,  1970, \mn@doi [\apss]
  {10.1007/BF00653471}, \href
  {https://ui.adsabs.harvard.edu/abs/1970Ap&SS...7....3S} {7, 3}

\bibitem[\protect\citeauthoryear{{Takahashi} et~al.,}{{Takahashi}
  et~al.}{2009}]{Takahashi2009}
{Takahashi} R.,  et~al., 2009, \mn@doi [\apj] {10.1088/0004-637X/700/1/479},
  \href {https://ui.adsabs.harvard.edu/abs/2009ApJ...700..479T} {700, 479}

\bibitem[\protect\citeauthoryear{{Tamone} et~al.,}{{Tamone}
  et~al.}{2020}]{Tamone2020}
{Tamone} A.,  et~al., 2020, \mn@doi [\mnras] {10.1093/mnras/staa3050}, \href
  {https://ui.adsabs.harvard.edu/abs/2020MNRAS.499.5527T} {499, 5527}

\bibitem[\protect\citeauthoryear{{Taruya}, {Nishimichi}, {Saito}  \&
  {Hiramatsu}}{{Taruya} et~al.}{2009}]{Taruya2009}
{Taruya} A.,  {Nishimichi} T.,  {Saito} S.,   {Hiramatsu} T.,  2009, \mn@doi
  [\prd] {10.1103/PhysRevD.80.123503}, \href
  {https://ui.adsabs.harvard.edu/abs/2009PhRvD..80l3503T} {80, 123503}

\bibitem[\protect\citeauthoryear{{Taruya}, {Nishimichi}  \& {Jeong}}{{Taruya}
  et~al.}{2021}]{Taruya2021}
{Taruya} A.,  {Nishimichi} T.,   {Jeong} D.,  2021, \mn@doi [\prd]
  {10.1103/PhysRevD.103.023501}, \href
  {https://ui.adsabs.harvard.edu/abs/2021PhRvD.103b3501T} {103, 023501}

\bibitem[\protect\citeauthoryear{{Tassev}}{{Tassev}}{2014a}]{Tassev2014b}
{Tassev} S.,  2014a, \mn@doi [\jcap] {10.1088/1475-7516/2014/06/008}, \href
  {https://ui.adsabs.harvard.edu/abs/2014JCAP...06..008T} {2014, 008}

\bibitem[\protect\citeauthoryear{{Tassev}}{{Tassev}}{2014b}]{Tassev2014}
{Tassev} S.,  2014b, \mn@doi [\jcap] {10.1088/1475-7516/2014/06/012}, \href
  {https://ui.adsabs.harvard.edu/abs/2014JCAP...06..012T} {2014, 012}

\bibitem[\protect\citeauthoryear{{Taylor} \& {Hamilton}}{{Taylor} \&
  {Hamilton}}{1996}]{Taylor1996}
{Taylor} A.~N.,  {Hamilton} A.~J.~S.,  1996, \mn@doi [\mnras]
  {10.1093/mnras/282.3.767}, \href
  {https://ui.adsabs.harvard.edu/abs/1996MNRAS.282..767T} {282, 767}

\bibitem[\protect\citeauthoryear{{Torrado} \& {Lewis}}{{Torrado} \&
  {Lewis}}{2021}]{Torrado2021}
{Torrado} J.,  {Lewis} A.,  2021, \mn@doi [\jcap]
  {10.1088/1475-7516/2021/05/057}, \href
  {https://ui.adsabs.harvard.edu/abs/2021JCAP...05..057T} {2021, 057}

\bibitem[\protect\citeauthoryear{{Vlah} \& {White}}{{Vlah} \&
  {White}}{2019}]{Vlah2019}
{Vlah} Z.,  {White} M.,  2019, \mn@doi [\jcap] {10.1088/1475-7516/2019/03/007},
  \href {https://ui.adsabs.harvard.edu/abs/2019JCAP...03..007V} {2019, 007}

\bibitem[\protect\citeauthoryear{{Vlah}, {Seljak}  \& {Baldauf}}{{Vlah}
  et~al.}{2015a}]{Vlah2015b}
{Vlah} Z.,  {Seljak} U.,   {Baldauf} T.,  2015a, \mn@doi [\prd]
  {10.1103/PhysRevD.91.023508}, \href
  {https://ui.adsabs.harvard.edu/abs/2015PhRvD..91b3508V} {91, 023508}

\bibitem[\protect\citeauthoryear{{Vlah}, {White}  \& {Aviles}}{{Vlah}
  et~al.}{2015b}]{Vlah2015}
{Vlah} Z.,  {White} M.,   {Aviles} A.,  2015b, \mn@doi [\jcap]
  {10.1088/1475-7516/2015/09/014}, \href
  {https://ui.adsabs.harvard.edu/abs/2015JCAP...09..014V} {2015, 014}

\bibitem[\protect\citeauthoryear{{Vlah}, {Seljak}, {Yat Chu}  \& {Feng}}{{Vlah}
  et~al.}{2016a}]{Vlah2016b}
{Vlah} Z.,  {Seljak} U.,  {Yat Chu} M.,   {Feng} Y.,  2016a, \mn@doi [\jcap]
  {10.1088/1475-7516/2016/03/057}, \href
  {https://ui.adsabs.harvard.edu/abs/2016JCAP...03..057V} {2016, 057}

\bibitem[\protect\citeauthoryear{{Vlah}, {Castorina}  \& {White}}{{Vlah}
  et~al.}{2016b}]{Vlah2016}
{Vlah} Z.,  {Castorina} E.,   {White} M.,  2016b, \mn@doi [\jcap]
  {10.1088/1475-7516/2016/12/007}, \href
  {https://ui.adsabs.harvard.edu/abs/2016JCAP...12..007V} {2016, 007}

\bibitem[\protect\citeauthoryear{{Wadekar} \& {Scoccimarro}}{{Wadekar} \&
  {Scoccimarro}}{2020}]{Wadekar2020}
{Wadekar} D.,  {Scoccimarro} R.,  2020, \mn@doi [\prd]
  {10.1103/PhysRevD.102.123517}, \href
  {https://ui.adsabs.harvard.edu/abs/2020PhRvD.102l3517W} {102, 123517}

\bibitem[\protect\citeauthoryear{{Wadekar}, {Ivanov}  \&
  {Scoccimarro}}{{Wadekar} et~al.}{2020}]{Wadekar2020b}
{Wadekar} D.,  {Ivanov} M.~M.,   {Scoccimarro} R.,  2020, \mn@doi [\prd]
  {10.1103/PhysRevD.102.123521}, \href
  {https://ui.adsabs.harvard.edu/abs/2020PhRvD.102l3521W} {102, 123521}

\bibitem[\protect\citeauthoryear{{Wang}, {Yu}, {Zhu}, {Yu}, {Pan}  \&
  {Pen}}{{Wang} et~al.}{2017}]{Wang2017}
{Wang} X.,  {Yu} H.-R.,  {Zhu} H.-M.,  {Yu} Y.,  {Pan} Q.,   {Pen} U.-L.,
  2017, \mn@doi [\apjl] {10.3847/2041-8213/aa738c}, \href
  {https://ui.adsabs.harvard.edu/abs/2017ApJ...841L..29W} {841, L29}

\bibitem[\protect\citeauthoryear{{Wang}, {Li}  \& {Cautun}}{{Wang}
  et~al.}{2020a}]{Wang2020}
{Wang} Y.,  {Li} B.,   {Cautun} M.,  2020a, \mn@doi [\mnras]
  {10.1093/mnras/staa2136}, \href
  {https://ui.adsabs.harvard.edu/abs/2020MNRAS.497.3451W} {497, 3451}

\bibitem[\protect\citeauthoryear{{Wang} et~al.,}{{Wang}
  et~al.}{2020b}]{Wangy2020}
{Wang} Y.,  et~al., 2020b, \mn@doi [\mnras] {10.1093/mnras/staa2593}, \href
  {https://ui.adsabs.harvard.edu/abs/2020MNRAS.498.3470W} {498, 3470}

\bibitem[\protect\citeauthoryear{{Weinberg}, {Mortonson}, {Eisenstein},
  {Hirata}, {Riess}  \& {Rozo}}{{Weinberg} et~al.}{2013}]{Weinberg2013}
{Weinberg} D.~H.,  {Mortonson} M.~J.,  {Eisenstein} D.~J.,  {Hirata} C.,
  {Riess} A.~G.,   {Rozo} E.,  2013, \mn@doi [\physrep]
  {10.1016/j.physrep.2013.05.001}, \href
  {https://ui.adsabs.harvard.edu/abs/2013PhR...530...87W} {530, 87}

\bibitem[\protect\citeauthoryear{{White}}{{White}}{2005}]{White2005}
{White} M.,  2005, \mn@doi [Astroparticle Physics]
  {10.1016/j.astropartphys.2005.07.007}, \href
  {https://ui.adsabs.harvard.edu/abs/2005APh....24..334W} {24, 334}

\bibitem[\protect\citeauthoryear{{White}}{{White}}{2014}]{White2014}
{White} M.,  2014, \mn@doi [\mnras] {10.1093/mnras/stu209}, \href
  {https://ui.adsabs.harvard.edu/abs/2014MNRAS.439.3630W} {439, 3630}

\bibitem[\protect\citeauthoryear{{White}}{{White}}{2015}]{White2015}
{White} M.,  2015, \mn@doi [\mnras] {10.1093/mnras/stv842}, \href
  {https://ui.adsabs.harvard.edu/abs/2015MNRAS.450.3822W} {450, 3822}

\bibitem[\protect\citeauthoryear{{Yu}, {Zhu}  \& {Pen}}{{Yu}
  et~al.}{2017}]{Yu2017}
{Yu} Y.,  {Zhu} H.-M.,   {Pen} U.-L.,  2017, \mn@doi [\apj]
  {10.3847/1538-4357/aa89e7}, \href
  {https://ui.adsabs.harvard.edu/abs/2017ApJ...847..110Y} {847, 110}

\bibitem[\protect\citeauthoryear{{Zel'dovich}}{{Zel'dovich}}{1970}]{Zeldovich1970}
{Zel'dovich} Y.~B.,  1970, \aap, \href
  {https://ui.adsabs.harvard.edu/abs/1970A&A.....5...84Z} {5, 84}

\bibitem[\protect\citeauthoryear{{Zhan}}{{Zhan}}{2011}]{Zhan2011SSPMA}
{Zhan} H.,  2011, \mn@doi [Scientia Sinica Physica, Mechanica \& Astronomica]
  {10.1360/132011-961}, \href
  {https://ui.adsabs.harvard.edu/abs/2011SSPMA..41.1441Z} {41, 1441}

\bibitem[\protect\citeauthoryear{{Zhan}}{{Zhan}}{2018}]{Zhan2018}
{Zhan} H.,  2018, in 42nd COSPAR Scientific Assembly. pp E1.16--4--18

\bibitem[\protect\citeauthoryear{{Zhan}}{{Zhan}}{2021}]{Zhan2021}
{Zhan} H.,  2021, Chinese Science Bulletin, 66, 1290

\bibitem[\protect\citeauthoryear{{Zhang}, {D'Amico}, {Senatore}, {Zhao}  \&
  {Cai}}{{Zhang} et~al.}{2022}]{Zhang2022}
{Zhang} P.,  {D'Amico} G.,  {Senatore} L.,  {Zhao} C.,   {Cai} Y.,  2022,
  \mn@doi [\jcap] {10.1088/1475-7516/2022/02/036}, \href
  {https://ui.adsabs.harvard.edu/abs/2022JCAP...02..036Z} {2022, 036}

\bibitem[\protect\citeauthoryear{{Zhao} et~al.,}{{Zhao}
  et~al.}{2021}]{Zhao2021}
{Zhao} G.-B.,  et~al., 2021, \mn@doi [\mnras] {10.1093/mnras/stab849}, \href
  {https://ui.adsabs.harvard.edu/abs/2021MNRAS.504...33Z} {504, 33}

\bibitem[\protect\citeauthoryear{{Zhao} et~al.,}{{Zhao}
  et~al.}{2023}]{zhao2023}
{Zhao} R.,  et~al., 2023, \mn@doi [arXiv e-prints] {10.48550/arXiv.2308.06206},
  \href {https://ui.adsabs.harvard.edu/abs/2023arXiv230806206Z} {p.
  arXiv:2308.06206}

\bibitem[\protect\citeauthoryear{{Zhu}, {Yu}, {Pen}, {Chen}  \& {Yu}}{{Zhu}
  et~al.}{2017}]{Zhu2017}
{Zhu} H.-M.,  {Yu} Y.,  {Pen} U.-L.,  {Chen} X.,   {Yu} H.-R.,  2017, \mn@doi
  [\prd] {10.1103/PhysRevD.96.123502}, \href
  {https://ui.adsabs.harvard.edu/abs/2017PhRvD..96l3502Z} {96, 123502}

\bibitem[\protect\citeauthoryear{{d'Amico}, {Gleyzes}, {Kokron}, {Markovic},
  {Senatore}, {Zhang}, {Beutler}  \& {Gil-Mar{\'\i}n}}{{d'Amico}
  et~al.}{2020}]{dAmico2020}
{d'Amico} G.,  {Gleyzes} J.,  {Kokron} N.,  {Markovic} K.,  {Senatore} L.,
  {Zhang} P.,  {Beutler} F.,   {Gil-Mar{\'\i}n} H.,  2020, \mn@doi [\jcap]
  {10.1088/1475-7516/2020/05/005}, \href
  {https://ui.adsabs.harvard.edu/abs/2020JCAP...05..005D} {2020, 005}

\bibitem[\protect\citeauthoryear{{de Mattia} et~al.,}{{de Mattia}
  et~al.}{2021}]{deMattia2021}
{de Mattia} A.,  et~al., 2021, \mn@doi [\mnras] {10.1093/mnras/staa3891}, \href
  {https://ui.adsabs.harvard.edu/abs/2021MNRAS.501.5616D} {501, 5616}

\bibitem[\protect\citeauthoryear{{von Hausegger}, {L{\'e}vy}  \&
  {Mohayaee}}{{von Hausegger} et~al.}{2022}]{Hausegger2022}
{von Hausegger} S.,  {L{\'e}vy} B.,   {Mohayaee} R.,  2022, \mn@doi [\prl]
  {10.1103/PhysRevLett.128.201302}, \href
  {https://ui.adsabs.harvard.edu/abs/2022PhRvL.128t1302V} {128, 201302}

\makeatother
\end{thebibliography}






\bsp	
\label{lastpage}
\end{document}